\documentclass[sn-basic,iicol]{sn-jnl}% Default with double column layout

\jyear{2023 }%

\theoremstyle{thmstyleone}%
\theoremstyle{thmstyletwo}%

\theoremstyle{thmstylethree}%
\usepackage{subcaption}
\usepackage{graphicx}
\usepackage{amsmath,amssymb}
\DeclareMathOperator*{\argmin}{arg\,min}

\DeclareMathOperator{\E}{\mathop{\mathbb{E}}}
\DeclareMathOperator{\Var}{\mathop{\mathbb{V}ar}}
\DeclareMathOperator{\Cov}{\mathop{\mathbb{C}ov}}

\DeclareMathOperator{\vecth}{vech}
\DeclareMathOperator{\diago}{diag}
\newcommand{\argminD}{\argmin} 
\algblock[Initialization]{Initialization}{End}
\algblock[Andif]{an}{otherwise}
\usepackage{bbm}
\usepackage{float}

\begin{document}

\title[Article Title]{Stochastic Variational Inference for GARCH Models}

%%=============================================================%%
%% Prefix	-> \pfx{Dr}
%% GivenName	-> \fnm{Joergen W.}
%% Particle	-> \spfx{van der} -> surname prefix
%% FamilyName	-> \sur{Ploeg}
%% Suffix	-> \sfx{IV}
%% NatureName	-> \tanm{Poet Laureate} -> Title after name
%% Degrees	-> \dgr{MSc, PhD}
%% \author*[1,2]{\pfx{Dr} \fnm{Joergen W.} \spfx{van der} \sur{Ploeg} \sfx{IV} \tanm{Poet Laureate} 
%%                 \dgr{MSc, PhD}}\email{iauthor@gmail.com}
%%=============================================================%%

\author*[1]{\fnm{Hanwen} \sur{Xuan}}\email{t.xuan@unsw.edu.au}

\author[2]{\fnm{Luca} \sur{Maestrini}}\email{luca.maestrini@anu.edu.au}
\equalcont{These authors contributed equally.}

\author[1,3]{\fnm{Feng} \sur{Chen}}\email{feng.chen@unsw.edu.au}
\equalcont{These authors contributed equally.}

\author[4,5]{\fnm{Clara} \sur{Grazian}}\email{clara.grazian@sydney.edu.au}
\equalcont{These authors contributed equally.}

\affil*[1]{\orgdiv{School of Mathematics and Statistics}, \orgname{University of New South Wales}, \orgaddress{\street{Anita B. Lawrence Centre}, \city{Kensington}, \postcode{2052}, \state{New South Wales}, \country{Australia}}}

\affil[2]{\orgdiv{Research School of Finance, Actuarial Studies and Statistics}, \orgname{The Australian National University}, \orgaddress{\street{Building 26C, Kingsley Street}, \city{Canberra}, \postcode{2601}, \state{ACT}, \country{Australia}}}

\affil[3]{\orgdiv{UNSW Data Science Hub}, \orgname{University of New South Wales}, \orgaddress{\street{Anita B. Lawrence Centre}, \city{Kensington}, \postcode{2052}, \state{NSW}, \country{Australia}}}

\affil[4]{\orgdiv{School of Mathematics and Statistics}, \orgname{University of Sydney}, \orgaddress{\street{Carslaw Building}, \city{Camperdown}, \postcode{2006}, \state{NSW}, \country{Australia}}}

\affil[5]{\orgdiv{ARC Training Centre in Data Analytics for Resources and Environments}, \orgname{University of Sydney}, \orgaddress{\street{ Biomedical Building}, \city{South Eveleigh}, \postcode{2015}, \state{NSW}, \country{Australia}}}

\abstract{Stochastic variational inference algorithms are derived for fitting various heteroskedastic time series models. We examine Gaussian, t, and skew-t response GARCH models and fit these using Gaussian variational approximating densities. We implement efficient stochastic gradient ascent procedures based on the use of control variates or the reparameterization trick and demonstrate that the proposed implementations provide a fast and accurate alternative to Markov chain Monte Carlo sampling. Additionally, we present sequential updating versions of our variational algorithms, which are suitable for efficient portfolio construction and dynamic asset allocation. }


%%\pacs[JEL Classification]{D8, H51}

%%\pacs[MSC Classification]{35A01, 65L10, 65L12, 65L20, 65L70}

\maketitle

\section{Introduction}\label{sec1}

Most financial decisions are based on the trade-off between risk and return (\citeauthor{10.2307/2975974}, \citeyear{10.2307/2975974}). Modeling volatility, which is often used to measure risk, is therefore crucial in risk management. Traditional econometric models assume that the variance of asset returns is constant, but this assumption may be unrealistic as it overlooks the presence of heteroskedasticity in financial time series (\citeauthor{engle2001garch}, \citeyear{engle2001garch}). To address this issue, \cite{engle1982autoregressive} introduced auto-regressive conditional heteroskedastic (ARCH) models, which utilize lagged disturbance terms to explain a time-varying conditional variance process. ARCH models were extended to the generalized ARCH (GARCH) class by \cite{bollerslev1986generalized} to allow for a more flexible lag structure where the current conditional variance depends not only on the past conditional variances but also on the past disturbance terms. Currently, GARCH models are widely employed to describe the dynamics of volatility processes; however, conventional ARCH and GARCH models assume Gaussian errors, which fail to capture common features of financial time series such as heavy-tailedness and skewness. To improve both goodness-of-fit and prediction accuracy, \cite{bollerslev1987conditionally} and \cite{lambert2001modelling} respectively studied the use of t and skew-t distributions for modeling error terms (see also \citeauthor{degiannakis2004volatility}, \citeyear{degiannakis2004volatility}).

Maximum likelihood estimation is typically used to estimate the parameters of GARCH models \citep{gonzalez1999efficiency}, as it guarantees ease of implementation and asymptotic efficiency \citep{silvennoinen2021consistency}. However, uncertainty quantification of maximum likelihood estimates is usually performed by taking advantage of their asymptotic normality \citep{nakatsuma2000bayesian}, which may lead to underestimating uncertainty and model risk as a consequence. Bayesian inference offers an attractive alternative as it naturally expresses the uncertainty around the estimation process by returning a posterior probability distribution for the model parameters. In a Bayesian context, \cite{nakatsuma2000bayesian} proposed a Markov chain Monte Carlo (MCMC) method with Metropolis-Hasting steps for fitting GARCH models. \cite{ardia2008financial} used MCMC to fit regime-switching GARCH models with t innovations. A fully automated MCMC procedure for Bayesian estimation in GARCH models with t innovations is available in the \texttt{R} package \texttt{bayesGARCH} \citep{ardia2010bayesian}. \cite{li2021efficient} propose a sequential Monte Carlo algorithm for Student t GARCH models. While several MCMC methods are available for GARCH models \citep{gerlach2006mcmc, henneke2011mcmc, virbickaite2015bayesian, contino2017bayesian, xaba2021performance}, these may suffer from slow convergence and be computationally intensive, in particular for long time series \citep{shiferaw2019time, mozumder2021option,livingston2022bayesian}.

Variational Bayes (VB) methods \citep{jordan1998introduction} have emerged as a more computationally efficient alternative to MCMC. Early research on variational inference focused on deriving algorithms with closed-form updates for conjugate models -- see, for example, \cite{bishop2006pattern}, \cite{ormerod2010explaining} or \cite{maestrini2018variational} for variational inference in skew-t regression through conjugate model representations. Over recent years, stochastic gradient methods for variational inference (e.g., \citeauthor{paisley2012variational}, \citeyear{paisley2012variational}) have found wider interest, especially for non-conjugate models and as an alternative to avoid lengthy mathematical derivations.

The performance of stochastic optimization algorithms, including stochastic VB, greatly depends on the variance of gradients and this has led to the development of computational solutions to reduce it. \cite{ranganath2014black} introduced a technique based on \textit{control variates} to improve the convergence of stochastic VB. This method requires minimal analytic calculations and can be applied to a variety of models. \cite{kingma2013auto} proposed an alternative variance reduction technique named \textit{reparametrization trick}, which consists in rewriting an expectation so that the distribution with respect to which the gradient is taken is independent of the model parameters. Another way of efficiently using the gradient of the model joint probability density to increase computational speed is presented in \cite{titsias2014doubly} and other approaches are discussed in \cite{tran2021practical}.

Variational inference has found limited attention in the literature on GARCH models. This is partially due to the characteristics of time series, which represent an obstacle to scalable algorithm implementations. \cite{hoffman2013stochastic} developed a scalable stochastic variational inference algorithm to analyze massive datasets for complex topic models by subsampling the data to form noisy estimates. However, there is no clear and unbiased way to subsample time series data for GARCH  models \citep{politis2003impact, alonso2006comparison}. \cite{tran2021variational} proposed a manifold-based variational Bayes approach for GARCH models with Gaussian error terms that maps the variational parameters from the Euclidean space to a Riemannian manifold, but its extension to models with non-Gaussian errors is not immediate.

In recent years there have been several attempts to develop sequential variational inference approaches for time series models, i.e., implementations of variational algorithms that update inference as soon as new data points of the series are available. \cite{lambert2022recursive} proposed recursive Gaussian variational approximations for models assuming observations are independent, which are therefore not suitable for GARCH models and time-dependent data. \cite{tomasetti2022updating} developed an updating VB (UVB) method as a sequential extension of \textit{batch stochastic variational approximations} for streaming data \citep{broderick2013streaming, foti2014stochastic}. This method is based on deriving a sequence of posterior distributions that are used as priors for the following steps through updates resulting from Bayes' theorem. Each posterior update only requires data observed in the previous step. While very efficient, UVB targets the approximating pseudo-posterior rather than the true posterior, resulting in reduced accuracy. An alternative sequential implementation is the one proposed by \cite{gunawan2021variational}, which targets the true posterior distribution instead. Their approach is named sequential stochastic VB (Seq-SVB) and it is used to perform prediction with factor stochastic volatility models. A potential drawback of Seq-SVB is that the likelihood function is computed for all the time series, affecting computational efficiency.

This work includes two main contributions. First, we develop stochastic variational Bayes (SVB) algorithms to approximate the posterior distributions of parameters in GARCH models with Gaussian, t and skew t errors (also known as innovations in the econometric literature). Our variational approximations are compared to an MCMC target. We assess the performance of variational inference  using control variates or the reparametrization trick in terms of computational speed and estimation accuracy. Results demonstrate that our stochastic variational implementations are accurate while being faster than MCMC.
Our second contribution is the adaptation of the stochastic VB approaches to allow for sequential updates. We implement the procedures developed by \cite{gunawan2021variational} and \cite{tomasetti2022updating} and compare them for the first time, in the specific case of GARCH models. Our results demonstrate that both UVB and Seq-SVB are faster than the base SVB and show the presence of a trade-off between accuracy and computational costs.

The remainder of the paper is organized as follows. Section~ \ref{sec2} reviews GARCH models, variational Bayes approximations and their stochastic versions. Section~\ref{sec3} develops VB algorithms for GARCH models with different innovation terms. Section~\ref{sec4} proposes sequential versions of the variational algorithms from the previous section for efficiently updating posterior distributions when new observations are available. Section~\ref{sec:simu} presents the results of our simulation experiments and Section~\ref{sec6} illustrates the use of VB for real-world financial time series. Section~\ref{sec:concl} concludes the paper with a discussion.

\section{Stochastic Variational Bayes for GARCH Models}
\label{sec2}

\subsection{GARCH Models}
 
Let $y_1,\ldots,y_T$ be a time series, for instance the log-difference of stock prices at two consecutive time points, i.e., $y_t = \log P_t - \log P_{t-1}$. The GARCH$(p,q)$ model for the $t$th observation, $t=1,\ldots,T$, takes the form
\begin{equation}
\label{eqn:Garch}
\begin{array}{c}
    y_t = \sigma_t\varepsilon_t,\quad\varepsilon_t \sim \mathcal{N}(0,1),\\[1.5ex]
    \sigma_t^2 = \omega + \sum_{i=1}^p \alpha_i y_{t-i}^2+ \sum_{j=1}^q \beta_j \sigma_{t-j}^2.
\end{array}
\end{equation}
The time-varying volatility process for $\sigma_t^2$ follows a deterministic autoregressive structure and here we implicitly assume the dependence of this structure on the $\alpha_i$ and $\beta_j$ parameters writing $\sigma_t^2$ instead of $\sigma_t^2(\alpha_1,\ldots,\alpha_p,\beta_1,\ldots,\beta_q)$. The success of this class of models lies in the fact that the log-likelihood function $\ell(\theta) = \log L(\theta;y)$ of $\theta = (\omega,\alpha_1, \ldots, \alpha_p, \beta_1, \ldots, \beta_q)$ is tractable and has the following expression for model \eqref{eqn:Garch}:
\begin{align}
    \ell(\theta)=-\frac{1}{2}\sum_{t=1}^T \left\{\log (2\pi)+\log(\sigma_t^2)+\frac{y_t^2}{\sigma_t^2}\right\}.
    \label{eqn: llk Gaussian}
\end{align}

In real applications it has been widely recognized that financial returns exhibit significant deviations from normality, and typically have heavier tails than the normal distribution \citep{shirvani2020stock}. To address this issue, GARCH models have been extended to include t error terms \citep{bollerslev1987conditionally, baillie1989common, beine2002central}. This is achieved by replacing the normal distribution assumption in \eqref{eqn:Garch} with $\varepsilon_t \sim \mathcal{T}(0,1;\nu)$, where $\mathcal{T}(0,1;\nu)$ represents a standardized t distribution with degrees of freedom $\nu$. A random variable $x$ has a  standardized t distribution $\mathcal{T}(0,1;\nu)$ with degrees of freedom $\nu>2$ if its probability density function is
$$f(x)=\frac{\Gamma\Big(\frac{\nu+1}{2}\Big)}{\sqrt{\pi(\nu-2)}\Gamma\Big(\frac{\nu}{2}\Big)}\left(1+\frac{x^2}{\nu-2}\right)^{-\frac{\nu+1}{2}}.$$
The log-likelihood function for the t GARCH$(p,q)$ model defined by the parameter vector $\theta = (\nu,\omega,\alpha_1, \ldots, \alpha_p, \beta_1, \ldots, \beta_q)$ is
\begin{align}
\begin{split}
    \ell(\theta)&=  T\bigg\{\log \Gamma\Big(\frac{\nu+1}{2}\Big)-\log\Gamma\Big(\frac{\nu}{2}\Big)\\
    & \quad-\frac{1}{2}\log\big\{\pi(\nu-2)\big\}\bigg\} -\frac{1}{2}\sum_{t=1}^T\bigg[\log(\sigma_t^2) \\  
    & \quad+ (\nu+1)\log\left\{1+\frac{y_t^2}{(\nu-2)\sigma_t^2}\right\}\bigg].
\end{split}
    \label{eqn:llk std}
\end{align}

Skewness is a pervasive feature observed in empirical data which can be taken into account in financial time series modeling and inference \citep{jondeau2019average}. \cite{lambert2001modelling} incorporated the skew t distribution introduced by \cite{fernandez1998bayesian} in GARCH models to account for skewness.
Skew-t distributions are highly flexible and introduce asymmetry in the data model by adding only one parameter to the t distribution. In the corresponding model, the normal distribution assumption in \eqref{eqn:Garch} is replaced by $\varepsilon_t \sim \mathcal{ST}(0,1;\nu,\xi)$, where the skewness parameter $\xi$ is a positive real number. A value of $\xi<1$ indicates negative skewness, while a value of $\xi>1$ indicates positive skewness. We follow the framework and definition provided by \cite{fernandez1998bayesian}: a random variable $x$ has a skewed t distribution $\mathcal{ST}(0,1;\nu,\xi)$ with degrees of freedom $\nu>2$ and skewness parameter $\xi>0$ if its probability density function is
\begin{align*}
    f(x)&=\frac{2s\Gamma\Big(\frac{\nu+1}{2}\Big)}{\Big(\xi+\frac{1}{\xi}\Big)\sqrt{\pi(\nu-2)}\Gamma\Big(\frac{\nu}{2}\Big)}\\
    &\quad\times\left\{1+\frac{(s x+m)^2}{\nu-2}\xi^{-2I_t}\right\}^{-\frac{\nu+1}{2}},
\end{align*}
where 
\begin{align*}
    I_t &= \begin{cases}
        1,\quad \text{if} \quad x \geq -m/s,\\
        -1, \quad \text{if} \quad x < -m/s,
    \end{cases} \\
        m &= \Gamma\left(\frac{\nu-1}{2}\right)\Big\{\Gamma\left(\frac{\nu}{2}\right)\Big\}^{-1}\frac{\sqrt{\nu-2}}{\sqrt{\pi}}\left(\xi-1/\xi\right)\\
    \mbox{and }s &= \sqrt{\left(\xi^2+1/\xi^2-1\right)-m^2}.
\end{align*} 
Note that the density function has been standardized to have a zero mean and unit variance. Defining the vector of parameters as $\theta = (\nu,\xi,\omega,\alpha_1, \ldots, \alpha_p, \beta_1, \ldots, \beta_q)$, the log-likelihood for the skewed t GARCH model with $T$ observations is given by
\begin{align}
\begin{split}
    \ell (\theta) &= T\bigg\{\log \Gamma\Big(\frac{\nu+1}{2}\Big)-\log\Gamma\Big(\frac{\nu}{2}\Big) \\
    & \quad-\frac{1}{2}\log\big\{\pi(\nu-2)\big\}+\log\left(\frac{2}{\xi+1/\xi}\right) \\
    & \quad+\log(s)\bigg\} -\frac{1}{2}\sum_{t=1}^T\bigg[\log(\sigma_t^2)+(1+\nu) \\ &\quad\log\left\{1+\frac{(sy_t/\sigma_t+m)^2}{\nu-2}\xi^{-2I_t}\right\}\bigg].
\end{split}
    \label{eqn: llk sstd}
\end{align}
While the likelihood is still tractable, inference based on MCMC may be computationally costly in this case, and convergence slow \citep{iqbal2021bayesian}. 

\subsection{Variational Inference}

Let $y$ denote a vector of observed data, and let $\theta$ be a vector of parameters. In a Bayesian setting, given a prior distribution $p(\theta)$ and a likelihood function $L(\theta; y)$, the prior beliefs about $\theta$ are updated after collecting the data to obtain the posterior distribution:
\begin{align*}
   p({\theta}\vert{y}) = \frac{L(\theta; y)p({\theta})}{p({y})} \propto L(\theta; y)p({\theta}),
\end{align*}
where $p({y})$ is the marginal likelihood, or evidence. In practice, the posterior distribution in Bayesian inference may be intractable, as its normalizing factor, or evidence, $p({y})$, is often not available in closed form. Consequently, simulation-based Markov chain Monte Carlo (MCMC) methods are typically employed to estimate the posterior distribution through sampling \citep{robert1999monte}. However, MCMC sampling can be computationally expensive or impractical for complex models. An alternative approach is provided by variational Bayes (VB), which aims to approximate the intractable posterior distribution using a variational approximating density $q(\theta)$ selected from a family of densities $\mathcal{Q}$. In VB, the approximating density is chosen to minimize the Kullback-Leibler (KL) divergence between the variational approximating density and the true posterior density.
\begin{align*}
    q^{*}_{\lambda}(\theta)&= \argminD_{q_{\lambda} \in \mathcal{Q}, \lambda\in\mathcal{M}}\text{KL}\big\{q_{\lambda}(\theta) \| p(\theta \vert y)\big\} \\
    &= \argminD_{q_{\lambda} \in \mathcal{Q}, \lambda\in\mathcal{M}} \int q_{\lambda}(\theta) \log \frac{q_{\lambda}(\theta)}{p(\theta\vert y)} d\theta ,
\end{align*}
where $\lambda$ denotes a vector of parameters of the variational approximating density in the parameter space $\mathcal{M}$. Minimization of the KL divergence is equivalent to maximizing the evidence lower bound (ELBO)  $\mathcal{L}(\lambda)$ (see, e.g., \citeauthor{bishop2006pattern}, \citeyear{bishop2006pattern} or \citeauthor{ranganath2014black}, \citeyear{ranganath2014black}), which is defined as 
\begin{align}
    \label{eqn:ELBO}
    \mathcal{L}(\lambda) = \int q_{\lambda}(\theta) \log \frac{L(\theta;y)p(\theta)}{q_{\lambda}(\theta)} d\theta.
\end{align}
In general, $\mathcal{L}(\lambda)$ does not have a closed-form solution in non-conjugate models. Stochastic variational inference approaches based on stochastic gradient methods (\citeauthor{paisley2012variational}, \citeyear{paisley2012variational}; \citeauthor{nott2012regression}, \citeyear{nott2012regression};
\citeauthor{hoffman2013stochastic}, \citeyear{hoffman2013stochastic}; \citeauthor{kingma2013auto}, \citeyear{kingma2013auto}; \citeauthor{ranganath2014black}, \citeyear{ranganath2014black}; \citeauthor{rezende2014stochastic}, \citeyear{rezende2014stochastic}) allow for efficient optimization of $\mathcal{L}(\lambda)$, even when the ELBO cannot be evaluated analytically. In stochastic variational inference an initial value of the variational parameters $\lambda^{(0)}$ is stated and iteratively updated through
\begin{equation*}
    \lambda^{(i+1)} = \lambda^{(i)}+a_i \widehat{\nabla_\lambda \mathcal{L}(\lambda^{(i)})}
\end{equation*}
until a stopping rule is satisfied. Here $\widehat{\nabla_\lambda \mathcal{L}(\lambda^{(i)})}$ denotes an unbiased estimator of the gradient vector of $\mathcal{L}(\lambda)$ computed with respect to $\lambda$ and $\{a_i,i \geq 1\}$ is a sequence of vector-valued learning rates satisfying the Robbins-Monro conditions $\sum_i a_i = \infty$ and $\sum_i a^2_{i} < \infty$ \citep{robbins1951stochastic}. The topic of adaptive learning rates has been widely discussed in the literature (e.g., \citeauthor{polyak1964some}, \citeyear{polyak1964some}; \citeauthor{duchi2011adaptive}, \citeyear{duchi2011adaptive}; \citeauthor{tieleman2012lecture}, \citeyear{tieleman2012lecture}; \citeauthor{zeiler2012adadelta}, \citeyear{zeiler2012adadelta}). In this work, we adopt ADAM adaptive learning rates \citep{kingma2014adam} and a moving average stopping criterion (see \citeauthor{tran2021practical}, \citeyear{tran2021practical} for more details).

It is possible to show that the gradient vector can be written in an expectation form computed with respect to the variational density $q_\lambda(\theta)$ as 
\begin{align}
\label{eqn:gradient elbo}
\nabla_{\lambda} \mathcal{L}(\lambda) = \E_{q_\lambda}\big\{h_{\lambda}(\theta)\nabla_\lambda \log q_{\lambda}(\theta)\big\},
\end{align}
where $h_{\lambda}(\theta)= \log\big\{ p(\theta)L(\theta;y)\big/q_{\lambda}(\theta)\big\}$ (e.g., see \citeauthor{tran2021practical}, \citeyear{tran2021practical}). A simple estimator of the $j$-th element of the ELBO gradient is then given by
\begin{align}
\label{eqn:gradientElboNaiveEstimator}
\widehat{\nabla_{\lambda_j} \mathcal{L}(\lambda)} = \frac{1}{S}\sum^{S}_{s=1}h_{\lambda}(\theta_s)\nabla_{\lambda_j} \log q_{\lambda}(\theta_s),
\end{align}
where $s=1,\ldots,S$ are samples from the variational distribution $q_\lambda(\theta)$.

In this work, we consider a popular form of VB known as Gaussian variational approximation, which assumes $q_\lambda(\theta)$ to be a multivariate Gaussian distribution with mean $\mu$ and covariance matrix $\Sigma$. Several authors have studied and discussed the parameterization of Gaussian variational approximations; see, for example, \citeauthor{opper2009variational}, \citeyear{opper2009variational}; \citeauthor{ormerod2012gaussian}, \citeyear{ormerod2012gaussian}; \citeauthor{ong2018gaussian}, \citeyear{ong2018gaussian}; \citeauthor{quiroz2022gaussian}, \citeyear{quiroz2022gaussian}. We adopt the Cholesky factorization suggested by \cite{titsias2014doubly} and factorize the covariance matrix as $\Sigma = LL^\top$ where $L$ is a lower triangular matrix. Consequently, the true posterior distribution $p(\theta\vert y)$ can be approximated by a $\mathcal{N}_{d}(\mu, LL^\top)$ distribution; in this case $\lambda=(\mu^\top,\mbox{vech}(L)^\top)^\top$, where $\mbox{vech}(\cdot)$ is the half-vectorization operator. Introducing this factorization reduces the computational cost to reach convergence while preserving the positive definiteness of the covariance matrix $\Sigma$. %Samples can be easily generated from a standard normal distribution and transformed through a deterministic function giving
%$\theta=g_\lambda(\epsilon)=g_{\mu,L}(\epsilon)=\mu+L\epsilon$, where $\epsilon \sim \mathcal{N}_{d}(0,I_d)$ and $I_d$ is the $d\times d$ identity matrix. 

The performance of stochastic variational inference greatly depends on the variance of the gradient estimator \citep{liu2019variance}. \cite{paisley2012variational} and \cite{ranganath2014black} adopt a variance reduction technique based on the use of control variates, where the gradient estimator function is replaced by another function with equivalent expectation but smaller variance. For any arbitrary number $c_j$, the $j$-th element of the estimator version of \eqref{eqn:gradientElboNaiveEstimator} can be slightly modified to
\begin{small}
\begin{equation}
    \widehat{\nabla_{\lambda_j} \mathcal{L}(\lambda)} = \frac{1}{S}\sum^{S}_{s=1}\left(h_{\lambda}(\theta_s)-c_j\right)\nabla_{\lambda_j} \log q_{\lambda}(\theta_s), 
    \label{eqn:gradient elbo with cv}
\end{equation}
\end{small}
which is still an unbiased estimator of the gradient of $\mathcal{L}(\lambda)$ (\citeauthor{ranganath2014black}, \citeyear{ranganath2014black}; \citeauthor{tran2021practical}, \citeyear{tran2021practical}).
The optimal control variate $c_j$ that minimizes the variance can be calculated by taking the derivatives of the variance of the gradient estimator with respect to $c_j$:
\begin{small}
\begin{align}
    c_j = \frac{\Cov\big\{\nabla_{\lambda_j} \log q_{\lambda}(\theta),h_{\lambda}(\theta)\nabla_{\lambda_j}\log q_{\lambda}(\theta)\big\}}{\Var\big\{\nabla_{\lambda_j} \log q_{\lambda}(\theta)\big\}},
    \label{eqn: CV}
\end{align}
\end{small}
where $j = 1, \ldots, d$.

The reparametrization trick (\citeauthor{kingma2014adam}, \citeyear{kingma2014adam}; \citeauthor{rezende2014stochastic}, \citeyear{rezende2014stochastic}; \citeauthor{titsias2014doubly}, \citeyear{titsias2014doubly}) offers an attractive alternative to the control variates approach. Suppose samples are meant to be generated from $\theta \sim q_\lambda(\theta)$, then it is often possible to write $\theta=g_\lambda(\epsilon)$, where $\epsilon$ is an auxiliary variable with an independent marginal distribution $p(\epsilon)$ and $g_\lambda(\epsilon)$ is a deterministic function parameterized by $\lambda$. Using this deterministic transformation, it is possible to rewrite $\mathcal{L} (\lambda)$ in \eqref{eqn:ELBO} as an expectation with respect to the alternative measure $p(\epsilon)$, i.e. as
\begin{align*}
    \mathcal{L}(\lambda)=\E_{q_\lambda(\theta)}\big\{h_\lambda(\theta)\big\}=\E_{p(\epsilon)}\Big[h_\lambda\big\{g_\lambda(\epsilon)\big\}\Big].
\end{align*}
Then the gradient of the evidence lower bound $\nabla_\lambda\mathcal{L}(\lambda)$ can be calculated with respect to the measure $p(\epsilon)$ 
\begin{align}
    \label{eqn:grad ELBO different measure}
    \nabla_\lambda\mathcal{L}(\lambda) = \E_{p(\epsilon)}\left
    \{\nabla_\lambda g_\lambda(\epsilon)^\top \nabla_\theta h_\lambda(\theta)\right\}
\end{align}
(see \citeauthor{tran2021practical}, \citeyear{tran2021practical} for details). The reparametrization trick is used to express the gradient of the expectation as an expectation of a gradient. Provided that $g_{\lambda}$ is differentiable, then samples can be generated from $p(\varepsilon)$ to estimate $\nabla_{\lambda} \mathcal{L}(\lambda)$ via Monte Carlo.

When using the reparametrization trick and Gaussian variational approximations with a Cholesky factorization, the estimates of the $\mathcal{L}(\lambda)$ gradient depend on the gradient of the deterministic transformation function $\nabla_\lambda g_\lambda(\epsilon)$, as shown in \eqref{eqn:grad ELBO different measure}, and are computed with respect to the variational parameters $\mu$ and $L$. The gradient of $\mathcal{L}(\lambda)$ with respect to the mean vector $\mu$ can be estimated via Monte Carlo sampling from the alternative probability measure $p(\epsilon)$ by noting that
\begin{align}
    \nabla_\mu\mathcal{L}(\lambda)
    &=\E_{p(\epsilon)}\big\{\nabla_\theta h_\lambda(\theta)\big\}.
    \label{eqn: grad ELBO with mu}
\end{align} 
Let $\vecth(A)$ denote the half-vectorization of a matrix $A$ of size $d\times d$, i.e. the vector of length $d(d+1)/2$ obtained by stacking the elements in the matrix lower triangle, including the elements along the diagonal of $A$, from left to right and top to bottom. The gradient of the half-vectorization of $L$ under $p(\epsilon)$ is
\begin{align}
   \nabla_{\vecth(L)}\mathcal{L}(\lambda) & =\E_{p(\epsilon)}\big[\vecth\big\{\nabla_\theta h_\lambda(\theta)\epsilon^T\big\}\big],
   \label{eqn: grad ELBO with vech}
\end{align} 
with $\theta=\mu+L\epsilon$.

\section{Batch Stochastic Variational Bayes for GARCH Models}\label{sec3}
\label{VB on GARCH}

In this section, we derive an algorithm for fitting GARCH models via SVB (Stochastic Variational Bayes). The algorithm provides approximate posterior densities for the parameters of interest in models with the distributional assumptions on the innovations of Section \ref{sec2}.

We adopt the notation of \cite{wand2014fully} and \cite{ong2018variational} and start by deriving the algorithm using the control variates approach first, followed by the reparametrization trick.

To define an SVB (Stochastic Variational Bayes) algorithm, it is necessary to have an unbiased estimator of the gradient vector of $\mathcal{L}(\lambda)$ with respect to the variational parameter $\lambda$, denoted as $\widehat{\nabla_\lambda \mathcal{L}(\lambda)}$.

Considering Equation \eqref{eqn:gradientElboNaiveEstimator}, this estimate depends on:
\begin{align}
\begin{split}
    h_{\lambda}(\theta) &= \log \frac{L(\theta;y)p(\theta)}{q_{\lambda}(\theta)}\\
    & = \ell(\theta) + \log p(\theta)  - \log q_{\lambda}(\theta).
    \label{eqn: h lambda}
\end{split}
\end{align}
where $\ell(\theta)$ is the log-likelihood function, $p(\theta)$ is the prior distribution and $q_{\lambda}(\theta)$ is the variational distribution. It is easy to notice that $\log q_{\lambda}(\theta)$ and, consequently, $\nabla_\lambda \log q_{\lambda}(\theta)$ depend only on the variational density $q_\lambda(\theta)$, and they do not rely on the specific model under examination. Here, we take $q_{\lambda}(\theta)$ to be a multivariate Gaussian distribution. The simplest approximation assumed the covariance matrix $\Sigma$ in the Gaussian variational density $q_\lambda(\theta)$ has a fully independent structure, meaning it is diagonal (mean-field approximation). However, this approximation lacks the ability to capture the dependence structure of the target posterior density and tends to underestimate the uncertainties of the model parameters \citep{ong2018gaussian}. We will compare results using a dense or a diagonal covariance matrix. 
%i.e., the covariance matrix off-diagonal entries are zeros, so that $\Sigma = \diago(\exp(2\varphi_1),\ldots,\exp(2\varphi_d))$, with $\varphi_i = \log(\sigma_i)$ and $\sigma_i = \Sigma_{ii}^{{1}/{2}}$ \citep{xu2019variance}. The mean-field variational distribution is then defined as
%\begin{align*}
%    q_\lambda(\theta) = \prod_{i=1}^d \mathcal{N}_d(\mu_i,\exp(2\varphi_i))
%\end{align*}
%with variational parameters $\lambda = (\mu^\top, \varphi^\top)^\top$, where $\mu= (\mu_1, \ldots, \mu_d)^\top$ and $\varphi = (\varphi_1, \ldots, \varphi_d)^\top$. In the case where $\theta$ is unidimensional, the gradient of the log variational density with respect to the transformed variational parameters $(\mu_1,\varphi_1)$ can be expressed as, 
%\begin{align*}
%    \nabla_\lambda \log q_{\lambda}(\theta) = \Bigg(& \frac{\theta-\mu_1}{\exp{(2\varphi_1})},\frac{(\theta-\mu_1)^2}{\exp{(2\varphi_1})}-1\Bigg).
%\end{align*}
%It can be easily extended to an independent multivariate Gaussian variational density. 
%This simplified mean-field assumption does
%increase the computational speed 

According to \cite{tan2018gaussian}, faster convergence in variational approximations for state-space models can be achieved by parametrizing the precision matrix instead of the covariance matrix. In our simulations, we have observed that, for GARCH models, variational algorithms using control variates converge faster when the approximation is based on a parameterization of the precision matrix. However, the opposite holds true when using the reparametrization trick. We define the precision matrix as $\Omega = \Sigma^{-1} = CC^\top$, where $C$ is a lower triangular matrix.
%It is, therefore, often more appropriate to consider an unrestricted covariance matrix $\Sigma$. \cite{tan2018gaussian} argue that it is computationally faster to parameterize the precision matrix in a Gaussian variational approximation in terms of the Cholesky factor, i.e, $\Omega = \Sigma^{-1} = CC^\top$.  It imposes a sparsity property in the precision matrix and exhibits the conditional independence structure in the posterior distribution appropriately which is particularly useful for the models such as the latent state-space model where the number of the local variables is significantly larger than the number of the global variables. However, for GARCH-type models, our SVB algorithm does converge faster by parameterizing the Cholesky factors in terms of the precision matrix rather than the covariance matrix $\Sigma=LL^\top$ in the control variates method but the opposite for the reparametrization trick case. Parameterization with a precision matrix does not offer clear computational gain over the Cholesky factors of a covariance matrix. This might be because by applying different parameterizations, sometimes we are able to remove most of the matrix inversions in the Monte Carlo step which reduces some computational burden. 

To implement an algorithm based on control variates, the variational parameters to be optimized are $\lambda=(\mu^\top,\vecth(C)^\top)^\top$, where $\mu$ is a $d$-dimensional vector, and $\vecth(C)$ is the half-vectorization of $C$, i.e. a vector of $d(d+1)/2$ elements obtained by vectorizing the lower triangular part of $C$. The log-density of the multivariate Gaussian distribution is
\begin{align}
\begin{split}
    \log q_\lambda(\theta) = &-\frac{d}{2} \log 2\pi + \log \lvert C \rvert \\ 
    &- \frac{1}{2}(\theta-\mu)^\top C C^\top (\theta-\mu)
\end{split}
    \label{eq:logq_CV_sec3}
\end{align}
and the gradient of the log-multivariate Gaussian density with respect to $\mu$ and $\vecth(C)$ is 
$$\nabla_\lambda \log q_\lambda(\theta) = (\nabla_\mu \log q_\lambda(\theta)^\top,\nabla_{\vecth{(C)}} \log q_\lambda(\theta)^\top)^\top,$$ 
where
\begin{align*}
    \nabla_\mu \log q_\lambda(\theta) = & \enskip C C^\top (\theta-\mu)\\
    \mbox{and }\nabla_{\vecth{(C)}} \log q_\lambda(\theta) = &\enskip \vecth\big(\diago(1/C) \\& -(\theta-\mu)(\theta-\mu)^\top C\big).
\end{align*}
Here $\diago(1/C)$ denotes a diagonal matrix whose $i$-th entry is $1/C_{ii}$ and all the elements of $\mu$ and $\vecth{(C)}$ are not restricted. To ensure that $C$ is positive-definite, we parameterize the optimization problem in terms of the lower triangular matrix $C'$, where $C'_{ij}=C_{ij}$ if $i\neq j$ and $C'_{ii}=\exp(C_{ii})$. 

While producing an unbiased estimator of the ELBO with smaller variance, the use of control variates can slow down the variational algorithm implementation \citep{tran2021practical} and the reparametrization trick can be used as an alternative to derive faster algorithms. This approach uses a reparametrized gradient estimator that is adapted to the specific model.  From \eqref{eqn: h lambda}, the generic expression for the gradient is
\begin{align}
\begin{split}
    \nabla_\theta & h_\lambda(\theta) = \nabla_\theta \log\frac{L(\theta;y)p(\theta)}{ q_\lambda(\theta)}\\
    &= \nabla_\theta \ell(\theta) + \nabla_\theta \log p(\theta)- \nabla_\theta\log q_\lambda(\theta).
\end{split}
\label{eqn:modelspe}
\end{align}
Finally, the gradient of the log multivariate Gaussian variational density with respect to $\theta$ is 
\begin{align*}
    \nabla_\theta\log q_\lambda(\theta) = -\Sigma^{-1} (\theta-\mu).
\end{align*}

\subsection{GARCH Model with Gaussian  Innovations} 
\label{sec:Gaussian GARCH}

For simplicity, we will restrict our attention to the GARCH(1,1) model. Extensions to the general GARCH(p,q) as given in \eqref{eqn:Garch} are straightforward. 
The parameters of a GARCH(1,1) with Gaussian innovations are $\phi = (\omega,\alpha,\beta)^\top$, with $\omega > 0$, $\alpha>0$, $\beta>0$ and $\alpha+\beta <1$. \cite{tran2021variational} suggest the following transformations to work with unconstrained parameters: $\alpha = \psi_1\psi_2$, $\beta = \psi_1(1-\psi_2)$ with $0<\psi_i<1$, $i=1,2$, $\theta_{\omega} =\log(\omega)$, $\theta_{\psi_1} =\log\{\psi_1/(1-\psi_1)\}$ and $\theta_{\psi_2} =\log\{\psi_2/(1-\psi_2)\}$,
therefore $\theta_{\phi} = (\theta_\omega,\theta_{\psi_1},\theta_{\psi_2})^\top$, and $\theta=\theta_{\phi}$. 
%\begin{align*}
%    \omega &= \exp (\theta_\omega),\\
%    \alpha &= [\{1+\exp(-\theta_{\psi_1})\}\{1+\exp(-%\theta_{\psi_2})\}]^{-1},\\
%    \beta & = \frac{\exp(-\theta_{\psi_2})}{\{1+\exp(-%\theta_{\psi_1})\}\{1+\exp(-\theta_{\psi_2})\}}.
%\end{align*}

An inverse gamma prior distribution $IG(1,1)$ for $\omega$ and independent uniform prior distributions $U(0,1)$ for $\psi_1$ and $\psi_2$ are assumed. These prior distributions induces a negative log-gamma density $NLG(1,1)$ for $\theta_{\omega}$ and an independent standard logistic distribution for $\theta_{\psi_1}$ and $\theta_{\psi_2}$, i.e.
\begin{equation}
        \log p(\theta_\omega) = -\theta_\omega - \exp (-\theta_\omega), \label{eqn: Gaussian GARCH log prior 1}
\end{equation}
\begin{equation}
   \log p(\theta_{\psi_1}) = -\theta_{\psi_1} - 2 \log (1+ \exp (-\theta_{\psi_1})) \label{eqn: Gaussian GARCH log prior 2}
\end{equation}
and
\begin{equation}
    \log p(\theta_{\psi_2}) = -\theta_{\psi_2} - 2 \log (1+ \exp (-\theta_{\psi_2})).
    \label{eqn: Gaussian GARCH log prior 3}
\end{equation}
The log-likelihood function of the GARCH(1,1) model with Gaussian innovations is defined in \eqref{eqn: llk Gaussian} and using \eqref{eqn: h lambda} this gives rise to
\begin{align} \label{G htheta}
\begin{split}
    h_{\lambda}(\theta) =  & -\frac{1}{2}\sum_{t=1}^T \left\{\log (2\pi)+\log(\sigma_t^2)+\frac{y_t^2}{\sigma_t^2}\right\} \\& + \log p (\theta)+\frac{d}{2} \log 2\pi - \log \lvert C \rvert\\ &+ \frac{1}{2}(\theta-\mu)^\top C C^\top (\theta-\mu).
\end{split}
\end{align}
Using \eqref{eq:logq_CV_sec3}, it is then possible to implement the control variates approach. 

To implement a variational algorithm using the reparametrization trick, the gradients $\nabla_\theta \ell(\theta)$ and $\nabla_\theta \log p(\theta)= \left(\nabla_{\theta_\omega} \log p(\theta_\omega),\nabla_{\theta_{\psi_1}} \log p(\theta_{\psi_1}),\nabla_{\theta_{\psi_2}} \log p(\theta_{\psi_2})\right)^\top$ for the GARCH(1,1) model with Gaussian innovations are needed. The three components of the first gradient are
\begin{equation}
    \nabla_{\theta_\omega} \log p(\theta_\omega)=-1+\exp{(-\theta_\omega)},\label{eqn: Gaussian GARCH gradients of prior 1}
\end{equation}
\begin{equation}
    \nabla_{\theta_{\psi_1}} \log p(\theta_{\psi_1})=-1+\frac{2\exp(-\theta_{\psi_1})}{1+\exp(-\theta_{\psi_1})}\label{eqn: Gaussian GARCH gradients of prior 2}
\end{equation}
and
\begin{equation}
    \nabla_{\theta_{\psi_2}} \log p(\theta_{\psi_2}) =-1+\frac{2\exp(-\theta_{\psi_2})}{1+\exp(-\theta_{\psi_2})}.\label{eqn: Gaussian GARCH gradients of prior 3}
\end{equation}
The gradient of the log-likelihood function for the Gaussian GARCH(1,1) model (see also \citeauthor{fiorentini1996analytic}, \citeyear{fiorentini1996analytic}) in \eqref{eqn: llk Gaussian} with respect to the unconstrained parameters is
\begin{align}
    \nabla_\theta \ell(\theta)=\frac{1}{2}\sum_{t=1}^T \frac{1}{\sigma_t^2}\frac{\partial \sigma_t^2}{\partial \theta}\left(\frac{y_t^2}{\sigma_t^2}-1\right)
\end{align}
and the partial derivatives of $\sigma_t^2$ with respect to the unconstrained parameters are given in  Appendix \ref{Appen 1}. Finally, $\nabla_\theta{h_{\lambda}(\theta)}$ defined in \eqref{eqn:modelspe} takes the following form:
\begin{align} \label{Grad G htheta}
\begin{split}
    \nabla_\theta h_\lambda(\theta_{\phi}) = & \enspace \frac{1}{2}\sum_{t=1}^T \frac{1}{\sigma_t^2}\frac{\partial \sigma_t^2}{\partial \theta_{\phi}}\left(\frac{y_t^2}{\sigma_t^2}-1\right) \\& + \nabla_\theta \log p(\theta_{\phi})+\Sigma^{-1} (\theta_{\phi}-\mu),
\end{split}
\end{align}
where $\nabla_\theta \log p(\theta_{\phi})$ is stated in \eqref{eqn: Gaussian GARCH gradients of prior 1}--\eqref{eqn: Gaussian GARCH gradients of prior 3}. 

\subsection{GARCH Model with t Innovations} \label{sec:t GARCH}

To adapt SVB with control variates to the GARCH model with t innovations, it is necessary to derive the model-specific $\log p(\theta)$ and $\ell(\theta)$. The log-likelihood function for the t model is given in Equation \eqref{eqn:llk std}. The prior distributions for $\omega$, $\alpha$, and $\beta$ in the time-varying variance equation for $\sigma_t^2$ remain the same as in the Gaussian case. As for the degrees of freedom parameter $\nu$, we choose a translated exponential prior density \citep{ardia2008financial}.\begin{align*}
    p(\nu)= \exp\big\{-(\nu-2)\big\}\mathbbm{1}_{\{\nu>2\}}.
\end{align*}
Similarly to $\omega$, $\alpha$, and $\beta$, $\nu$ can also be mapped to an unrestricted space by setting $\nu=\log({\exp(\theta_\nu)+1})+2$. According to \cite{tan2018gaussian}, this specific transformation ensures greater numerical stability compared to the transformation $\nu=\exp(\theta_\nu)+2$. The resulting log-prior density of $\theta_{\nu}$ is then
\begin{align}
\begin{split}
    \log p(\theta_\nu) &= - \log \{\exp(\theta_\nu)+1\} \\
    & \quad -\log \{1+\exp(-\theta_\nu)\}.     \label{eqn: log p_nu}
\end{split}
\end{align}
The parameters to be optimized are then the unconstrained ones, $\theta = (\theta_{\phi},\theta_{\nu})$, where $\theta_{\phi}=(\theta_{\omega},\theta_{\psi_1},\theta_{\psi_2})^T$, and $\phi=(\omega,\alpha,\beta)^T$.

The expression of $h_{\lambda}(\theta)$ defined in \eqref{eqn: h lambda} takes the following form for the t GARCH(1,1) model: 
\begin{align} 
\begin{split}
h_{\lambda} & (\theta) =  T\bigg\{\log \Gamma\Big(\frac{\log\{\exp(\theta_\nu)+1\}+3}{2}\Big)\\ & -\log\Gamma\Big(\frac{\log\{\exp(\theta_\nu)+1\}+2}{2}\Big) \\
    & -\frac{1}{2}\log\big\{\pi(\log\{\exp(\theta_\nu)+1\})\big\}\bigg\} \\
    &-\frac{1}{2}\sum_{t=1}^T\bigg[\log(\sigma_t^2)  + (\log\{\exp(\theta_\nu)+1\}+3)\\
    &\log\left\{1+\frac{y_t^2}{\log\{\exp(\theta_\nu)+1\}\sigma_t^2}\right\}\bigg]+ \log p (\theta) \\
    & + \frac{d}{2} \log 2\pi - \log \lvert C \rvert \\ 
    &+ \frac{1}{2}(\theta-\mu)^\top C C^\top (\theta-\mu), \label{S htheta}
\end{split}
\end{align}
where $\log p(\theta)$ is given by the sum of \eqref{eqn: Gaussian GARCH log prior 1}--\eqref{eqn: Gaussian GARCH log prior 3} and \eqref{eqn: log p_nu}.

To implement SVB with the reparametrization trick, the gradients of the log-prior densities for the $\theta_{\phi}$ components remain the same as in Equations \eqref{eqn: Gaussian GARCH gradients of prior 1}--\eqref{eqn: Gaussian GARCH gradients of prior 3}. Additionally, the gradient of the log-prior density for $\theta_\nu$ is given by
\begin{align}
\begin{split}
   \nabla_{\theta_\nu} \log p(\theta_\nu)&=\frac{\exp(-\theta_\nu)}{1+\exp(-\theta_\nu)}\\&\quad-\frac{\exp(\theta_\nu)}{1+\exp(\theta_\nu)}.
   \label{eqn: t GARCH gradients of prior}
\end{split}
\end{align}
The derivational steps for the gradient of the log-likelihood function from the t GARCH model are provided in Appendix \ref{Appen 2} (also refer to \citeauthor{levy2003analytic}, \citeyear{levy2003analytic}). Therefore, the expression for $\nabla_\theta{h_{\lambda}(\theta)}$, as defined in Equation \eqref{eqn:modelspe} for the t GARCH(1,1) model, takes the following form:
\begin{align} \label{Grad S htheta}
\begin{split}
    \nabla_\theta h_\lambda(\theta) = & \enspace \nabla_\theta \ell(\theta) + \nabla_\theta \log p(\theta) \\&
    +\Sigma^{-1} (\theta-\mu),
\end{split}
\end{align}
where the expression of $\nabla_\theta \ell(\theta)=\left(\frac{\partial \ell(\theta)}{\partial \theta_{\phi}},\frac{\partial \ell(\theta)}{\partial \theta_\nu}\right)^\top$ can be found in the appendix \ref{Appen 2} and $\nabla_\theta \log p(\theta)$ is stated in \eqref{eqn: Gaussian GARCH gradients of prior 1}--\eqref{eqn: Gaussian GARCH gradients of prior 3} and \eqref{eqn: t GARCH gradients of prior}.

%{\color{red} The expression for $\nabla_\theta h_\lambda(\theta)$ in the t model is missing.}

\subsection{GARCH Model with Skewed t Innovations}
\label{sec:ssVB}

The log-likelihood for the GARCH(1,1) model with standardized skewed t innovations is presented in Equation \eqref{eqn: llk sstd}. Additionally, the skewness parameter $\xi$ is assumed to follow an inverse gamma $IG(1,1)$ prior distribution, according to the same reasoning in the definition of the prior distribution for $\omega$:
\begin{align*}
    p(\xi) = \xi^{-2}\exp(-\xi^{-1})\mathbbm{1}_{\{\xi>0\}}.
\end{align*}
Once again, the skewness parameter $\xi$ can be transformed into an unconstrained parameter using the transformation $\xi = \log{\exp(\theta_\xi)+1}$. As a result, the log-prior density of $\theta_\xi$ is given by
\begin{align}\label{eqn: log p_sk}
\begin{split}
    \log p(\theta_\xi) = & -2 \log\left(\log(\exp(\theta_\xi)+1)\right)\\
    &-\log(\exp(\theta_\xi)+1)^{-1} \\
    &-\log\left(1+\exp(-\theta_\xi)\right).
\end{split}
\end{align}
The original set of parameters $(\phi,\nu,\xi)^\top$ with $\phi = (\omega,\alpha,\beta)^\top$ is replaced by the unrestricted parameters $\theta=(\theta_{\phi},\theta_{\nu},\theta_{\xi})^\top$, where $\theta_{\phi} = (\theta_\omega,\theta_{\psi_1},\theta_{\psi_2})$. The prior densities for the constrained parameters $\omega$, $\alpha$, $\beta$, and $\nu$ remain the same as those of the t GARCH model described in Section \ref{sec:t GARCH}.

The expression for $h_{\lambda}(\theta)$ defined in Equation \eqref{eqn: h lambda} takes the following form for the skewed t GARCH(1,1) model:
\begin{align} 
\begin{split}
     h_{\lambda}&(\theta) = T\bigg\{\log \Gamma\Big(\frac{\log\{\exp(\theta_\nu)+1\}+3}{2}\Big)\\ & -\log\Gamma\Big(\frac{\log\{\exp(\theta_\nu)+1\}+2}{2}\Big) \\
    & -\frac{1}{2}\log\big\{\pi(\log\{\exp(\theta_\nu)+1\})\big\}\bigg\} \\
    & +\log\left(\frac{2}{\log\{\exp(\theta_\xi)+1\}+1/\log\{\exp(\theta_\xi)+1\}}\right)  \\
    & +\log(s)\bigg\} -\frac{1}{2}\sum_{t=1}^T\bigg[\log(\sigma_t^2)+(\log\{\exp(\theta_\nu)+1\} \\ 
    &+3)\log\left\{1+\frac{(sy_t/\sigma_t+m)^2}{\log\{\exp(\theta_\nu)+1\}}\xi^{-2I_t}\right\}\bigg] \\
    & + \log p (\theta)+\frac{d}{2} \log 2\pi - \log \lvert C \rvert \\ 
    & + \frac{1}{2}(\theta-\mu)^\top C C^\top (\theta-\mu), 
    \label{SS htheta}
\end{split}
\end{align}
where $\log p(\theta)$ is given by the sum of \eqref{eqn: Gaussian GARCH log prior 1}--\eqref{eqn: Gaussian GARCH log prior 3}, \eqref{eqn: log p_nu} and \eqref{eqn: log p_sk}.    

To implement SVB with the reparametrization trick, the gradients of the log-prior densities with respect to $\theta_\omega, \theta_\alpha, \theta_\beta, \theta_\nu$ are required. These gradients can be found in Appendix \ref{Appen 2}. Additionally, the gradient of the log-prior density for $\theta_\xi$ is
\begin{align}
    \nabla_{\theta_\xi} & \log p(\theta_\xi)= -\frac{2\exp(\theta_\xi)}{\big(1+\exp(\theta_\xi)\big)\log\big(\exp(\theta_\xi)+1\big)} \nonumber \\
    &\quad+\frac{\exp(\theta_\xi)}{1+\exp(\theta_\xi)}+\frac{\exp(-\theta_\xi)}{1+\exp(-\theta_\xi)}.
    \label{eqn: skewed t GARCH gradients of prior}
\end{align}
The gradient for the log-likelihood function of the skewed t GARCH model is given in Appendix \ref{Appen 3}. From \eqref{eqn:modelspe},
\begin{align} \label{Grad SS htheta}
\begin{split}
    \nabla_\theta h_\lambda(\theta) = & \enspace \nabla_\theta \log p(y\vert \theta) + \nabla_\theta \log p(\theta) \\&
    +\Sigma^{-1} (\theta-\mu),
\end{split}
\end{align}
where $\nabla_\theta \log p(y\vert \theta)$ is provided in Appendix \ref{Appen 3} and $\nabla_\theta \log p(\theta)$ is given by \eqref{eqn: Gaussian GARCH gradients of prior 1}--\eqref{eqn: Gaussian GARCH gradients of prior 3}, \eqref{eqn: t GARCH gradients of prior} and \eqref{eqn: skewed t GARCH gradients of prior}.

Algorithms \ref{algo1} and \ref{algo2} outline SVB procedures for GARCH-type models with different types of innovations, namely Gaussian, t, and skewed t. These algorithms utilize the Gaussian Cholesky factorization and incorporate the use of control variates and the reparametrization trick, respectively.

\begin{algorithm*}
\caption{SVB for GARCH(1,1) models with Gaussian, t and skewed t innovations using a Gaussian approximating density and the control variates.}
\label{algo1}
\begin{algorithmic}
\Require \\
\begin{itemize}
    \item initial $\lambda^{(0)}=(\mu^{(0)^{\top}},\vecth(C^{(0)})^\top)^\top$;
    \item adaptive learning weights $\beta_{1}$,$\beta_{2}\in(0,1)$
    \item  fixed learning rate $\eta_{0}$
    \item threshold $\tau$
    \item rolling window size $t_W$
    \item maximum patience parameter $P$
    \item prior distribution $p(\theta)$
    \item likelihood function $L(\theta;y)$
    \item Monte Carlo Samples $S$.
\end{itemize}
\vspace{0.5\baselineskip}
\hrule\\
\Initialization \textbf{:}\\
\begin{itemize}
    \item Set $C'^{(0)}=C^{(0)}$.
    \item Update $\diago(C'^{(0)}) = \exp\big(\diago(C'^{(0)})\big)$.
    \item Generate $\theta_{(s)} \sim \mathcal{N}\big(\mu^{(0)},(C'^{(0)}C'^{(0)^{\top}})^{-1}\big)$.
    \item Compute $\widehat{\nabla_{\lambda} \mathcal{L}(\lambda^{(0)})}$ in \eqref{eqn:gradientElboNaiveEstimator} where $h_\lambda(\theta)$ is given in \eqref{G htheta} for Gaussian, \eqref{S htheta} for t, and \eqref{SS htheta} for skewed t innovations.
    \item Calculate the $j$-th element of the control variate $c_j$ in \eqref{eqn: CV} using $\theta_s$.
    \item Set $g_0 := \widehat{\nabla_{\lambda} \mathcal{L}(\lambda^{(0)})},v_0 := (g_0)^2,\bar{g}:=g_0, \bar{v}:=v_0$, $t=0$, patience = 0.
\end{itemize}
\End \textbf{ Initialization} 
\vspace{0.5\baselineskip}
\\
\hrule
\While{\textbf{stop = FALSE}}
\begin{itemize}
    \item Set $C'^{(t)}=C^{(t)}$.
    \item Update $\diago(C'^{(t)}) = \exp\big(\diago(C'^{(t)})\big)$.
    \item Generate $\theta_s \sim \mathcal{N}\big(\mu^{(t)},(C'^{(t)}C'^{(t)^{\top}})^{-1}\big)$.
    \item Compute the unbiased estimator $g_t :=\nabla_{\lambda} \widehat{\mathcal{L}(\lambda^{(t)})}$.
    \item Estimate the new $j$-th entry of the control variate $c_j$ in \eqref{eqn: CV}.
    \item Calculate the coordinate-wise operator $v_t=(g_t)^2$ and set $\bar{g}=\beta_1\bar{g}+(1-\beta_1)g_t$, $\bar{v}=\beta_2\bar{v}+(1-\beta_2)v_t$.
    \item Compute the learning rate $\alpha_t=\min(\eta_0,\eta_0\frac{\tau}{t})$. 
    \item Update $\lambda^{(t+1)}=\lambda^{(t)}+\alpha_t\frac{\bar{g}}{\sqrt{\bar{v}}}.$
    \item Compute the ELBO estimator $\widehat{\mathcal{L}(\lambda^{(t)})}$ in \eqref{eqn:ELBO} using Monte Carlo samples $\theta_s$.
\end{itemize}
\If{$t \geq t_W$} 
\State Compute the moving average lower bound
$\overline{\mathcal{L}}_{t-t_W+1}(\lambda)=\frac{1}{t_W}\sum^{t_W}_{k=1}\widehat{\mathcal{L}(\lambda^{(t-k+1)})},$ 
\an \textbf{d if  }{$\overline{\mathcal{L}}_{t-t_W+1}(\lambda)\geq \max(\overline{\mathcal{L}}(\lambda))$ } \textbf{then} patience = 0,
\otherwise \quad patience = patience + 1.
\EndIf

\If{patience $\geq P$} \textbf{stop = TRUE}.
\Else \State {Set t = t+1.}
\EndIf
\EndWhile
\end{algorithmic}
\end{algorithm*}

\begin{algorithm*}
\caption{SVB for GARCH(1,1) models with Gaussian, t and skewed t innovations using a Gaussian approximating density and the reparametrization trick.}
\label{algo2}
\begin{algorithmic}
\Require\\
\begin{itemize}
    \item initial $\lambda^{(0)}=(\mu^{(0)^{\top}},\vecth(C^{(0)})^\top)^\top$;
    \item adaptive learning weights $\beta_{1}$,$\beta_{2}\in(0,1)$
    \item  fixed learning rate $\eta_{0}$
    \item threshold $\tau$
    \item rolling window size $t_W$
    \item maximum patience parameter $P$
    \item prior distribution $p(\theta)$
    \item likelihood function $L(\theta;y)$
    \item gradients of log-likelihood function $\nabla \log L(\theta;y)$
    \item Monte Carlo Samples $S$.
\end{itemize}
\vspace{0.5\baselineskip}
\hrule\\
\Initialization \textbf{:}\\
\begin{itemize}
    \item Generate $\epsilon_s \sim p_{\epsilon}(.)=\mathcal{N}_d(0,I), \quad s=1,\ldots,S$
    \item Compute the unbiased gradient estimator of the ELBO in \eqref{eqn: grad ELBO with mu} and \eqref{eqn: grad ELBO with vech} where $\nabla_\theta h_\lambda(\theta)$ are stated in \eqref{Grad G htheta}, \eqref{Grad S htheta} and \eqref{Grad SS htheta} for Gaussian, t and skewed t innovations respectively, 
$\nabla_{\lambda}\widehat{\mathcal{L}(\lambda^{(0)})}=\big(\nabla_\mu\widehat{\mathcal{L}(\lambda^{(0)})^T},\nabla_{\vecth(L)}\widehat{\mathcal{L}(\lambda^{(0)})^T}\big)^T$, 
with Monte Carlo samples $\theta_s=\mu^{(0)}+L^{(0)}\epsilon_s$.
\item Set $g_0 := \widehat{\nabla_{\lambda} \mathcal{L}(\lambda^{(0)})},v_0 := (g_0)^2,\bar{g}:=g_0, \bar{v}:=v_0$, $t=0$, patience = 0 and \textbf{stop = FALSE}.
\end{itemize}
\End \textbf{ Initialization}
\vspace{0.5\baselineskip}
\\
\hrule
\While{\textbf{stop = FALSE}}
\begin{itemize}
    \item Generate $\epsilon_s \sim p_{\epsilon}(.)=\mathcal{N}_d(0,I), s=1,\ldots,S$.
    \item Recalculate $\mu^{(t)},L^{(t)}$ from the updated $\lambda^{(t)}$.
    \item Update the unbiased gradient estimator $\nabla_{\lambda}\widehat{\mathcal{L}(\lambda^{(t)})}=\big(\nabla_\mu\widehat{\mathcal{L}(\lambda^{(t)})^T},\nabla_{\vecth(L)}\widehat{\mathcal{L}(\lambda^{(t)})^T}\big)^T$ in \eqref{eqn: grad ELBO with mu} and \eqref{eqn: grad ELBO with vech} where $\nabla_\theta h_\lambda(\theta)$ are stated in \eqref{Grad G htheta}, \eqref{Grad S htheta} and \eqref{Grad SS htheta} for Gaussian, t and skewed t innovations respectively,  
with new Monte Carlo samples $\theta_s=\mu^{(t)}+L^{(t)}\epsilon_s$.
    \item Calculate the coordinate-wise operator $v_t=(g_t)^2$ and set $\bar{g}=\beta_1\bar{g}+(1-\beta_1)g_t$, $\bar{v}=\beta_2\bar{v}+(1-\beta_2)v_t$.
    \item Compute the learning rate $\alpha_t=\min(\eta_0,\eta_0\frac{\tau}{t})$ and update $$\lambda^{(t+1)}=\lambda^{(t)}+\alpha_t\frac{\bar{g}}{\sqrt{\bar{v}}}.$$
    \item Compute the ELBO estimator $\widehat{\mathcal{L}(\lambda^{(t)})}$ in \eqref{eqn:ELBO} using Monte Carlo samples $\theta_s$.
\end{itemize}
\If{$t \geq t_W$} 
\State Compute the moving average lower bound
$\overline{\mathcal{L}}_{t-t_W+1}(\lambda)=\frac{1}{t_W}\sum^{t_W}_{k=1}\widehat{\mathcal{L}(\lambda^{(t-k+1)})},$ 
\an \textbf{d if  }{$\overline{\mathcal{L}}_{t-t_W+1}(\lambda)\geq \max(\overline{\mathcal{L}}(\lambda))$ } \textbf{then} patience = 0,
\otherwise \quad patience = patience + 1.
\EndIf
\If{patience $\geq P$} \textbf{stop = TRUE}.
\Else \State {Set t = t+1.}
\EndIf
\EndWhile
\end{algorithmic}
\end{algorithm*}

\section{Sequential Variational Inference}\label{sec4}

In the previous section, we introduced SVB approximations for GARCH models. One characteristic of time series data, such as those described by GARCH models, is that data points may become available sequentially. Ideally, inference should be updated sequentially as new information becomes available, avoiding the inefficient re-evaluation of the likelihood function for the entire dataset. In this section, we present implementations of the variational approximations from the previous section that allow for sequential updating. We adapt the strategies of \cite{gunawan2021variational} and \cite{tomasetti2022updating} for the sequential updating of variational approximations for GARCH models for the first time in the literature and compare their performance.

\subsection{Updating Variational Bayes}

In their work, \cite{tomasetti2022updating} propose a procedure called updating variational Bayes (UVB), which enables sequential updates of posterior distributions without the need to re-compute the likelihood function at each step.

Consider a sequence of time points $T_1, T_2, \ldots$, and suppose we want to update the posterior density $p(\theta \vert y_{1:T_n})$ after observing new data points in the interval $[T_{n+1},T_{n+h}]$ to derive $p(\theta \vert y_{1:T_{n+h}})$. The posterior distribution computed up to time $T_n$ can then be used as the prior distribution for the subsequent observation interval. As a result, the likelihood function for the interval $[T_{n+1},T_{n+h}]$ is based solely on $h$ data points:
\begin{align}
   p(\theta\vert y_{1:T_{n+h}}) \varpropto L (\theta ; y_{T_{n+1}:T_{n+h}})p(\theta\vert y_{1:T_n}).
    \label{eqn:Exact posterior}
\end{align}
Hence, the posteriors for new observations can be updated sequentially by recursively applying Bayes' theorem.

We now define the UVB procedure for GARCH-type models. To initialize the UVB algorithm, we can employ the batch stochastic variational approach described in Section \ref{sec3} to compute the first variational approximation $q_{\lambda_n}(\theta\vert y_{1:T_n})$ of the posterior density $p(\theta\vert y_{1:T_n})$. After obtaining the initial variational approximation and observing new data points in the interval $[T_{n+1},T_{n+h}]$, we can approximate the posterior distribution defined in Equation \eqref{eqn:Exact posterior} with a pseudo-posterior obtained using the variational approximation of the posterior distribution after the first $T_n$ points:
\begin{align}
    \Tilde{p}(\theta\vert y_{1:T_{n+h}}) \varpropto L (\theta;y_{T_{n+1}:T_{n+h}}\vert \theta)q_{\lambda_n}(\theta\vert y_{T_1:T_n}).
    \label{eqn:UVB}
\end{align}

Instead of minimizing the Kullback-Leibler divergence between the variational density $q_{\lambda_{n+h}}(\theta\vert y_{1:T_{n+h}})$ and the exact posterior density $p(\theta\vert y_{1:T_{n+h}})$, UVB modifies the objective function by substituting the exact posterior $p(\theta\vert y_{1:T_{n+h}})$ with the pseudo-posterior density $\Tilde{p}(\theta\vert y_{1:T_{n+h}})$. For each $n=1,2,\ldots$, the variational parameter $\lambda_{n+h}$ is optimized by maximizing the evidence lower bound (ELBO) using a batch stochastic variational approximation. This modification enables UVB to update the posterior density sequentially and efficiently. It is worth noting that in this article, we employ the faster reparametrization trick approach for UVB, whereas the original paper only considers the control variates case, which was found to be slower than the reparametrization trick approach in our numerical experiments. Algorithm \ref{algo: UVB} outlines the steps involved in implementing the UVB approach.
\begin{algorithm}
\caption{Updating Variational Bayes}
\label{algo: UVB}
\begin{algorithmic}
\Require
Prior, likelihood
\hrule\\
\Initialization \textbf{:}\\
Observe $y_{T_1:T_n}$. \\
Compute the optimized variational density $q_{\lambda_n}(\theta\vert y_{T_1:T_n})$ using SVB via Algorithm \ref{algo1} or \ref{algo2}.
\End \textbf{ Initialization}\\
\vspace{0.5\baselineskip}
\hrule
\For{$j = 1,\ldots,h$}
\State Collect new data points $y_{{T_{n+j-1}+1}:T_{n+j}}$.
\State Use $q_{\lambda_{n+j-1}}$ and \eqref{eqn:UVB} to construct the UVB pseudo-posterior up $\tilde{p}(\theta\vert y_{T_1:T_{n+j}})$ up to proportionality.
\State Minimize $\text{KL}\big\{q_{\lambda_{n+j}}(\theta\vert y_{T_1:T_{n+j}}) \| \tilde{p}(\theta\vert y_{T_1:T_{n+j}})\big\}$ using SVB via Algorithm \ref{algo1} or \ref{algo2}.
\EndFor
\end{algorithmic}
\end{algorithm}

\subsection{Sequential Variational Bayes Approximation}

The alternative approach proposed by \cite{gunawan2021variational}, called sequential SVB (Seq-SVB), differs from UVB in that it targets the true posterior distribution instead of the pseudo-posterior distribution \eqref{eqn:UVB}.

Instead of sequentially using variational approximations to the posterior distribution, Seq-SVB optimizes the variational parameters $\lambda_n$ at time $T_n$ and employs them as initial values for the stochastic variational approximations of the variational parameter $\lambda_{n+1}$ at time $T_{n+1}$. Seq-SVB achieves greater accuracy than UVB as it directly targets the true posterior distribution. However, it comes at a higher computational cost since it requires evaluating the likelihood function at each time in the interval $[T_1, T_{n+1}]$.

Now, we define the Seq-SVB algorithm for GARCH models. In the initialization step, batch SVB, as outlined in Algorithm \ref{algo1} or \ref{algo2}, is utilized for the data up to time $T_n$ to obtain the optimal variational parameters $\lambda$ up to time $T_n$. Subsequently, the first variational approximation $q_{\lambda_n}(\theta \vert y_{T_1:T_n})$ of the true posterior $p(\theta \vert y_{T_1:T_n})$ is obtained. After observing new data points up to time $T_{n+1}$, another SVB algorithm can be implemented using all the available data in the interval $[T_1,T_{n+1}]$, with the optimal $\lambda_n$ from the initialization serving as the initial value for the variational parameter. This procedure is repeated for the instances $(n+1), (n+2), \ldots$. The implementation of Seq-SVB is summarized in Algorithm \ref{algo: Seq-SVB}.
\begin{algorithm}
\caption{Sequential Stochastic Variational Bayes}
\label{algo: Seq-SVB}
\begin{algorithmic}
\Require
Prior, likelihood
\hrule\\
\Initialization \textbf{:}\\
Observe $y_{T_1:T_n}$. \\
Compute $q_{\lambda_n}(\theta\vert y_{T_1:T_n})$ using SVB Algorithm \ref{algo1} or \ref{algo2}.
\End \textbf{ Initialization}\\
\vspace{0.5\baselineskip}
\hrule
\For{$j = 1, 2, \ldots, h$}
\State Define the new vector of observations $y_{T_1:T_{n+j}}$.
\State Initialise SVB using $\lambda_{n+j-1}$ 
\State Compute $q_{\lambda_{n+j}}(\theta\vert y_{T_1:T_{n+j})}$ using SVB via Algorithm \ref{algo1} or \ref{algo2}.
\EndFor
\end{algorithmic}
\end{algorithm}

\section{Simulation Study}
\label{sec:simu}

\subsection{Batch Stochastic Variational Bayes}

We first assess the accuracy of the batch SVB approximation discussed in Section \ref{VB on GARCH} by applying Algorithms \ref{algo1} and \ref{algo2} to GARCH models with the innovation types under examination. All computations were performed in \texttt{R} \citep{r2023} on the Katana computational cluster with 8-CPU cores and 4GB memory per core of the University of New South Wales (\url{https://docs.restech.unsw.edu.au/using_katana/about_katana/}).

We generated time series of lengths 1,000 and 5,000 from GARCH(1,1) models with Gaussian, t, and skew t innovations using the \texttt{R} package \texttt{fGarch} \citep{wuertz2022package}. For each setting, we simulated 1,000 replicates. The GARCH models were generated using parameter values similar to those estimated from the S\&P500 return index time series analyzed in Section \ref{sec6} with MCMC. Specifically, we used $\omega=0.1$, $\alpha=0.2$, and $\beta=0.75$ for Gaussian, t, and skew t innovations. We set $\nu=4$ for the t and skew t degrees of freedom, and $\xi=0.8$ as the skew t asymmetry parameter.

The algorithm parameters of Algorithms \ref{algo1} and \ref{algo2} were set to the default values employed in \cite{tran2021practical}, with some slight adjustments made to improve convergence. Specifically, the maximum patience parameter was set to $P=100$, and the moving average window size was set to $t_W=25$. Both adaptive learning weights were set to $\beta_1=\beta_2=0.9$, and the adaptive learning rate was set to $\eta_0=0.02$. We used $\tau=1,000$ as the threshold. In order to investigate the impact of the number of Monte Carlo samples on the efficiency and computational time, we conducted experiments using different values of $S$ for both the control variates and reparametrization methods. Specifically, we fixed the number of MC samples at $S=5, 10,$ and $50$ for both methods. By comparing the results obtained with these different values of $S$, we can assess the influence of the number of MC samples on the estimation accuracy and computational efficiency of the two approaches. In this paper, we focus on presenting the results obtained using the control variates method with $S=10$, and the reparametrization trick method with $S=5$. These particular choices for the number of MC samples were selected based on similar computational time, ensuring a fair comparison between the two approaches (See table \ref{tab:garch_times}). The additional results obtained using different values of $S$ can be found in the appendix \ref{secB1}, providing a comprehensive analysis of the sensitivity of the results to the choice of MC sample size.

\begin{figure}[htbp]
\centering
 \begin{minipage}{220pt}  
  \includegraphics[width=\columnwidth]{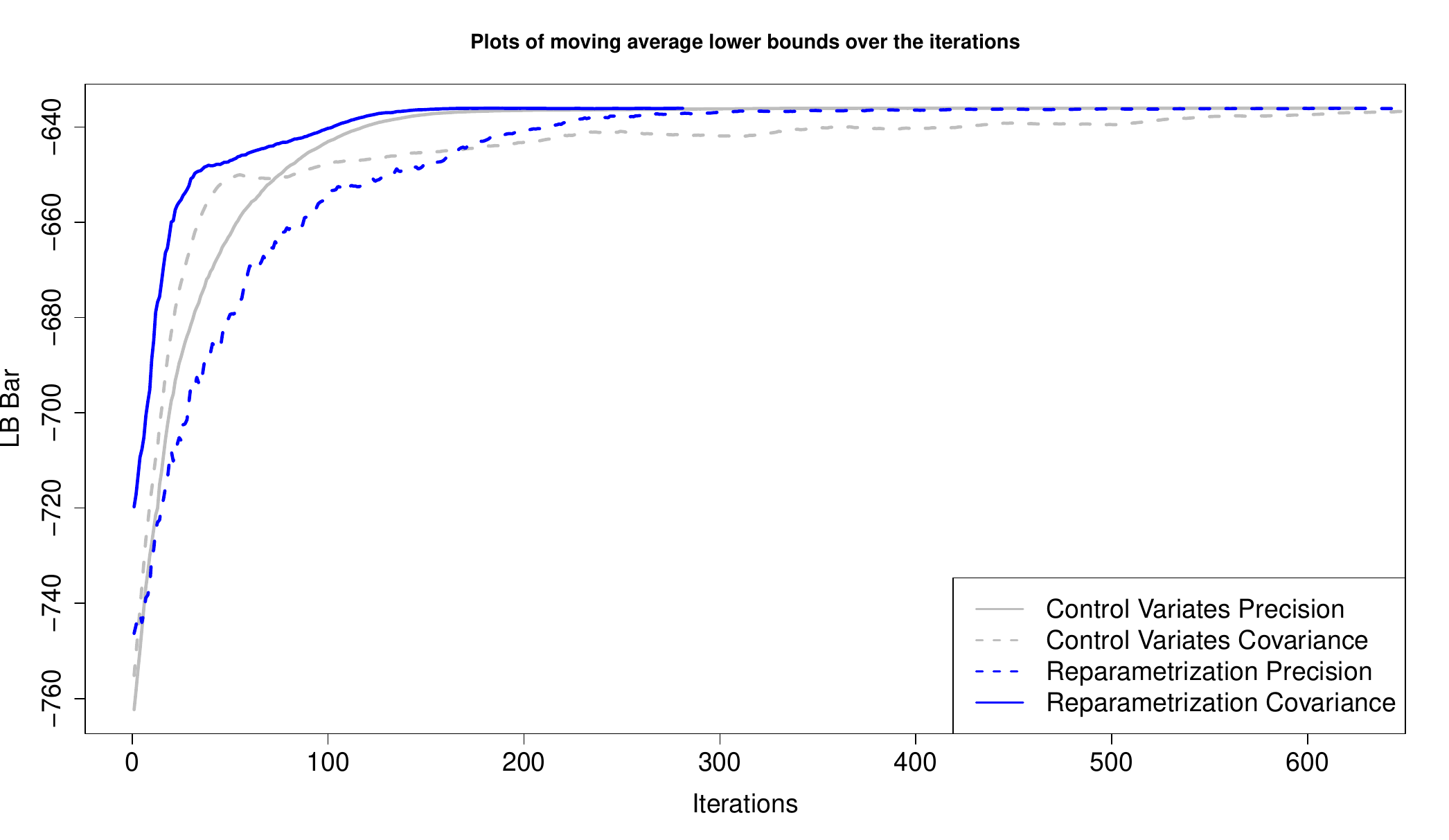}
  \caption{Moving average lower bounds over iterations for the control variates and reparametrization trick methods with different Cholesky factorizations (i.e., precision matrix or covariance matrix). Results refer to a single replicate.}
  \label{fig:ELBO Converance rates}
  \end{minipage}
\end{figure}

Figure \ref{fig:ELBO Converance rates} shows the ELBO values with respect to the number of iterations in implementations based on control variates and the reparametrization trick, using both the Cholesky factorization of the precision matrix $\Omega$ and the covariance matrix $\Sigma$. Interestingly, convergence appears to be faster when using the precision matrix with control variates, and when using the covariance matrix with the reparametrization trick. In particular, when using the covariance matrix with control variates, SVB may fail to reach convergence in some cases.

The accuracy of the approximation is analyzed by comparing it to the posterior distribution obtained through MCMC with 1,000,000 iterations, after discarding the first 100,000 samples as burn-in. MCMC was implemented using random walk Metropolis-Hastings with Gaussian proposal distributions. The mean and variance of the proposal distributions were set to the maximum likelihood estimates, following \cite{chan2016modeling}. For comparison with the MCMC target, we computed the variational approximation accuracy with respect to the posterior density as 
\begin{align*}
    \mbox{accuracy}\equiv 100\left(1-\frac{1}{2}\int_{-\infty}^{\infty}\big\vert q_{\lambda}(\theta)-p(\theta \vert y)\big\vert d\theta\right)\%,
\end{align*} 
with 100\% indicating perfect matching between the variational density $q_{\lambda}(\theta)$ and the posterior distribution $p(\theta \vert y)$ obtained via MCMC.

\begin{figure}[htbp]
\centering
 \begin{minipage}{220pt}
  \includegraphics[width=\columnwidth]{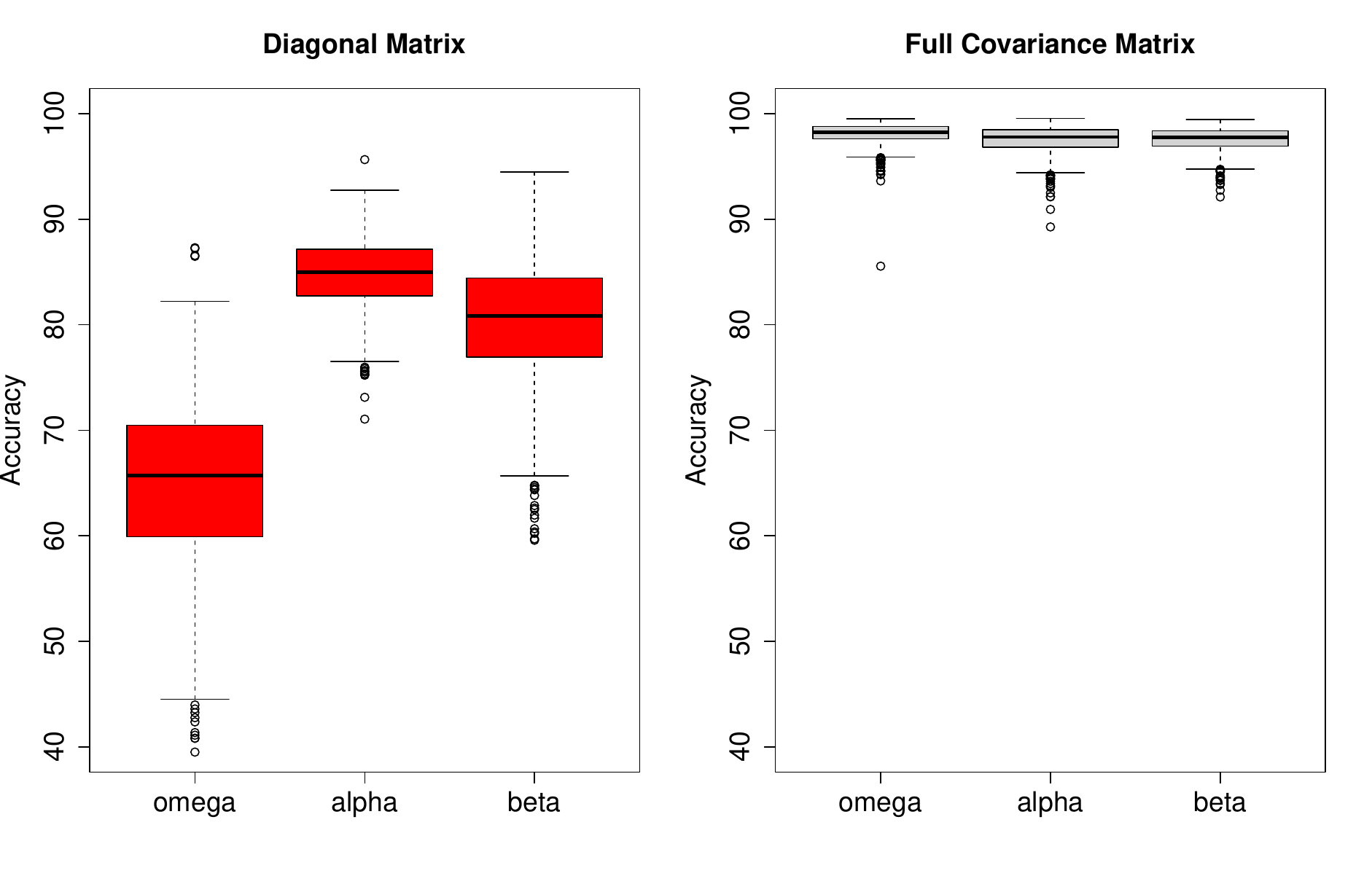}
  \caption{Boxplots of the accuracy of the SVB approximate posterior densities implemented with control variates for the parameters of the GARCH model with Gaussian innovations. The boxplots for SVB using a  full covariance matrix of the Gaussian approximating density are shown in the left panel; boxplots for SVB using a diagonal covariance matrix of the Gaussian approximating density are shown in the right panel.}
  \label{fig:boxplots of Gaussian GARCH accuracy}
  \end{minipage}
\end{figure}

Figure \ref{fig:boxplots of Gaussian GARCH accuracy} compares the accuracy of the variational approximations obtained when using SVB (on the left) and mean-field SVB (on the right) with control variates for time series of length 1,000 from a GARCH(1,1) model with Gaussian innovations. As expected, the mean-field SVB approach, which uses a fully diagonal covariance matrix for the variational approximating density, results in significantly lower accuracy for all the parameters. On the other hand, the implementation based on a full covariance matrix allows to achieve high accuracy compared to the MCMC target.

\begin{figure*}[htbp]
  \centering
  \begin{subfigure}[t]{0.24\textwidth}
    \centering
    \includegraphics[width=\textwidth]{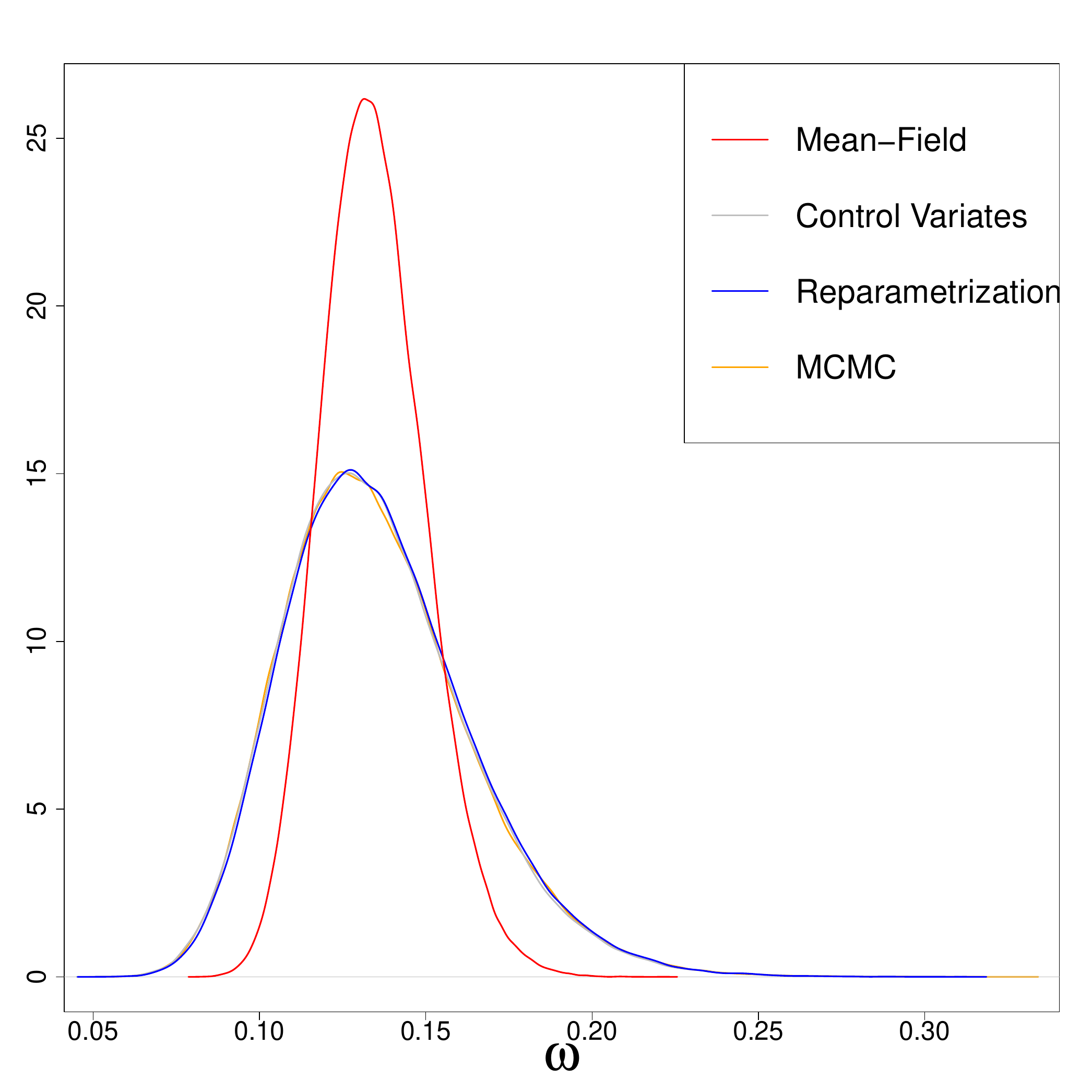}
    \label{fig:plot1}
  \end{subfigure}
  \hfill
  \begin{subfigure}[t]{0.24\textwidth}
    \centering
    \includegraphics[width=\textwidth]{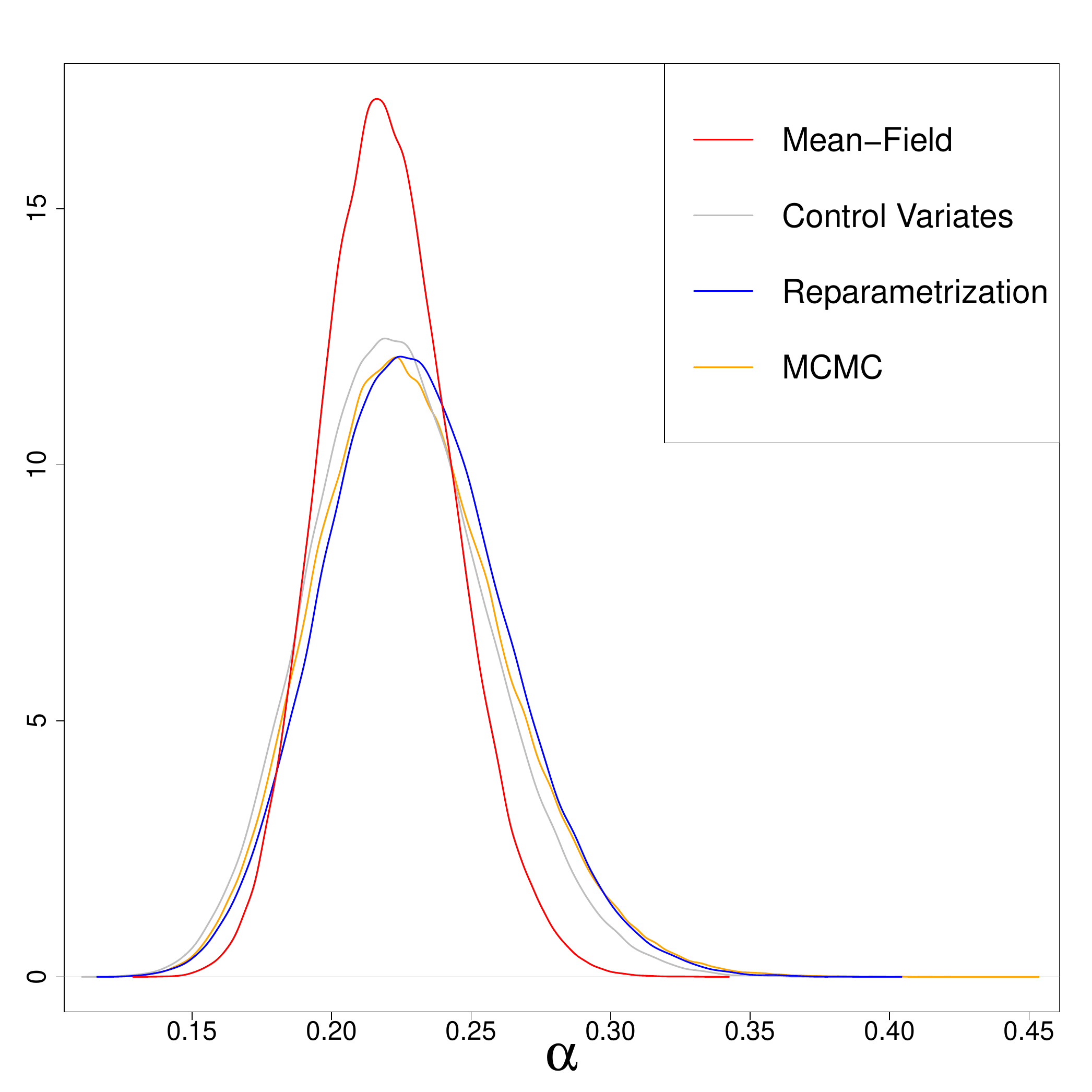}
    \label{fig:plot2}
  \end{subfigure}
  \hfill
  \begin{subfigure}[t]{0.24\textwidth}
    \centering
    \includegraphics[width=\textwidth]{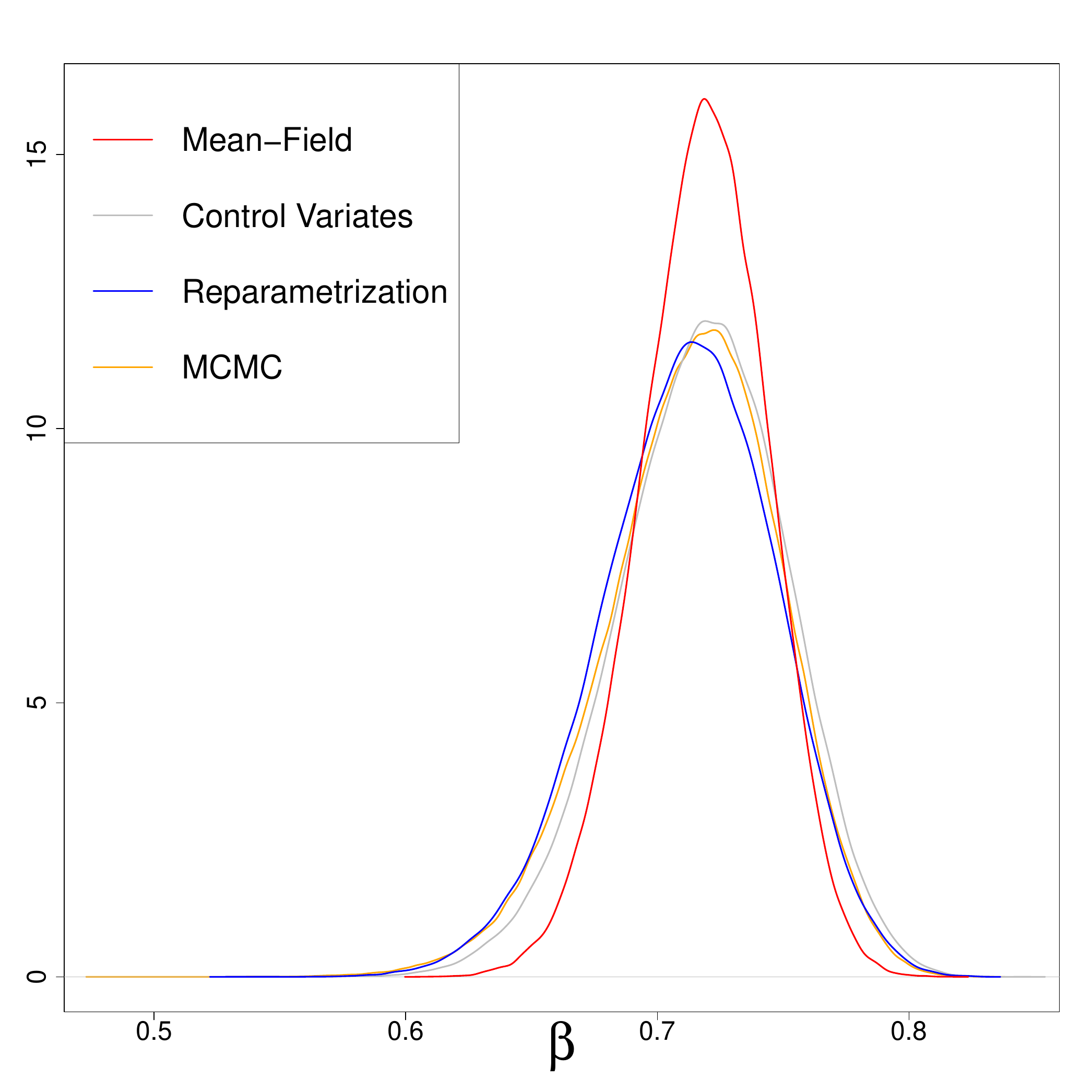}
    \label{fig:plot2}
  \end{subfigure}
  \hfill
  \begin{subfigure}[t]{0.24\textwidth}
    \centering
    \includegraphics[width=\textwidth]{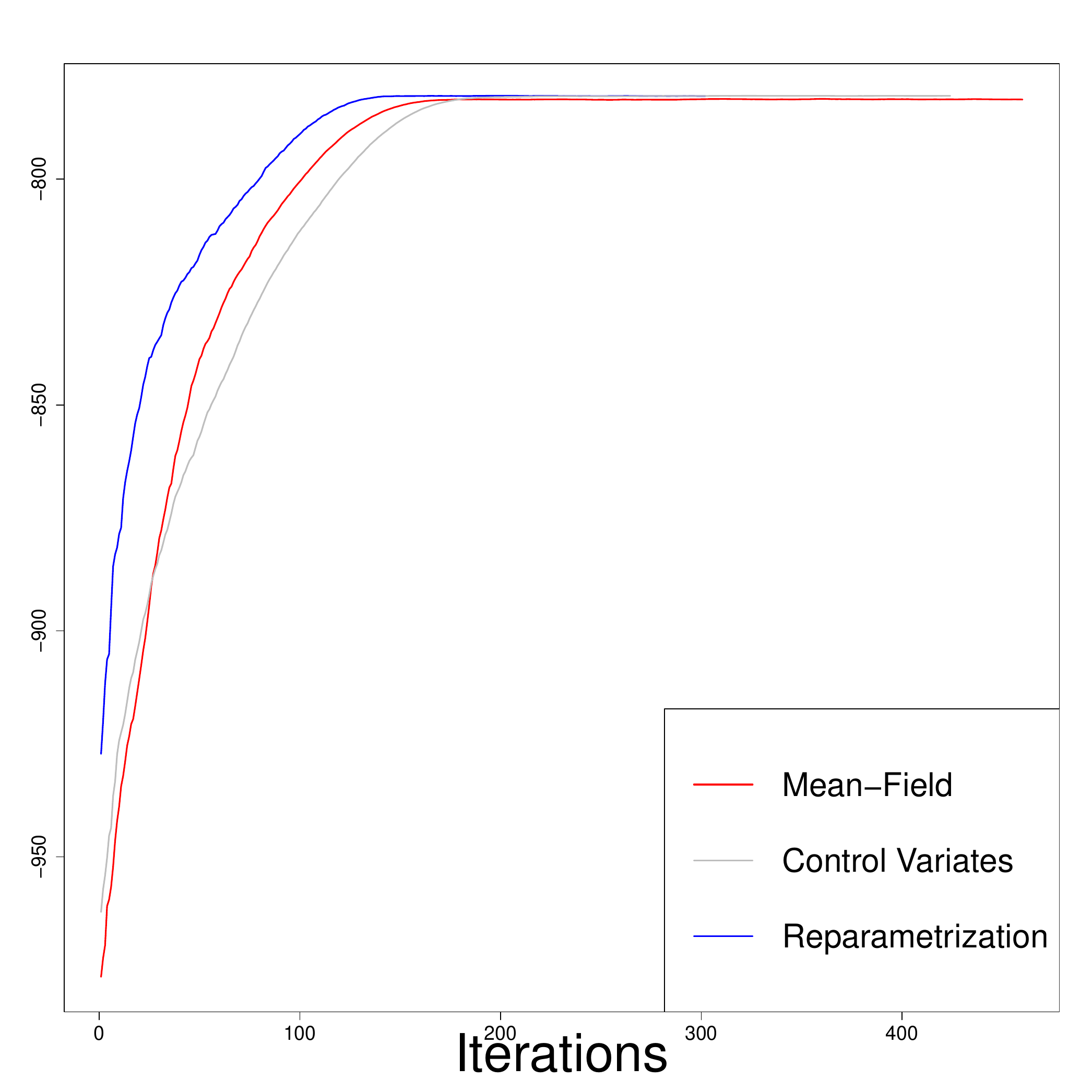}
    \label{fig:plot2}
  \end{subfigure}
  \caption{Posteriors estimations using mean-field VB (with control variates), SVB (with control variates), and SVB (with reparametrization trick) compared with the posterior distributions obtained via MCMC. The last plot on the right shows the moving average ELBO over iterations for a randomly selected simulated Gaussian GARCH-type dataset.}
  \label{fig:Posteriors of GARCH}
\end{figure*}

Figure \ref{fig:Posteriors of GARCH} presents a comparison of the approximated posterior distributions obtained using SVB with a diagonal (mean-field) and non-diagonal covariance matrix, with control variates, or with the reparametrization trick and a non-diagonal covariance matrix, along with the MCMC posterior. On the right, the ELBO is illustrated as a function of the number of iterations. The results are based on time series of length 1,000 with Gaussian innovations. Once again, it is evident that the mean-field SVB approach tends to underestimate the posterior uncertainty. On the other hand, both control variates and the reparametrization trick produce highly accurate approximations. Control variates and the reparametrization trick lead to a higher average ELBO ($-781.57$ for the control variates approach and $-781.56$ for the reparametrization trick implementation) than the mean-field approximation ($ -782.26$), and the reparametrization trick algorithm achieved faster convergence in general.

\begin{figure*}[htbp]
  \centering
  \includegraphics[width=\textwidth]{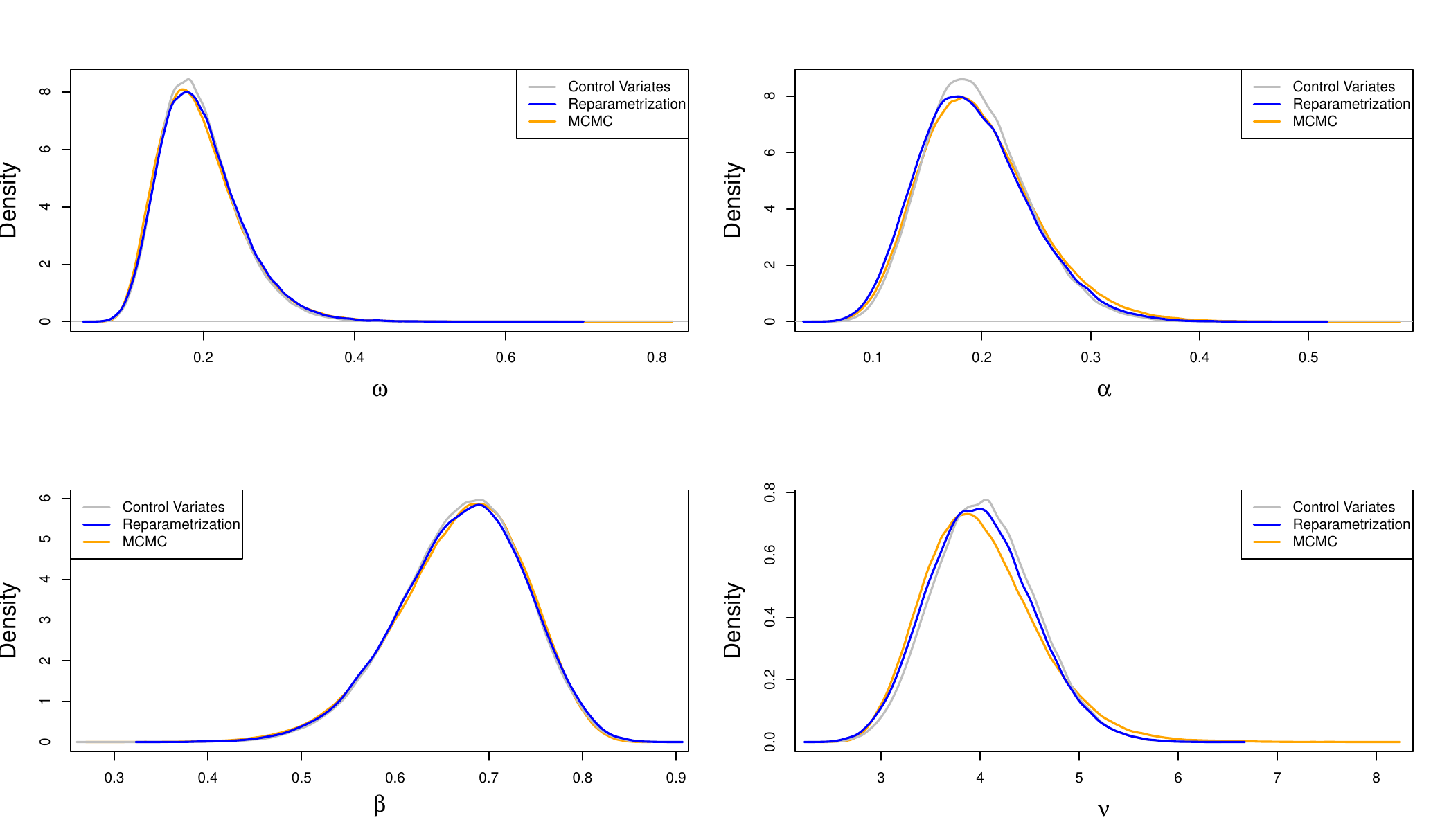}
  \caption{Approximated posterior densities from the SVB methods and MCMC estimates, as well as the moving average ELBO values over iterations, for a randomly selected t GARCH-type dataset.}
  \label{fig: t posteriors}
\end{figure*}
Figure \ref{fig: t posteriors} shows the efficiency of our SVB method in approximating the posterior densities of the parameters for a randomly selected T GARCH-type dataset. Both the control variates and reparametrization trick approaches exhibit an impressive level of agreement with the MCMC approximation, which utilized 1 million Monte Carlo samples. The overlapping of the posterior densities suggests that our SVB methods provide accurate and reliable estimates of the model parameters. 

%Table \ref{tab:posterior-quantiles} displays the 1st quartile, median, and 3rd quartile of the accuracies across the 1,000 replicates. All implementations ensure a high level of accuracy for the approximate posterior densities of the model parameters for the shorter time series, however, the control variates method could not maintain the same level of accuracy if the time series length increases. These findings highlight the superior performance of the reparametrization trick approach in capturing the underlying dynamics of longer time series data.

%\begin{table*}[htbp]
%\centering
%\caption{Parameter posterior quantiles for control variates and reparametrization trick methods over 1,000 replications of simulated t GARCH(1,1) datasets. Values in parentheses are for reparametrization trick.}
%\label{tab:posterior-quantiles}
%\begin{tabular}{c c c c c c c}
%\hline
%& \multicolumn{6}{c}{\textbf{Control Variates (Reparametrization Trick)}} \\
%\hline
%& \multicolumn{3}{c}{1,000 observations} & \multicolumn{3}{c}{5,000 observations} \\
%\hline
%& 1st quartile & median & 3rd quartile & 1st quartile & median & 3rd quartile \\
%\hline
%$\omega$ & 94.40 (94.73) & 96.13 (96.32) & 97.33 (97.58) & 87.83 (94.53) & 91.52 (96.15) & 94.39 (97.51) \\
%$\alpha$ & 93.69 (92.19) & 95.45 (94.56) & %97.09 (96.54) & 88.88 (92.72) & 92.76 (94.83) & 95.50 (96.58) \\
%$\beta$ & 94.47 (93.26) & 96.12 (95.30) & 97.34 (96.75) & 90.75 (93.12) & 93.72 (95.18) & 95.95 (97.07) \\
%$\nu$ & 90.64 (90.28) & 92.50 (92.53) & 94.20 (94.42) & 92.39 (92.90) & 94.55 (95.09) & 96.22 (96.72) \\
%\hline
%\end{tabular}
%\end{table*}

\begin{figure*}[htbp]
  \centering
  \includegraphics[width=\textwidth]{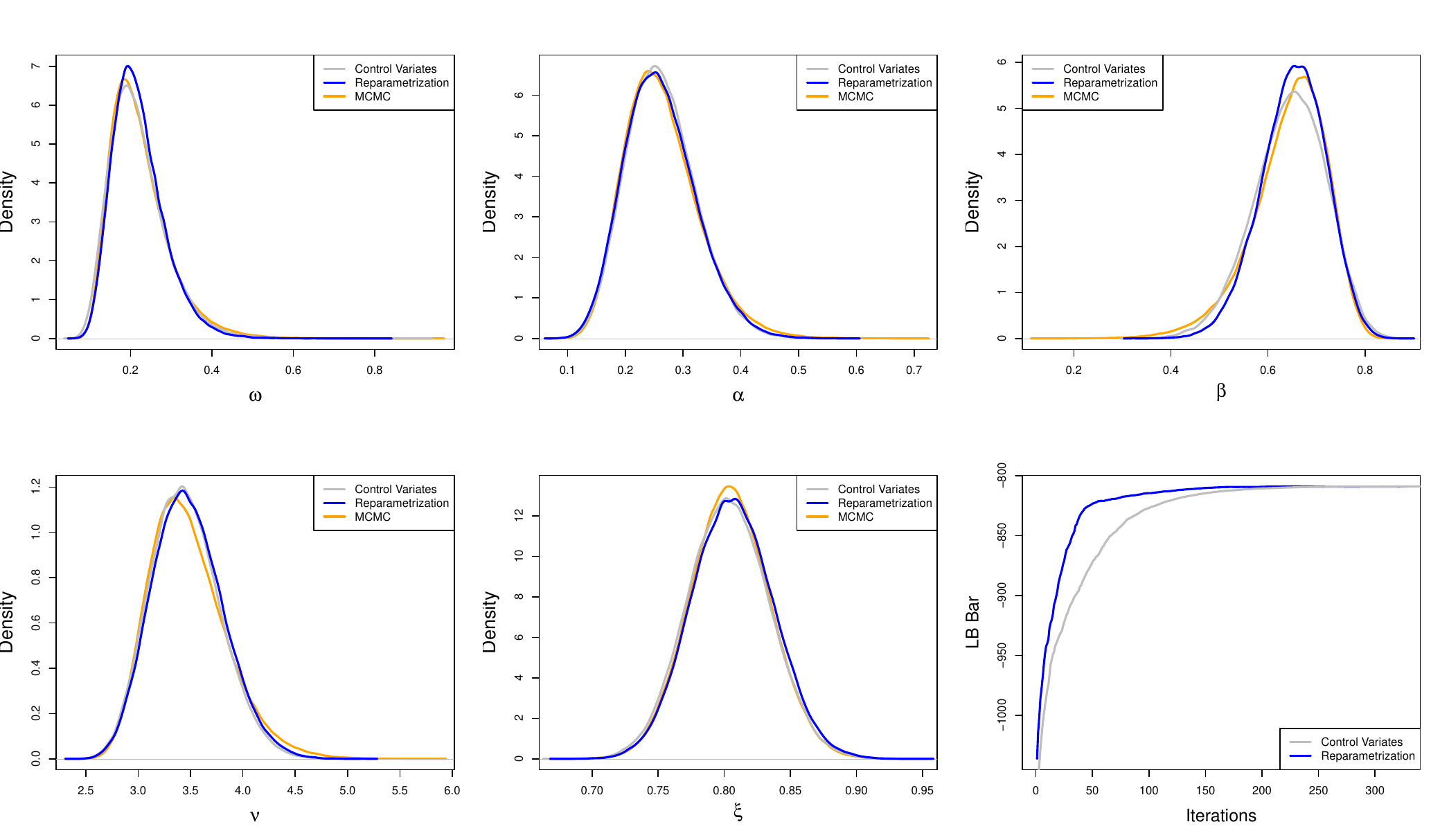}
  \caption{Approximated posterior densities from the SVB methods and MCMC estimates, as well as the moving average ELBO values over iterations, for a randomly selected skewed t GARCH-type dataset.}
  \label{fig:skewed t posteriors}
\end{figure*}

Figure \ref{fig:skewed t posteriors} illustrates the approximate posterior densities for SVB applied to one replicate of the simulation study generated from the GARCH model with skew t innovations. The figure also includes a plot showing the SVB lower bound as a function of the number of iterations for both the control variates and reparametrization trick implementations. All approximations appear to provide accurate results. The convergence rate suggests that the reparametrization trick reaches convergence faster compared to the algorithm using control variates.

Figure \ref{fig:Boxplot Gaussian GARCH} - \ref{fig:Boxplot Skewed T GARCH} (in appendix \ref{secB1}) present boxplots comparing the accuracy results of SVB algorithms for a GARCH(1,1) model with different innovations. For shorter time series, we did not observe a clear advantage in terms of accuracy between the control variates and reparametrization trick approaches. However, as the length of the time series increases, the reparametrization trick method consistently outperforms the control variates method in terms of accuracy. This is evident from the boxplots, where the accuracy results obtained using the control variates method are consistently lower and exhibit a wider spread compared to the reparametrization trick method. 

Table \ref{tab:accuracy-results} provides a summary of the average accuracy levels of the posterior approximations for all the parameters in the three considered models. The table also includes a comparison between time series with 1,000 observations and the longer ones with 5,000 observations. It is evident that the mean-field approximation underestimates the uncertainty of the posterior distribution and exhibits the lowest accuracy. Across all datasets, the accuracy is consistently below 89\%, with $\omega$ being the most challenging parameter to estimate. This issue persists across datasets of different lengths.

On the other hand, both the control variates and reparametrization trick methods achieve higher accuracy results when using an unrestricted covariance matrix. For both methods, the accuracy results for most parameters are consistently above 88\% across different time series lengths except for the skewness parameter $\xi$ in the longer time series with the reparametrization trick approach. Interestingly, the performance of the control variates and reparametrization trick methods varies depending on the length of the time series. The control variates approach demonstrates higher accuracy for shorter time series, while the reparametrization trick method performs better for longer time series. This finding suggests that the choice of method may depend on the specific characteristics and length of the time series under consideration.

\begin{table*}[htbp]
\centering
\caption{Average accuracy for the posterior distributions of each parameter of different GARCH models and methods with 1000 and 5000 observations across 1000 simulated datasets.}
\label{tab:accuracy-results}
\begin{tabular}{l c c c c c c c c c c}
\hline
& \multicolumn{5}{c}{1000 observations} & \multicolumn{5}{c}{5000 observations} \\
\hline
& $\omega$ & $\alpha$ & $\beta$ & $\nu$ & $\xi$ & $\omega$ & $\alpha$ & $\beta$ & $\nu$ & $\xi$ \\
%\hline
Gaussian GARCH MF & 66.19 & 88.08 & 85.75 & & & 58.81 & 84.24 & 78.30 & & \\
%\hline
Gaussian GARCH CV & 96.62 & 96.35 & 96.52 & & & 91.04 & 94.08 & 92.99 & & \\
%\hline
Gaussian GARCH RT & 95.93 & 94.76 & 95.00 & & & 95.35 & 94.64 & 94.61 & & \\
%\hline
t GARCH CV & 95.57 & 95.12 & 95.59 & 92.23 & & 90.88 & 91.79 & 93.05 & 94.07 & \\
%\hline
t GARCH RT & 95.92 & 93.97 & 94.65 & 91.81 & & 95.64 & 94.35 & 94.79 & 94.42 & \\
%\hline
Skewed t GARCH CV & 95.20 & 94.42 & 95.40 & 91.22 & 93.04 & 88.80 & 89.49 & 91.69 & 92.78 & 88.70 \\
%\hline
Skewed t GARCH RT & 95.73 & 94.19 & 94.87 & 90.35 & 91.32 & 94.41 & 93.65 & 93.79 & 92.99 & 82.34 \\
\hline
\end{tabular}
\end{table*}

\begin{table*}[htbp]
\centering
\caption{Average computational running time (in seconds) of SVB algorithms for various GARCH models, with time series of length 1,000 and 5,000 across 1000 simulated datasets.}
\label{tab:garch_times}
\begin{tabular}{lcccccc}
\hline
& \multicolumn{2}{c}{Gaussian GARCH} & \multicolumn{2}{c}{T GARCH} & \multicolumn{2}{c}{Skewed T GARCH} \\
& 1000 obs & 5000 obs & 1000 obs & 5000 obs & 1000 obs & 5000 obs \\
\hline
CV $S=5$ & 2.83 & 6.08 & 3.23 & 7.13 & 5.93 & 17.34 \\
CV $S=10$ & 4.40 & 11.90 & 4.75 & 13.03 & 9.13 & 31.10 \\
CV $S=50$ & 17.35 & 60.70 & 17.95 & 64.44 & 35.46 & 167.56 \\
RT $S=5$ & 4.26 & 16.93 & 4.69 & 17.81 & 8.42 & 32.34 \\
RT $S=10$ & 8.03 & 33.26 & 8.29 & 35.11 & 15.32 & 63.66 \\
RT $S=50$ & 38.78 & 165.92 & 39.55 & 175.37 & 74.46 & 324.71 \\
MCMC & 23.66 & 99.28 & 26.47 & 109.39 & 75.23 & 326.98 \\
\hline
\end{tabular}
\end{table*}

Table \ref{tab:garch_times} presents the computational running times of all the considered implementations, including a comparison with standard MCMC. The MCMC was stopped after 50,000 iterations, as suggested in \cite{tan2018gaussian} and \cite{ong2018gaussian}, and convergence was also checked using the Gelman-Rubin diagnostics and autocorrelations \citep{brooks1998general}.

The table presents the average computational times for various approximation methods applied to 1,000 datasets with different numbers of Monte Carlo samples $S$. It is observed that the computational time increases as the number of MC samples increases. Comparing the same number of MC samples, the control variates approach exhibits faster computational times than the reparametrization trick method. However, it is important to note that the accuracy results of the control variates approach may not meet the same standards, particularly for longer time series. By selecting MC samples of $S=10$ for the control variates approach and $S=5$ for the reparametrization trick approach, we are able to achieve reasonable accuracy results while significantly reducing the computational time compared to the traditional MCMC sampling approximation method. 

For time series with 1,000 observations, SVB with control variates is 4.37, 4.57, and 7.24 times faster than the traditional MCMC method for Gaussian, t, and skewed t GARCH models, respectively. The reparametrization trick method performs even faster, precisely 4.55, 4.64, and 7.93 times faster than MCMC. For the longer time series with 5,000 observations, SVB with control variates is 7.34, 7.40, and 9.51 times faster than MCMC in terms of average computational time, while SVB with the reparametrization trick is 4.86, 5.14, and 9.11 times faster for Gaussian, t, and skewed t GARCH models, respectively.

While offering a notable reduction in computational time compared to traditional MCMC methods, our SVB methods maintain reasonable accuracy in approximating the posterior densities in GARCH-type models. This combination of accuracy and computational speed makes SVB an attractive alternative for analyzing GARCH models in various applications.
\subsection{Sequential Variational Bayes}

We now investigate the performance of the sequential alternatives to SVB described in Section \ref{sec4} and compare the accuracy and computational times of UVB and Seq-SVB. We generate 1,000 datasets from the GARCH model with skewed t innovations and 1,000 observations and an equal number of replicates from the same model but with 5,000 observations. We perform this comparison focusing on the skewed t case, which comprises both Gaussian and t GARCH models as special cases.

For the sequential VB method, we exclusively employ the reparametrization trick approach with a full covariance matrix. This choice is motivated by the fact that the mean-field batch version offers significantly lower accuracy compared to the unrestricted case, as shown in the previous simulated experiments. 
The decision to choose the reparametrization trick approach over the control variates method for our sequential variational Bayes implementation was based on the availability of prior discussions and implementations as the control variates method has already been extensively discussed and implemented in \cite{tomasetti2022updating}. On the other hand, the reparametrization trick for UVB has not been explored in previous work. Therefore, to expand the scope of our study and explore alternative approaches, we opted to implement the reparametrization trick for the sequential version.  

\begin{figure}[htbp]
\centering
 \begin{minipage}{220pt}
  \includegraphics[width=\columnwidth]{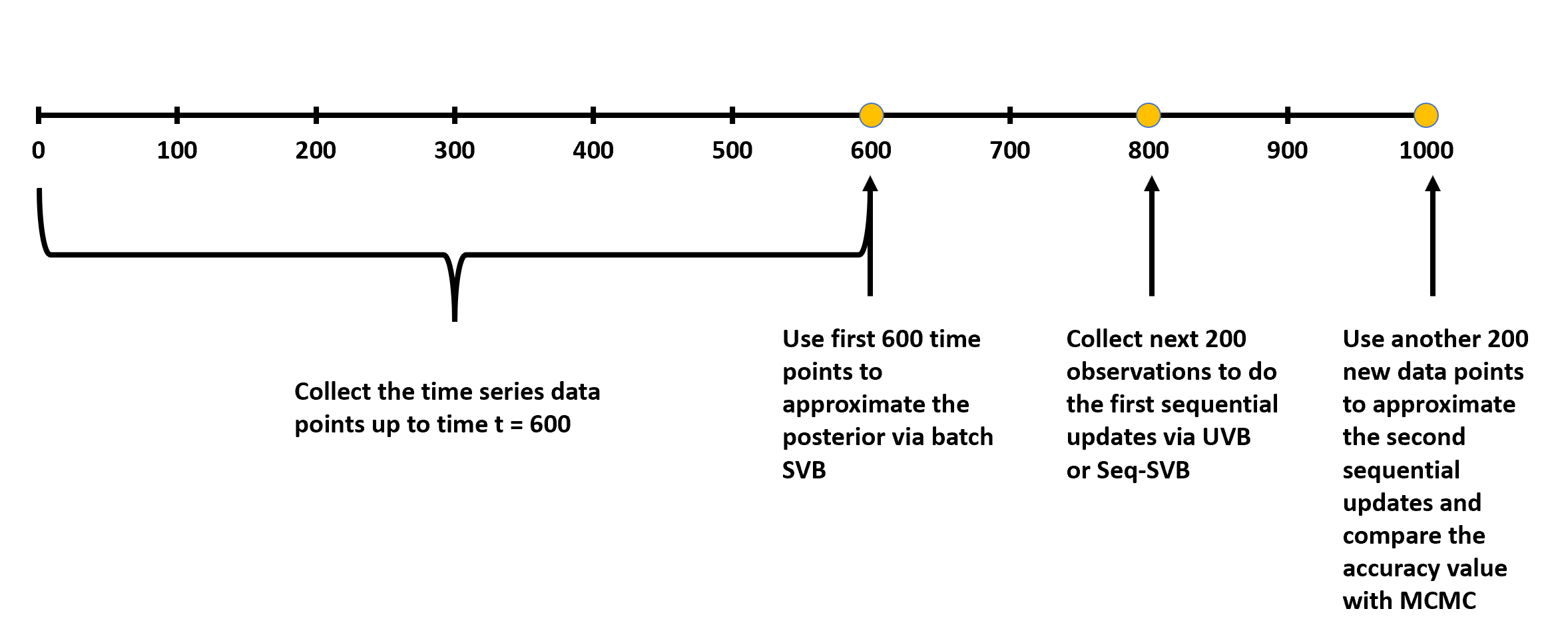}
  \caption{Demonstration of sequential SVB method, i.e., $T_n = 600$ and $c = 2$, the posterior is estimated for the first $600$ time points via batch SVB, and the posteriors are updated twice as new observations arrive up to $T_N = 1000$.}
  \label{fig:demonstration of sequential svb}
  \end{minipage}
\end{figure}

For the time series with 1,000 observations, i.e., $T_N=1,000$, we choose ten different starting points $T_n = {500,550,600,\ldots,950}$ and perform sequential updates every $(T_N-T_n)/c$ observations, with $c={1,2,5,10}$. For example, if $T_N=1,000$, $T_n=600$ and $c=2$, the posterior distribution is first approximated using batch SVB on the first 600 observations, and then the posterior distribution is updated twice as new data in blocks of 200 observations arrive (see Figure \ref{fig:demonstration of sequential svb}).

%Since UVB targets the approximated posterior distributions rather than the true densities, in order to achieve more stable convergence results of sequential SVB methods, we introduced stricter stopping criteria: the maximum patience parameter is set to $P=200$, and the moving average window size parameter is set to $t_W=50$. We maintained a fixed number of Monte Carlo samples at $S=10$. 

\begin{figure*}[htbp]
    \centering
    \includegraphics[width=\textwidth]{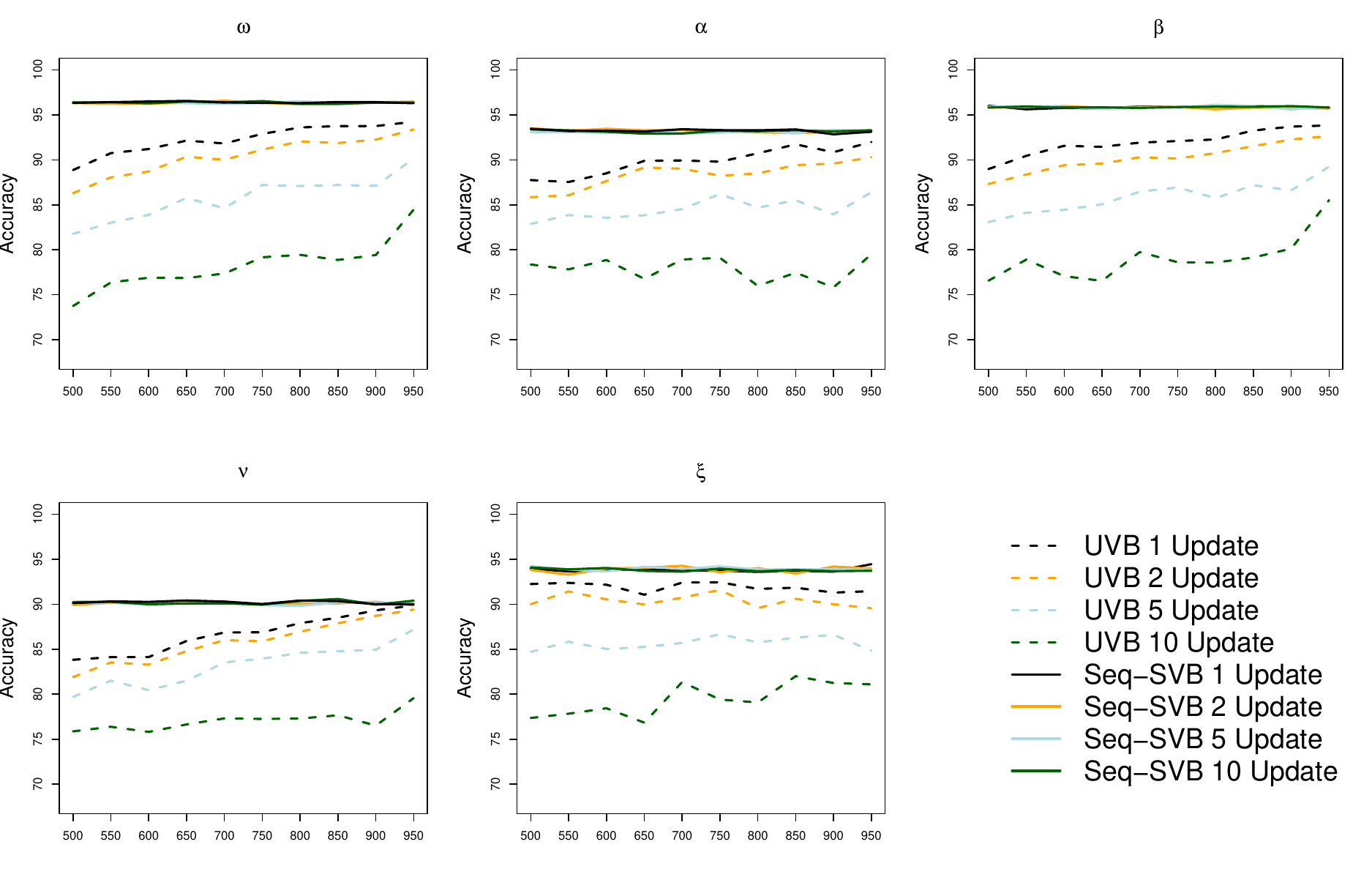}
    \caption{Average accuracy over 1,000 simulated skew t GARCH datasets with 1,000 observations for different starting values and different numbers of sequential updates.}
    \label{Accuracy 1000}
\end{figure*}

Figure \ref{Accuracy 1000} shows the average accuracy obtained across 1,000 simulated skew t GARCH datasets, each containing 1,000 observations. The accuracy is measured for different starting values and various numbers of sequential updates. Seq-SVB methods consistently outperform UVB in terms of accuracy. This can be attributed to the fact that Seq-SVB directly targets the true posteriors instead of aiming to approximate the pseudo-posterior distributions. Remarkably, Seq-SVB methods exhibit nearly identical accuracy regardless of the number of updates or the choice of starting points. This result is expected, as the only difference among Seq-SVB approaches is the initial value of the variational parameters for the SVB algorithm, which ultimately should converge to the same value. On the other hand, UVB approaches produce accuracy results that depend on the number of observations in the updates. As the number of sequential updates increases, the accuracy tends to decrease since the posterior approximation at each update is based on fewer observations, and each sequential update introduces a slight deviation from the true posterior. If the initial SVB approximation is based on more observations, the overall approximation is more accurate.

\begin{figure}[htbp]
\centering
 \begin{minipage}{220pt}
  \includegraphics[width=\columnwidth]{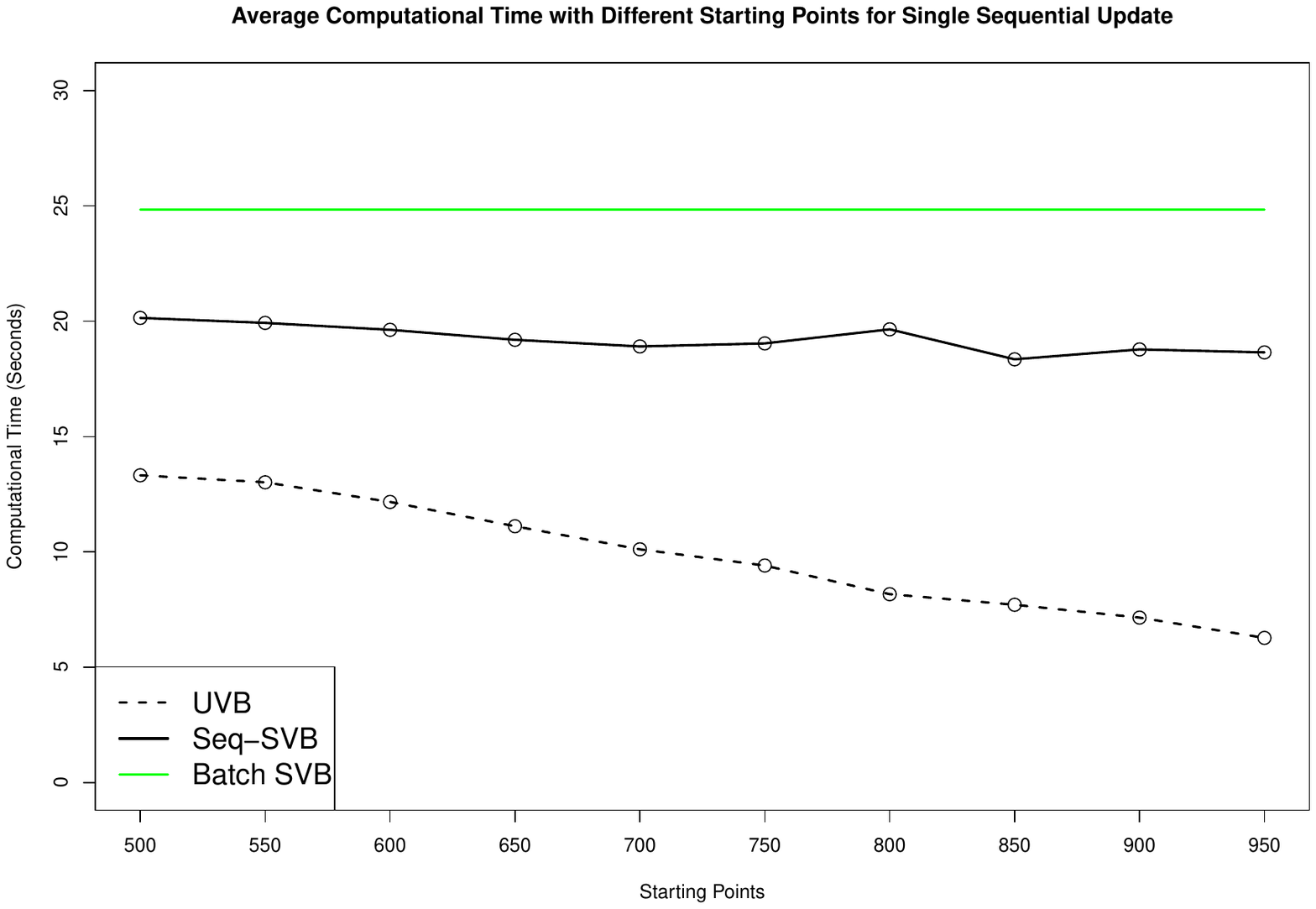}
  \caption{Both UVB and Seq-SVB outperform batch SVB in terms of computational speed. Seq-SVB shows a small downward slope and UVB exhibits a more noticeable downward slope indicating different computational gain.}
  \label{fig:computational time online}
  \end{minipage}
\end{figure}

Figure \ref{fig:computational time online} shows the average computational time with different starting points for a single sequential update. Both UVB and Seq-SVB outperform batch SVB in terms of computational speed. Seq-SVB exhibits a small downward slope, indicating a slight decrease in computation time as the number of new data points decreases. UVB, on the other hand, shows a more noticeable downward slope, indicating a more significant reduction in computational time with fewer newly arrived data points.

In Seq-SVB, the initial values used in the optimization algorithm can have an impact on convergence and computational time. When fewer new data points are available, the initial values might already be relatively close to the final converged values at time $T_N$. As a result, the optimization algorithm requires fewer iterations to reach convergence, leading to a slight decrease in computational time.

Nevertheless, the significant reduction in computation time observed in the UVB approach can be attributed to the lower number of observations that need to be recomputed in the log-likelihood calculations. UVB only needs to incorporate the likelihood of the newly arrived data points rather than the entire dataset, resulting in faster computations.

Table \ref{tab:online computational time} presents the average computational time for a single update using both UVB and Seq-SVB with 1,000 and 5,000 observations. Regardless of the time series length, the trends remain consistent. UVB consistently outperforms Seq-SVB in terms of computational speed per update, and the computational gains are more significant when there are fewer newly arrived data points.

This advantage of UVB becomes more evident with longer time series lengths as Seq-SVB requires processing the entire data series for each update, while UVB only needs to incorporate the incremental time series.
\begin{table*}[htbp]
        \centering
        \resizebox{\textwidth}{!}{
        \begin{tabular}{l c c c c c c c c c c}
        \hline
       & $t=500$ & $t=550$ &$t=600$ &$t=650$ &$t=700$ &$t=750$ &$t=800$ &$t=850$ & $t=900$& $t=950$\\\hline
         UVB 1000 observations &13.32& 13.02 &12.17& 11.11 &10.11 & 9.41  &8.17 & 7.71 & 7.15&  6.27 \\
          Seq-SVB 1000 observations  & 20.14 &19.93 &19.63 &19.19& 18.91 &19.04& 19.65& 18.35& 18.78 &18.65 \\
          \hline
          & & & & & & & & & & \\
          \hline
          & $t=2500$ & $t=2750$ &$t=3000$ &$t=3250$ &$t=3500$ &$t=3750$ &$t=4000$ &$t=4250$ & $t=4500$& $t=4750$\\
          \hline
          UVB 5000 observations &46.16 &42.81 &38.33 &33.00 &30.14 &26.30 &22.59 &17.27 &13.81  &9.06 \\
          Seq-SVB 5000 observations  & 80.31 &84.80 &83.74 &83.16 &83.36 &78.27 &85.03 &79.21 &81.60 &78.99 \\
          \hline
        \end{tabular}}
        \caption{Average computational time for one single update will decrease as the number of new observations decreases. UVB is caused by the greater number of the likelihood function computed. Seq-SVB is due to worse starting values for SVB.}
        \label{tab:online computational time}
    \end{table*}
    
\section{Standard \& Poor Data}\label{sec6}

We apply our batch and sequential variational Bayes methods to a real financial stock return dataset, specifically a time series of the Standard \& Poor 500 (S\&P500) stock index. After performing the data clearance procedure, which involved removing any missing or incomplete data, as well as calculating the log-returns of the time series, we obtained a dataset comprising 1,000 observations. These observations were recorded over a period spanning from 2nd January 2015 to 4th January 2019.
The log-returns of the time series exhibit a range between -4.18\% and 5.69\%, with a median log return of 0.03\%. The distribution of the log returns is negatively skewed, as evidenced by a skewness value of -0.33. Additionally, the kurtosis of the distribution is higher than that of a normal distribution, indicating excess kurtosis, with a kurtosis value of 4.59. We implemented the SVB approaches using Gaussian, T, and skewed T models on the provided dataset. Model selection criteria, namely the Akaike Information Criterion (AIC) and Bayesian Information Criterion (BIC), were used to assess the performance of the models. The AIC and BIC values for each model are reported in Table \ref{tab:aic-bic}.
%\begin{minipage}{220pt}
\begin{table}[htbp]
\centering
\begin{tabular}{lcc}
\hline
Model & AIC & BIC \\
\hline
Gaussian GARCH & 2262.46 & 2277.18 \\
T GARCH & 2170.88 & 2190.51 \\
Skewed T GARCH & 2165.44 & 2189.98 \\
\hline
\end{tabular}
\caption{AIC and BIC values for GARCH models}
\label{tab:aic-bic}
\end{table}
%\end{minipage}
The skewed T GARCH model returns the lowest values for both AIC and BIC. With an AIC value of 2165.44 and a BIC value of 2189.98, the skewed T GARCH model is selected as the best-fitting model for the dataset.

\begin{figure*}[htbp]
  \centering
  \includegraphics[width=\textwidth]{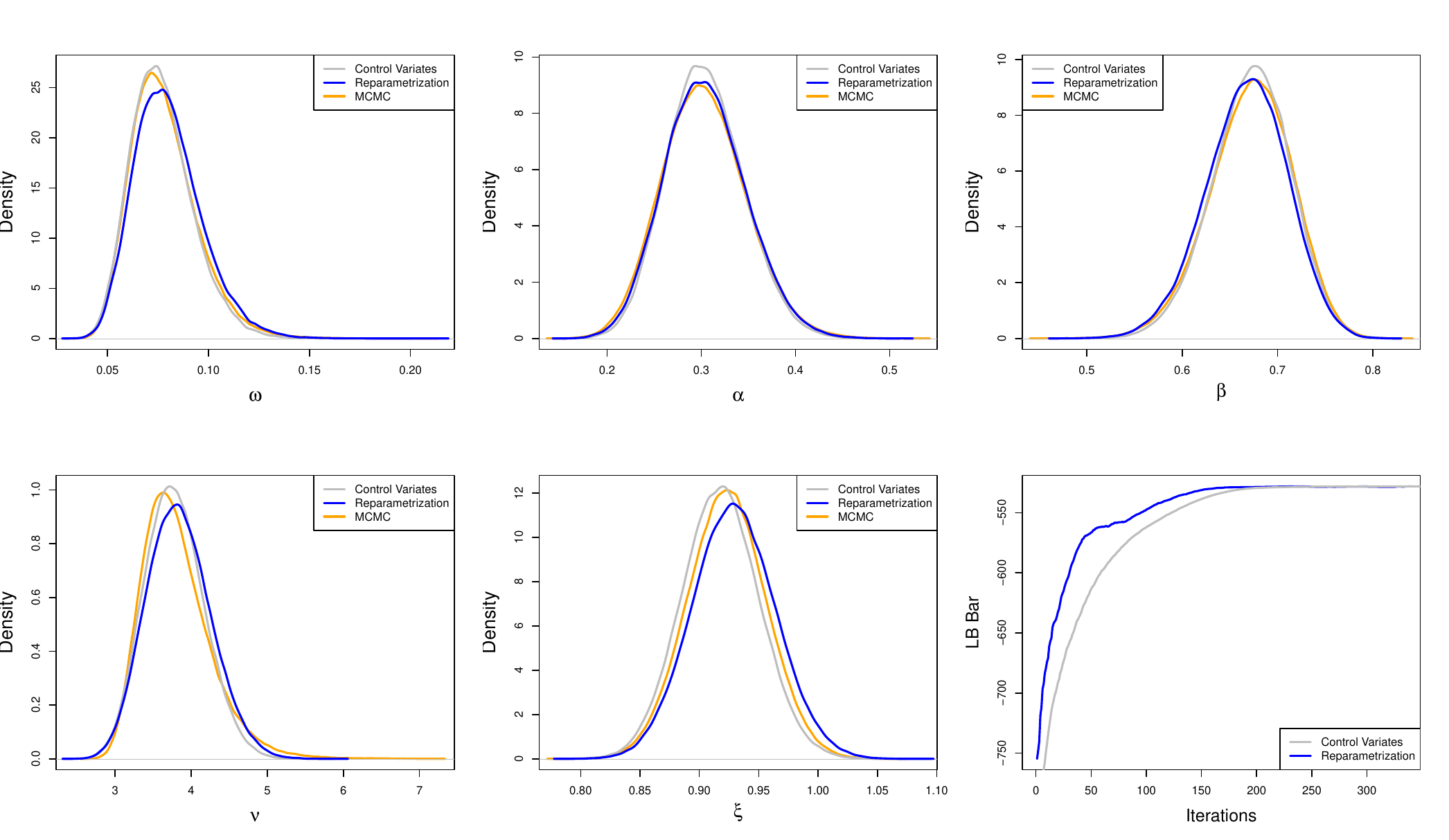}
  \caption{S\&P500 dataset. The posterior density plots of the skewed t GARCH model parameters $\omega,\alpha,\beta,\nu,\xi$ using MCMC and batch SVB with control variates and reparametrization trick methods.}
  \label{fig:skewed t real data}
\end{figure*}

Figure \ref{fig:skewed t real data} showcases the estimated marginal posterior density plots for the parameters $\omega, \alpha, \beta, \nu,$ and $\xi$ of the skewed t GARCH model, using MCMC and batch SVB with control variates and the reparametrization trick. The density plots obtained from both the control variates method and the reparametrization trick closely resemble the estimates obtained from the MCMC method, indicating a high level of accuracy in approximating the posterior densities.

The accuracy results for the control variates method are 97.56\% for $\omega$, 95.86\% for $\alpha$, 97.42\% for $\beta$, 93.23\% for $\nu$ and 93.44\% for $\xi$. The reparametrization trick implementation yields similar accuracy results: 94.06\% for $\omega$, 98.13\% for $\alpha$, 95.42\% for $\beta$, 90.49\% for $\nu$ and 92.80\% for $\xi$. These accuracy percentages demonstrate the effectiveness of both implementations in providing reasonable approximations of the posterior densities from the skewed t GARCH model when applied to a real S\&P500 dataset.

\begin{figure*}[htbp]
    \centering
    \includegraphics[width=\textwidth]{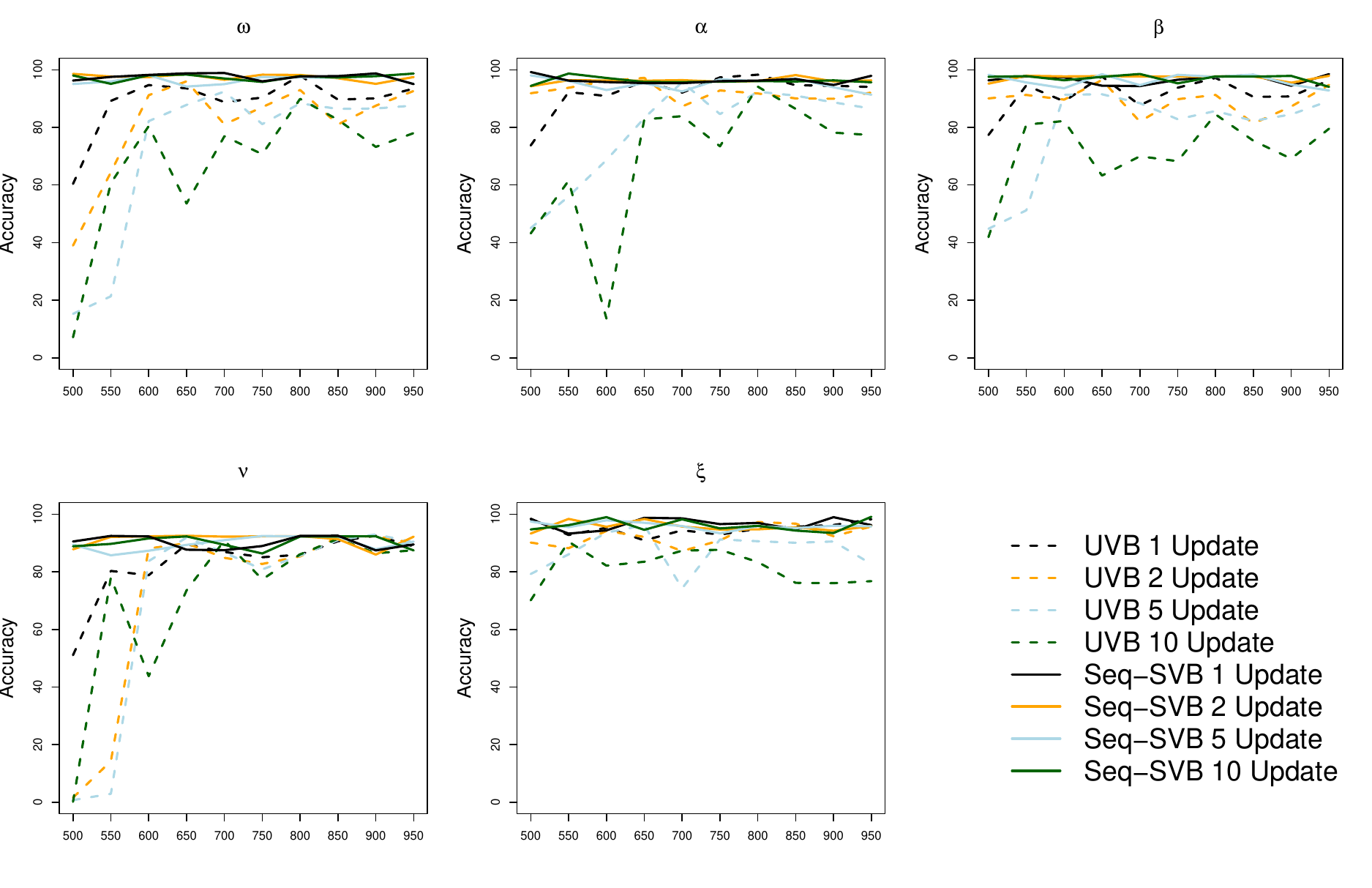}
    \caption{Accuracy results of a real S\&P 500 dataset with different starting values and different numbers of sequential updates under the skewed T GARCH model.}
    \label{fig:RealDataSeqVB}
\end{figure*}

Figure \ref{fig:RealDataSeqVB} illustrates the accuracy results obtained from applying the skewed t GARCH model to the real S\&P 500 dataset using the Seq-SVB and UVB methods. Different starting values and varying numbers of sequential updates are considered.

Consistent with previous findings, the Seq-SVB approaches consistently outperform the UVB methods in terms of accuracy. The accuracy results in the UVB methods are significantly influenced by the choice of starting values and the number of updates. On the other hand, the Seq-SVB approach demonstrates robust accuracy regardless of the starting values and the number of updates.

The upward slope observed in the plot indicates that using fewer newly arrived data points leads to more accurate approximations. This suggests that incorporating a smaller portion of the newly observed data in each update results in improved accuracy. The Seq-SVB approach benefits from this property, as it only updates the variational parameters with a subset of the newly arrived data, leading to more accurate posterior approximations.

These results provide further evidence that the Seq-SVB methods are well-suited for sequential inference tasks, such as analyzing real financial datasets. They offer accurate approximations of the posterior distributions, and the accuracy is relatively insensitive to the choice of starting values and the number of updates.

\begin{table}[htbp]
\centering
 \begin{minipage}{220pt}
\begin{tabular}{cccccc}
\hline
&SVB&1update& 2update&5update&10update\\
\hline
$\omega$ & 96.06 & 88.82 & 81.29 & 72.91 & 67.33 \\
$\alpha$ & 99.18 & 92.39 & 92.29 & 79.19 & 69.45 \\
$\beta$ & 97.80 & 91.51 & 89.39 & 79.18 & 71.52 \\
$\nu$ & 90.60 & 82.96 & 72.12 & 70.66 & 71.51 \\
$\xi$ & 96.26 & 94.98 & 92.48 & 87.45 & 81.35 \\
\hline
\end{tabular}
\caption{Comparison of the average accuracy results for the skewed t GARCH model parameters using batch SVB and UVB with ten different starting values, $T_n = \{500,550,600,\ldots,950\}$.}
\label{tab:accuracyresults2}
\end{minipage}
\end{table}

\begin{table}[htbp]
\centering
 \begin{minipage}{220pt}
\begin{tabular}{cccccc}
\hline
&SVB&1update& 2update&5update&10update\\
\hline
$\omega$ & 96.06 & 97.50 & 97.49 & 96.40 & 97.40 \\
$\alpha$ & 99.18 & 96.37 & 96.20 & 95.00 & 96.17 \\
$\beta$ & 97.80 & 96.56 & 97.29 & 96.25 & 97.05 \\
$\nu$ & 90.60 & 90.13 & 91.12 & 89.80 & 90.24 \\
$\xi$ & 96.26 & 96.65 & 95.63 & 95.98 & 96.05 \\
\hline
\end{tabular}
\caption{Comparison of the average accuracy results for the skewed t GARCH model parameters using batch SVB and Seq-SVB with ten different starting values, $T_n = \{500,550,600,\ldots,950\}$.}
\label{tab:accuracyresults3}
\end{minipage}
\end{table}

Table \ref{tab:accuracyresults2} and \ref{tab:accuracyresults3} summarize the mean accuracy results for the parameters of the skewed t GARCH model using UVB and Seq-SVB approaches, compared to the batch SVB method. The accuracy results are obtained by considering ten different starting values ranging from $T_n = 500$ to $T_n = 950$.

The findings indicate that the accuracy of the UVB method decreases as the number of updates increases. This suggests that the UVB method may struggle to provide accurate results when a larger number of updates are applied. On the other hand, the Seq-SVB method exhibits more consistent accuracy results across different starting values, indicating its robustness and insensitivity to the number of updates.

However, it is important to note that the Seq-SVB method requires a longer computational time compared to UVB. The mean computational time for Seq-SVB, considering ten different starting values, is 47.62 seconds for a single update, while UVB only requires 17.11 seconds. Therefore, the choice between Seq-SVB and UVB should consider the trade-off between computational time and accuracy. Seq-SVB may provide higher accuracy but requires more computational resources, while UVB offers a more time-efficient solution at the expense of slightly lower accuracy.

In summary, the Seq-SVB method demonstrates robust accuracy across different starting values and update frequencies, while UVB is more sensitive to the number of updates but computationally faster. Researchers should consider these trade-offs based on their specific requirements and constraints when selecting between Seq-SVB and UVB for their applications.

\section{Conclusion}
\label{sec:concl}

This paper introduces and compares batch stochastic variational Bayes (SVB) algorithms for approximating posterior densities in GARCH-type models, including Gaussian, t, and skewed t GARCH models. The control variates and reparametrization trick methods are implemented and compared with traditional MCMC estimation. Both SVB methods demonstrate high accuracy in approximating the posterior densities of GARCH time series models.

For the models examined in this study, both the control variates and reparametrization trick approaches demonstrate notable improvements in computational efficiency compared to traditional MCMC methods. The computational time is influenced by the number of Monte Carlo samples $S$. In general, the control variates method exhibits faster computation times compared to the reparametrization trick method when the same number of MC samples is employed. However, the control variates method may struggle to maintain the same level of accuracy when using a small number of MC samples, especially for the longer time series.

We have provided sequential versions of the SVB approaches, namely UVB and Seq-SVB, to handle streaming data or scenarios with frequent updates. Both UVB and Seq-SVB outperform the batch SVB methods in terms of computational speed. UVB achieves significant time savings by only evaluating the likelihood of new observations in each update. However, UVB may sacrifice some accuracy compared to Seq-SVB, especially with a high number of sequential updates or a substantial number of new observations.

A promising approach is to run UVB and Seq-SVB in parallel, leveraging the strengths of both methods. Additionally, refreshing the pseudo-posterior in UVB using the posterior obtained from Seq-SVB can potentially improve accuracy while still benefiting from the computational efficiency of UVB.

In summary, stochastic variational inference methods, particularly the reparametrization trick approach, provide an effective and efficient framework for posterior approximation in GARCH time series models. The choice between UVB and Seq-SVB depends on the specific trade-off between computational speed and accuracy requirements, making them valuable tools for researchers and practitioners in various applications involving time series analysis.

\backmatter

\bmhead{Acknowledgments}
This research includes computations using the computational cluster Katana supported by Research Technology Services at UNSW Sydney.
\begin{appendices}

\section{}\label{secA1}

\subsection{Analytical Derivatives of Time-varying Variance Series in Gaussian GARCH Model} \label{Appen 1}
By using the chain rule,
\begin{align}
    \frac{\partial \sigma_t^2}{\partial \theta_{\phi}}=\frac{\partial \sigma_t^2}{\partial\phi}\frac{\partial\phi}{\partial 
    \theta_{\phi}},
    \label{eqn:chain rule}
\end{align}
where $\phi = (\omega,\alpha,\beta)^\top$ is the original parameters for the classical GARCH(1,1) model. Taking the partial derivatives with respect to $\phi$, we can derive the following recursive formulas:
\begin{align*}
    \frac{\partial \sigma_t^2}{\partial\omega}&=1+\beta\frac{\partial \sigma_{t-1}^2}{\partial \omega}\\
    \frac{\partial \sigma_t^2}{\partial\alpha} &=y_{t-1}^2+\beta\frac{\partial \sigma_{t-1}^2}{\partial \alpha}\\
    \frac{\partial \sigma_t^2}{\partial\beta} &= \sigma_{t-1}^2+\beta\frac{\partial \sigma_{t-1}^2}{\partial \beta};
\end{align*}
the derivatives of the original parameters $\phi$ with respect to the unconstrained parameters $\theta_{\phi} = (\theta_\omega,\theta_{\psi_1},\theta_{\psi_2})^\top$, $\frac{\partial\phi}{\partial \theta_\phi}$, are:
\begin{align*}
    \frac{\partial \omega}{\partial \theta_\omega} &= \exp(\theta_\omega)\\
    \frac{\partial \alpha}{\partial \theta_{\psi_1}} &= \frac{\exp(-\theta_{\psi_1})}{(1+\exp(-\theta_{\psi_1}))^2}\frac{1}{1+\exp(-\theta_{\psi_2})}\\
    \frac{\partial \alpha}{\partial \theta_{\psi_2}} &= \frac{\exp(-\theta_{\psi_2})}{(1+\exp(-\theta_{\psi_2}))^2}\frac{1}{1+\exp(-\theta_{\psi_1})}\\
     \frac{\partial \beta}{\partial \theta_{\psi_1}} &= \frac{\exp(-\theta_{\psi_1})}{(1+\exp(-\theta_{\psi_1}))^2}\frac{\exp(-\theta_{\psi_2})}{1+\exp(-\theta_{\psi_2})}\\
     \frac{\partial \beta}{\partial \theta_{\psi_2}} &= - \frac{\exp(\theta_{\psi_2})}{(1+\exp(\theta_{\psi_2}))^2}\frac{1}{1+\exp(-\theta_{\psi_1})} \\
     \frac{\partial \omega}{\partial \theta_{\psi_1}} & = \frac{\partial \omega}{\partial \theta_{\psi_2}} = \frac{\partial \alpha}{\partial \theta_{\omega}} = \frac{\partial \beta}{\partial \theta_{\omega}} = 0.
\end{align*}
Combining all the parts together, we can derive an analytical form of \eqref{eqn:chain rule}, 
\begin{align*}
    \frac{\partial \sigma_t^2}{\partial \theta_\omega} & = \frac{\partial \sigma_t^2}{\partial\omega} \frac{\partial \omega}{\partial \theta_\omega} = \omega + \beta\frac{\partial \sigma_{t-1}^2}{\partial \theta_\omega} \\
    \frac{\partial \sigma_t^2}{\partial \theta_{\psi_1}} & = \frac{\partial \sigma_t^2}{\partial\alpha} \frac{\partial \alpha}{\partial \theta_{\psi_1}} + \frac{\partial \sigma_t^2}{\partial\beta} \frac{\partial \beta}{\partial \theta_{\psi_1}} \\
    & = \mathcal{D}_1 y_{t-1}^2 + \mathcal{D}_2 \sigma_{t-1}^2 + \beta \frac{\partial \sigma_{t-1}^2}{\partial \theta_{\psi_1}} \\
    \frac{\partial \sigma_t^2}{\partial \theta_{\psi_2}} & = \frac{\partial \sigma_t^2}{\partial\alpha} \frac{\partial \alpha}{\partial \theta_{\psi_2}} + \frac{\partial \sigma_t^2}{\partial\beta} \frac{\partial \beta}{\partial \theta_{\psi_2}} \\
    & = \mathcal{D}_3 y_{t-1}^2 + \mathcal{D}_4 \sigma_{t-1}^2 + \beta \frac{\partial \sigma_{t-1}^2}{\partial \theta_{\psi_2}},
\end{align*}
where 
\begin{align*}
    \mathcal{D}_1 &= \alpha \frac{\exp(-\theta_{\psi_1})}{1+\exp(-\theta_{\psi_1})} \\
    \mathcal{D}_2 &= \beta \frac{\exp(-\theta_{\psi_1})}{1+\exp(-\theta_{\psi_1})} \\
    \mathcal{D}_3 &= \alpha \frac{\exp(-\theta_{\psi_2})}{1+\exp(-\theta_{\psi_2})} \\
    \mathcal{D}_4 &= -\beta \frac{\exp(\theta_{\psi_2})}{1+\exp(\theta_{\psi_2})}. 
\end{align*}

\subsection{Analytical Derivatives of the Log-likelihood Function of t GARCH Model} \label{Appen 2}

The gradient of the log-likelihood function for the t GARCH model is given by 
\begin{align}
    \nabla_\theta \ell(\theta) = \frac{\partial \ell(\theta)}{\partial \theta} =\left(\frac{\partial \ell(\theta)}{\partial \theta_{\phi}},\frac{\partial \ell(\theta)}{\partial \theta_\nu}\right)^\top.
\end{align}
Differentiating the log-likelihood function with respect to $\theta_\phi$, we obtain
\begin{align*}
   \frac{\partial \ell(\theta_{\phi},\theta_{\nu})}{\partial \theta_{\phi}}=\frac{1}{2}\sum_{t=1}^T \frac{1}{\sigma_t^2}\frac{\partial \sigma_t^2}{\partial \theta_\phi}\left[\frac{y_t^2}{\sigma_t^2}\mathcal{A}-1\right],
\end{align*}
where 
$$\mathcal{A}=\frac{\sigma_t^2(\nu+1)}{(v-2)\sigma_t^2+y_t^2}.
$$
Note that the recursive gradient $\frac{\partial \sigma_t^2}{\partial \theta_\phi}$ can be found as in Appendix \ref{Appen 1}.

Differentiating the log-likelihood function with respect to $\theta_\nu$, we obtain
\begin{align*}
   \frac{\partial \ell(\theta_{\phi},\theta_{\nu})}{\partial \theta_{\nu}}&= \frac{\partial \ell(\theta_{\phi},\theta_{\nu})}{\partial {\nu}}\frac{\partial \nu}{\partial \theta_\nu}\\
   \frac{\partial \ell(\theta_{\phi},\theta_{\nu})}{\partial {\nu}} &= \sum_{t=1}^T\Big\{\mathcal{B}+\mathcal{A}\frac{y_t^2}{2(\nu-2)\sigma_t^2}\\&-\frac{1}{2}\log \Big(1+\frac{y_t^2}{(\nu-2)\sigma_t^2}\Big)\Big\}\\
  \frac{\partial \nu}{\partial \theta_\nu} &= \frac{\exp(\theta_\nu)}{1+\exp(\theta_\nu)},
\end{align*}
where 
\begin{align*}
    \mathcal{B}=\frac{1}{2}\Psi\left(\frac{\nu+1}{2}\right)-\frac{1}{2}\Psi\left(\frac{\nu}{2}\right)-\frac{1}{2(\nu-2)},
\end{align*}
and $\Psi(.)$ is defined as the derivative of the logarithmic gamma function.

\subsection{Analytical Derivatives of the Log-likelihood Function of skewed t GARCH Model} \label{Appen 3}
Similar to section \ref{Appen 2}, the analytical derivatives for the log-likelihood function $\nabla_\theta \log p(y\vert \theta)$ of the skewed t GARCH model can be decomposed as the following.
Recall $\phi = (\omega,\alpha,\beta)^\top$ and $\theta_\phi = (\theta_\omega,\theta_\alpha,\theta_\beta)^\top$,
\begin{align}
    \nabla_\theta \log p(y\vert\theta)=\frac{\partial \ell(\theta)}{\partial \theta} = \left(\frac{\partial \ell(\theta)}{\partial \theta_{\phi}},\frac{\partial \ell(\theta)}{\partial \theta_\nu},\frac{\partial \ell(\theta)}{\partial \theta_\xi}\right)^\top.
\end{align}
Differentiating the log-likelihood function with respect to the first part of the unconstrained parameters $\theta_\phi$, it yields,
\begin{align*}
   \frac{\partial l(\theta)}{\partial \theta_{\phi}}=\frac{1}{2}\sum_{t=1}^T \frac{1}{\sigma_t^2}\frac{\partial \sigma_t^2}{\partial \theta_\phi}\left(\mathcal{C}-1\right),
\end{align*}
where $$\mathcal{C}= \frac{s(\nu+1)\left(\frac{sy_t}{\sigma_t}+m\right)y_t}{(\nu-2)\xi^{2I_t}\left(1+\frac{(sz_t+m)^2}{\nu-2}\xi^{-2I_t}\right)\sigma_t}.$$
The recursive sequence $\frac{\partial \sigma_t^2}{\partial \theta_\phi}$ has been computed before and can be found in the Gaussian GARCH model reparametrization trick part in section \ref{Appen 1} which we will not restate here again. We can use the chain rule to derive the gradients of the log-likelihood function with respect to the transformed parameter $\theta_\nu$. We first let $h = (1+\nu)\log\big(1+\frac{\left(s\frac{y_t}{\sigma_t}+m\right)^2}{\nu-2}\xi^{-2I_t}\big)$ such that, 
\begin{align*}
    \frac{\partial \ell(\theta)}{\partial \theta_\nu}  = & \quad \frac{\partial \ell(\theta)}{\partial \nu} \frac{\partial \nu}{\partial \theta_\nu} \nonumber \\
     = & \quad T \bigg( \frac{1}{2}\Psi(\frac{\nu+1}{2})\frac{\partial \nu}{\partial \theta_\nu}-\frac{1}{2}\Psi(\frac{\nu}{2})\frac{\partial \nu}{\partial \theta_\nu} \\ & -\frac{1}{2(\nu-2)}\frac{\partial \nu}{\partial \theta_\nu}+\frac{\partial \log(s)}{\partial \theta_\nu}\bigg)-
     \frac{1}{2}\sum_{t=1}^T\frac{\partial h}{\partial \theta_\nu},
\end{align*}
where $\Psi(.)$ is defined as the logarithmic derivative of the gamma function and 
\begin{align*}
    \frac{\partial \nu}{\partial \theta_\nu} = \frac{\exp(\theta_\nu)}{1+\exp(\theta_\nu)}.
\end{align*}
It is time to compute the partial derivatives of $\log(s)$ and the partial derivatives of the function $h$ both with respect to the unconstrained parameter $\theta_\nu$, 
\begin{align*}
    \frac{\partial \log(s)}{\partial \theta_\nu} & = \frac{\partial \log(s)}{\partial m}\frac{\partial m}{\partial \nu} \frac{\partial \nu}{\partial \theta_\nu} \\
    & = \frac{1}{2}\frac{\partial \log(s^2)}{\partial m}\frac{\partial m}{\partial \nu} \frac{\partial \nu}{\partial \theta_\nu},
\end{align*}
where 
\begin{align*}
    \frac{\partial \log(s^2)}{\partial m} &= \frac{-2m}{(\xi^2+\frac{1}{\xi^2}-1)-m^2}\\
    \label{dm_dv}
    \frac{\partial m}{\partial \nu}&= -\left(\xi-\frac{1}{\xi}\right)\frac{1}{\sqrt{\pi}}\Gamma(\frac{\nu-1}{2}) \\& \frac{\left(\Psi(\frac{\nu}{2})(\nu-2)+\Psi(\frac{\nu-1}{2})(2-\nu)-1\right)}{2\Gamma(\frac{\nu}{2})\sqrt{\nu-2}}\\
    \frac{\partial \nu}{\partial \theta_\nu} &= \frac{\exp(\theta_\nu)}{1+\exp(\theta_\nu)}.
\end{align*}
Let $f=\log\big(1+\frac{\left(s\frac{y_t}{\sigma_t}+m\right)^2}{\nu-2}\xi^{-2I_t}\big)$ and $g=\frac{\left(s\frac{y_t}{\sigma_t}+m\right)^2}{\nu-2}$, the partial derivative of function $f$ with respect to $\nu$ can be derived as the following, 
\begin{align*}
    \frac{\partial f}{\partial \nu} &= \frac{\xi^{-2I_t}\frac{\partial g}{\partial \nu} }{1+\frac{\left(s\frac{y_t}{\sigma_t}+m\right)^2}{\nu-2}\xi^{-2I_t}}.
\end{align*}
The gradients of function $g$ with respect to $\nu$ is, 
\begin{align*}
    \frac{\partial g}{\partial \nu}&= (\nu-2)^{-1}\frac{\partial \left((s\frac{y_t}{\sigma_t}+m^2)\right)}{\partial \nu} \\ & + \left(s\frac{y_t}{\sigma_t}+m\right)^2 \frac{\partial \left((\nu-2)^{-1}\right)}{\partial \nu},
\end{align*}
and the derivatives for both $\frac{\partial \left((s\frac{y_t}{\sigma_t}+m^2)\right)}{\partial \nu}$ and $\frac{\partial \left((\nu-2)^{-1}\right)}{\partial \nu}$ can be computed as,
\begin{align*}
    \frac{\partial \left((\nu-2)^{-1}\right)}{\partial \nu} &= -\frac{1}{(\nu-2)^2} \\
    \frac{\partial \left((s\frac{y_t}{\sigma_t}+m^2)\right)}{\partial \nu} &= 2\left(s\frac{y_t}{\sigma_t}+m\right)\frac{\partial \left(s\frac{y_t}{\sigma_t}+m\right)}{\partial \nu} \\
    & = 2\left(s\frac{y_t}{\sigma_t}+m\right)\left(\frac{y_t}{\sigma_t}\frac{\partial s}{\partial \nu}+\frac{\partial m}{\partial \nu}\right).
\end{align*}
Since $\frac{\partial m}{\partial \nu}$ is stated in equation \ref{dm_dv},  we are only required to find the partial derivatives for $\frac{\partial s}{\partial \nu}$, which is relatively easy to find,
\begin{align*}
    \frac{\partial s}{\partial \nu} &= \frac{1}{2\sqrt{\left(\xi^2+\frac{1}{\xi^2}-1\right)-m^2}} \frac{\partial (\xi^2+\frac{1}{\xi^2}-1-m^2)}{\partial \nu} \\
    & = \frac{1}{2\sqrt{\left(\xi^2+\frac{1}{\xi^2}-1\right)-m^2}} \left(-\frac{\partial (m^2)}{\partial \nu}\right) \\
    & = \frac{1}{2\sqrt{\left(\xi^2+\frac{1}{\xi^2}-1\right)-m^2}} \left(-2m\frac{\partial m}{\partial \nu}\right)\\
    & = -\frac{1}{\sqrt{\left(\xi^2+\frac{1}{\xi^2}-1\right)-m^2}} m\frac{\partial m}{\partial \nu}.
\end{align*}
Since function $h = (1+\nu)\log\big(1+\frac{\left(s\frac{y_t}{\sigma_t}+m\right)^2}{\nu-2}\xi^{-2I_t}\big)$, the first derivative with respect to $\nu$ can be computed using the chain rule,
\begin{align*}
    \frac{\partial h}{\partial \nu} &= (1+\nu)\frac{\partial f}{\partial \nu}+f
\end{align*}
where $f=\log\big(1+\frac{\left(s\frac{y_t}{\sigma_t}+m\right)^2}{\nu-2}\xi^{-2I_t}\big)$ and the first derivative with respect to the transformed parameter $\theta_\xi$ is,
\begin{align*}
    \frac{\partial h}{\partial \theta_\nu} = \frac{\partial h}{\partial \nu} \frac{\partial \nu}{\partial \theta_\nu},
\end{align*}
where we have worked out all the necessary parts above. \\
\\
The gradients of the log-likelihood function with respect to the transformed parameter $\theta_\xi$ can be calculated in a similar way. Since we define $h = (1+\nu)\log\big(1+\frac{\left(s\frac{y_t}{\sigma_t}+m\right)^2}{\nu-2}\xi^{-2I_t}\big)$ such that, 
\begin{align*}
    \frac{\partial \ell(\theta)}{\partial \theta_\xi}  = & \quad \frac{\partial \ell(\theta)}{\partial \xi} \frac{\partial \xi}{\partial \theta_\xi} \nonumber \\
     = & \quad T \left(-\frac{\xi^2-1}{\xi^3+\xi}\frac{\partial \xi}{\partial \theta_\xi} +\frac{\partial \log(s)}{\partial \theta_\xi}\right)-
     \frac{1}{2}\sum_{t=1}^T\frac{\partial h}{\partial \theta_\xi},
\end{align*}
where $$\frac{\partial \xi}{\partial \theta_\xi} = \frac{\exp(\theta_\xi)}{1+\exp(\theta_\xi)}.$$ Differentiating $\log(s)$ with the unrestricted skewness parameters $\theta_\xi$, we have, 
\begin{align*}
    \frac{\partial \log(s)}{\partial \theta_\xi} & = \frac{\partial \log(s)}{\partial \xi}\frac{\partial \xi}{\partial \theta_\xi} \\
    & = \frac{1}{2}\frac{\partial \log(s^2)}{\partial \xi}\frac{\partial \xi}{\partial \theta_\xi} \\
    & = \frac{1}{2}\frac{\partial \log\left((\xi^2+\frac{1}{\xi^2}-1)-m^2\right)}{\partial \xi}\frac{\partial \xi}{\partial \theta_\xi} \\
    & = \frac{1}{2}\left( \frac{2\xi-\frac{2}{\xi^3}-\frac{\partial (m^2)}{\partial \xi}}{(\xi^2+\frac{1}{\xi^2}-1)-m^2}\right)\frac{\partial \xi}{\partial \theta_\xi}.
\end{align*}
With some further differentiation, we can get, 
\begin{align}
    \frac{\partial (m^2)}{\partial \xi}&=2m\frac{\partial m}{\partial \xi} \\
    \frac{\partial m}{\partial \xi} &= \left(1+\frac{1}{\xi^2}\right)\frac{\Gamma(\frac{\nu-1}{2})\sqrt{\nu-2}}{\sqrt{\pi}\Gamma(\frac{\nu}{2})}.
    \label{dmdxi}
\end{align}
Now, we are trying to find the partial derivative of the function $f=\log\big(1+\frac{\left(s\frac{y_t}{\sigma_t}+m\right)^2}{\nu-2}\xi^{-2I_t}\big)$ with respect to $\theta_\xi$,
\begin{align}
    \label{eqn:dfdthetaxi}
    \frac{\partial f}{\partial \theta_\xi} & = \frac{\partial f}{\partial \xi}\frac{\partial \xi}{\partial \theta_\xi},
\end{align}
where 
\begin{align*}
    \frac{\partial \xi}{\partial \theta_\xi} &= \frac{\exp(\theta_\xi)}{1+\exp(\theta_\xi)}\\
    \frac{\partial f}{\partial \xi} &= \frac{(\nu-2)^{-1}\frac{\partial }{\partial \xi}\left(\left(s\frac{y_t}{\sigma_t}+m\right)^2\xi^{-2I_t}\right) }{1+\frac{\left(s\frac{y_t}{\sigma_t}+m\right)^2}{\nu-2}\xi^{-2I_t}}
\end{align*}
We first let $r=\left(s\frac{y_t}{\sigma_t}+m\right)^2\xi^{-2I_t}$, hence,
\begin{align*}
    \frac{\partial r}{\partial \xi} & = \frac{\partial \left(s\frac{y_t}{\sigma_t}+m\right)^2}{\partial \xi}\xi^{-2I_t}+\left(s\frac{y_t}{\sigma_t}+m\right)^2\frac{\partial (\xi^{-2I_t})}{\partial \xi} \\
    & = 2\left(s\frac{y_t}{\sigma_t}+m\right)\left(\frac{y_t}{\sigma_t}\frac{\partial s}{\partial \xi}+\frac{\partial m}{\partial \xi}\right)\xi^{-2I_t} \\ & \enskip+\left(s\frac{y_t}{\sigma_t}+m\right)^2 \frac{\partial (\xi^{-2I_t})}{\partial \xi} \\
    & = 2\left(s\frac{y_t}{\sigma_t}+m\right)\left(\frac{y_t}{\sigma_t}\frac{\partial s}{\partial \xi}+\frac{\partial m}{\partial \xi}\right)\xi^{-2I_t} \\ & \enskip -\left(s\frac{y_t}{\sigma_t}+m\right)^2 \left(2I_t\xi^{-2I_t-1}\right)
\end{align*}
The mathematical derivation of $\frac{\partial m}{\partial \xi}$ is stated in equation \ref{dmdxi}, thus, it is only necessary to find the partial derivatives for $\frac{\partial s}{\partial \xi}$, 
\begin{align*}
    \frac{\partial s}{\partial \xi} &= \frac{1}{2\sqrt{\left(\xi^2+\frac{1}{\xi^2}-1\right)-m^2}} \frac{\partial (\xi^2+\frac{1}{\xi^2}-1-m^2)}{\partial \xi} \\
    &= \frac{1}{2\sqrt{\left(\xi^2+\frac{1}{\xi^2}-1\right)-m^2}} \left(2\xi-\frac{2}{\xi^3}-2m\frac{\partial m}{\partial \xi}\right).
\end{align*}
Now we have all the parts worked out to compute the partial derivative of function $f$ with respect to $\theta_\xi$ in equation \ref{eqn:dfdthetaxi}. \\
\\
Since $h = (1+\nu)\log\big(1+\frac{\left(s\frac{y_t}{\sigma_t}+m\right)^2}{\nu-2}\xi^{-2I_t}\big)$, the first derivative with respect to $\xi$ is computed as, 
\begin{align*}
    \frac{\partial h}{\partial \xi} &= (1+\nu)\frac{\partial f}{\partial \xi},
\end{align*}
and the first derivative with respect to the transformed parameter $\theta_\xi$ is the following, 
\begin{align*}
    \frac{\partial h}{\partial \theta_\xi} = \frac{\partial h}{\partial \xi} \frac{\partial \xi}{\partial \theta_\xi}.
\end{align*}
Therefore, the reparametrization model-specific gradient, $$\nabla_\theta h_\lambda(\theta) =\nabla_\theta \log p(\theta) + \nabla_\theta \log p(y\vert \theta) - \nabla_\theta\log q_\lambda(\theta),$$  can be fully computed now.  

\section{Additional Figures \& Tables}\label{secB1}

Tables~\ref{tab:accuracy-gaussian}-\ref{tab:accuracy-skewed-t} present the accuracy results for the parameters of the Gaussian, t, and skewed t GARCH model, respectively. The accuracy is measured as a percentage and higher values indicate a closer approximation to the true values, which are reported for both the control variates and reparametrization trick 
approaches with different numbers of Monte Carlo samples. The table is divided into two sections based on the number of observations, 1000 and 5000. Please note that the accuracy results in the table represent the mean accuracy over multiple simulations or datasets. 

\begin{figure*}[htbp]
  \centering
  \begin{subfigure}[t]{0.24\textwidth}
    \centering
    \includegraphics[width=\textwidth]{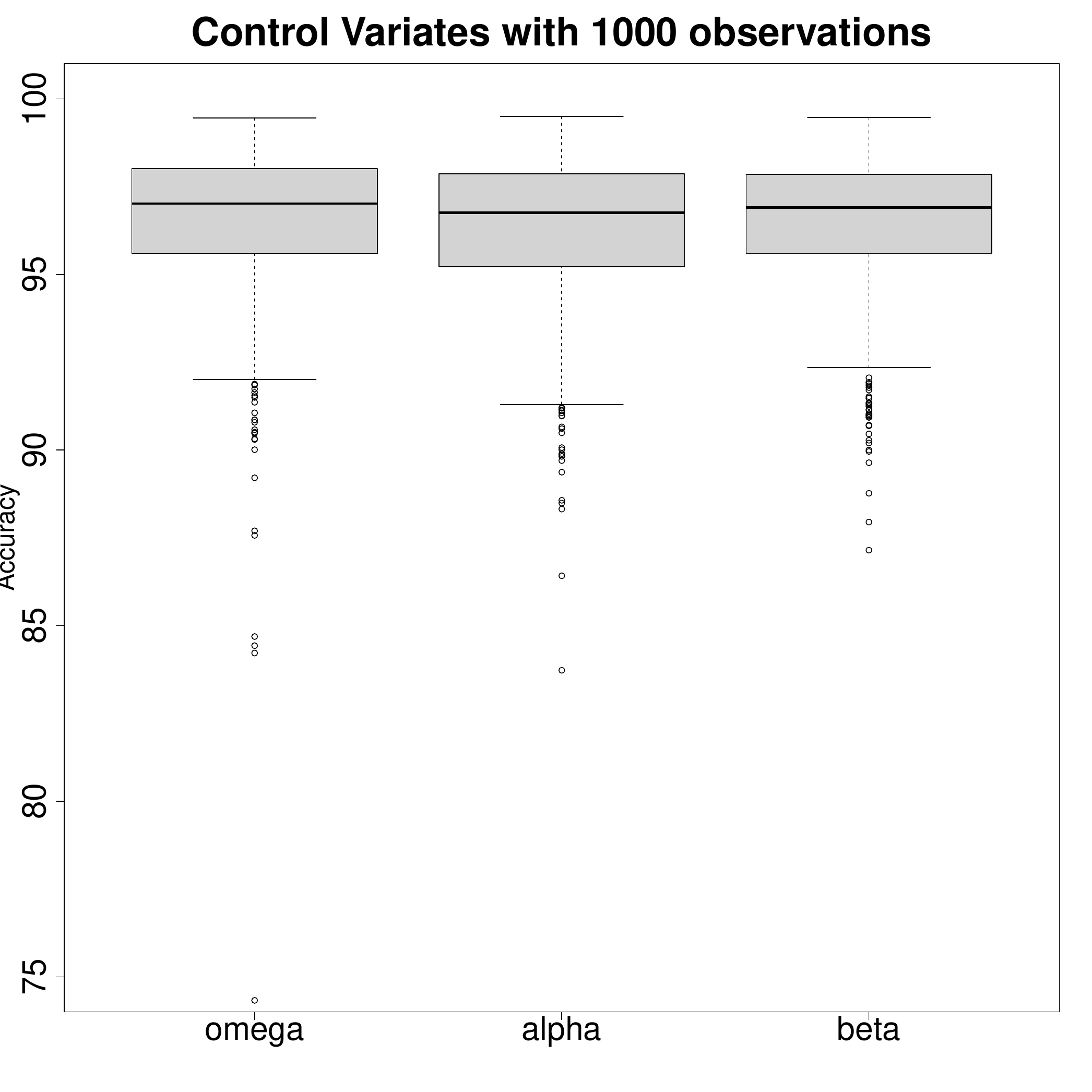}
    \label{fig:plot1}
  \end{subfigure}
  \hfill
  \begin{subfigure}[t]{0.24\textwidth}
    \centering
    \includegraphics[width=\textwidth]{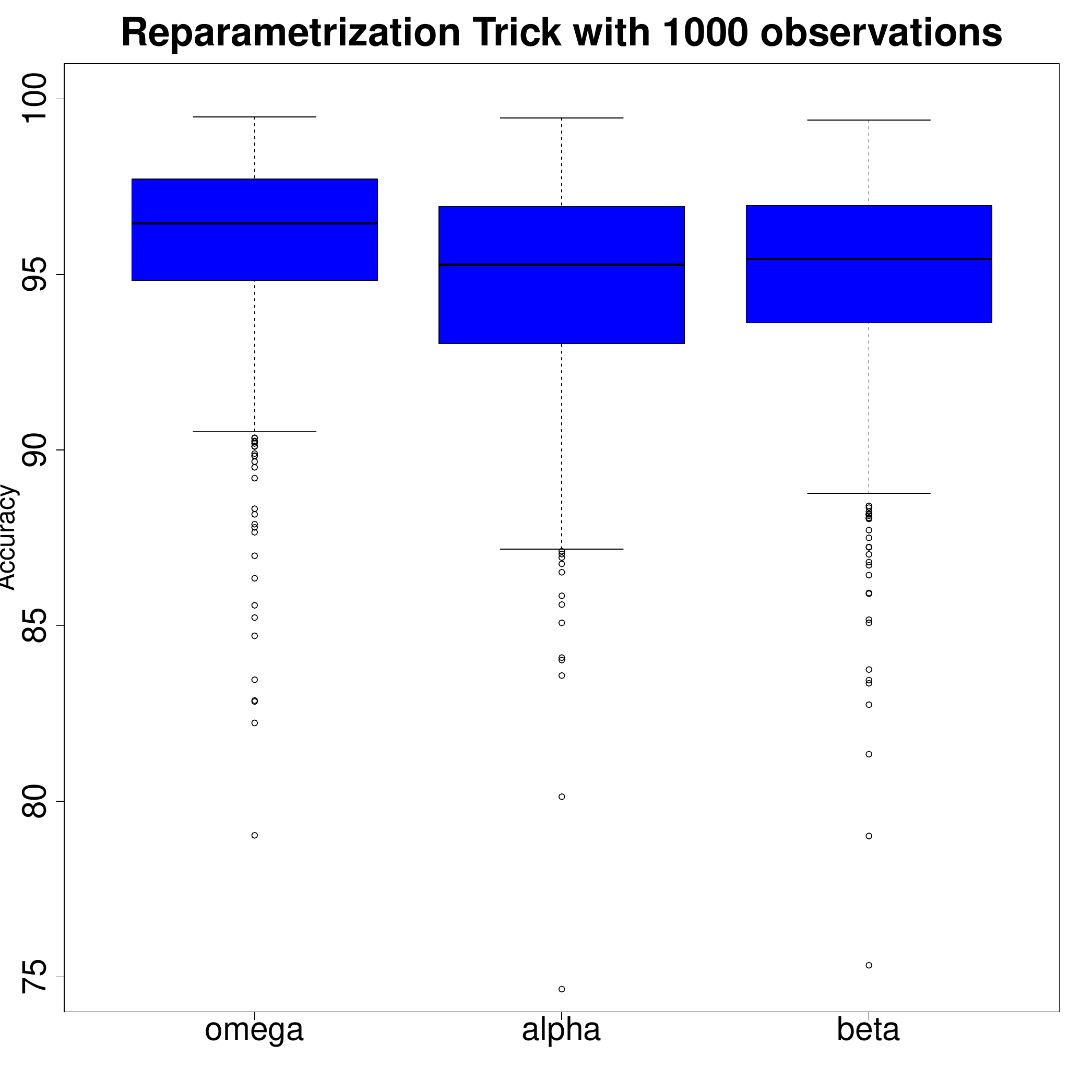}
    \label{fig:plot2}
  \end{subfigure}
  \hfill
  \begin{subfigure}[t]{0.24\textwidth}
    \centering
    \includegraphics[width=\textwidth]{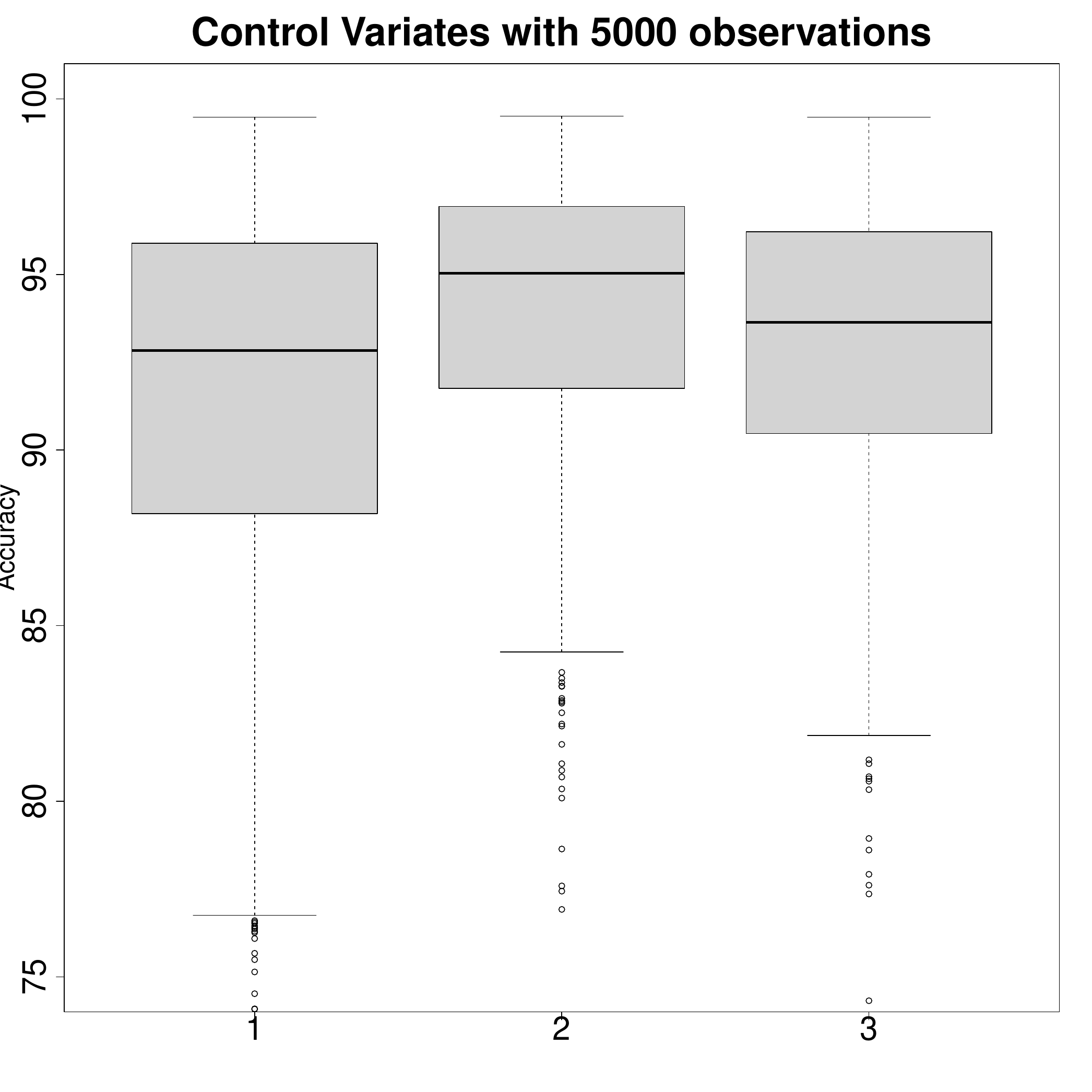}
    \label{fig:plot2}
  \end{subfigure}
  \hfill
  \begin{subfigure}[t]{0.24\textwidth}
    \centering
    \includegraphics[width=\textwidth]{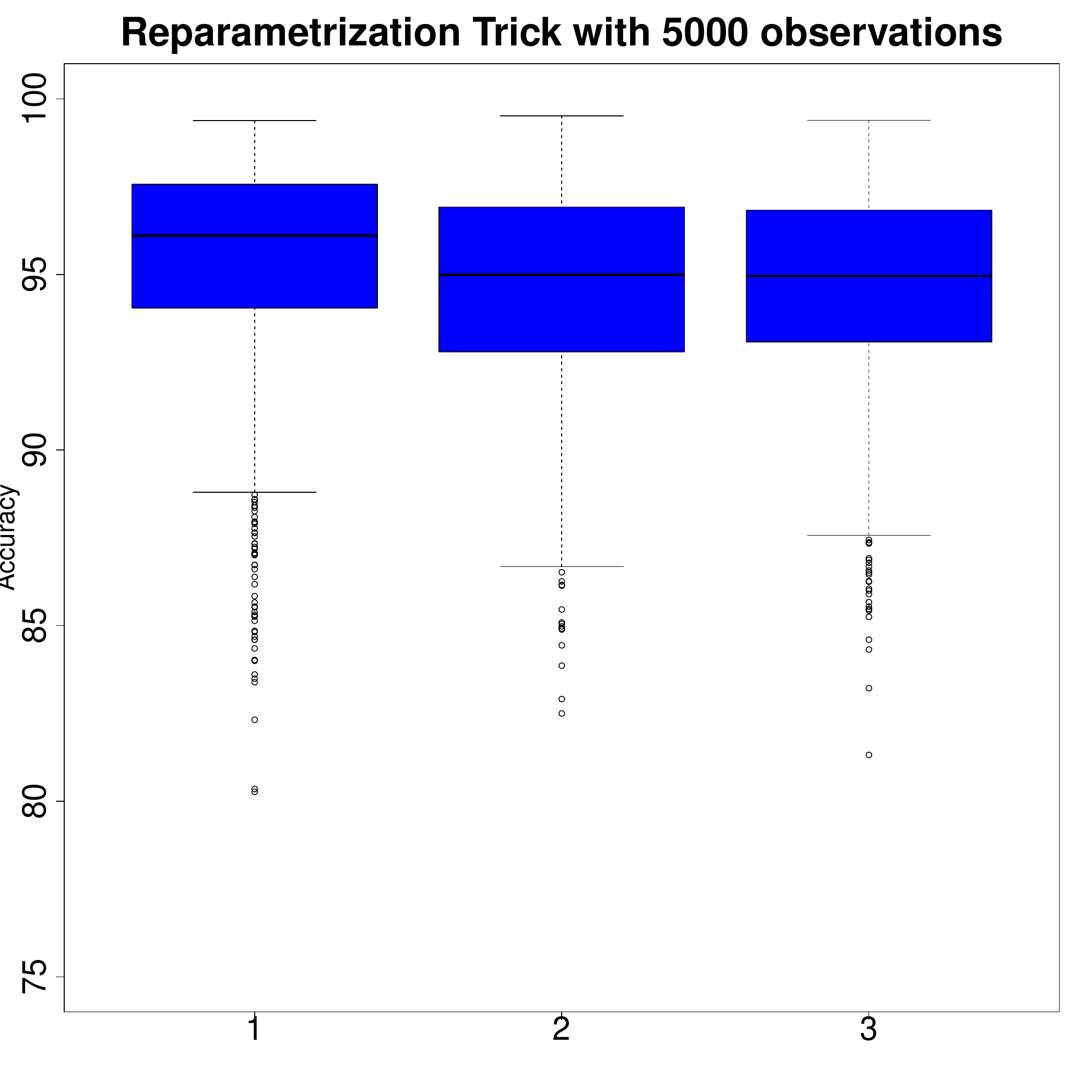}
    \label{fig:plot2}
  \end{subfigure}
  \caption{Boxplots for the accuracy results of the control variates and reparametrization trick methods for the Gaussian GARCH model, with 1,000 and 5,000 observations, based on 1,000 replications.}
  \label{fig:Boxplot Gaussian GARCH}
\end{figure*}

\begin{figure*}[htbp]
  \centering
  \begin{subfigure}[t]{0.24\textwidth}
    \centering
    \includegraphics[width=\textwidth]{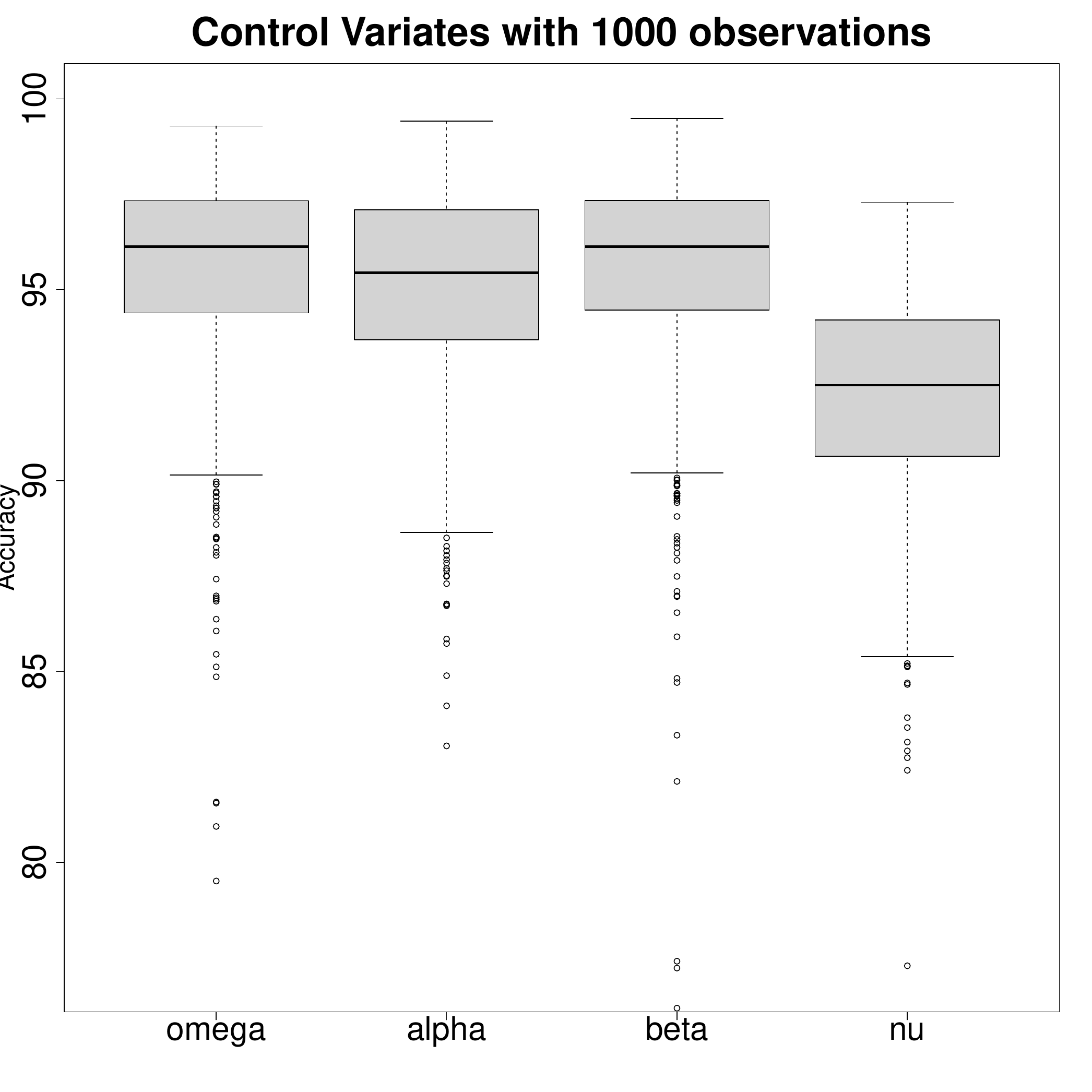}
    \label{fig:plot1}
  \end{subfigure}
  \hfill
  \begin{subfigure}[t]{0.24\textwidth}
    \centering
    \includegraphics[width=\textwidth]{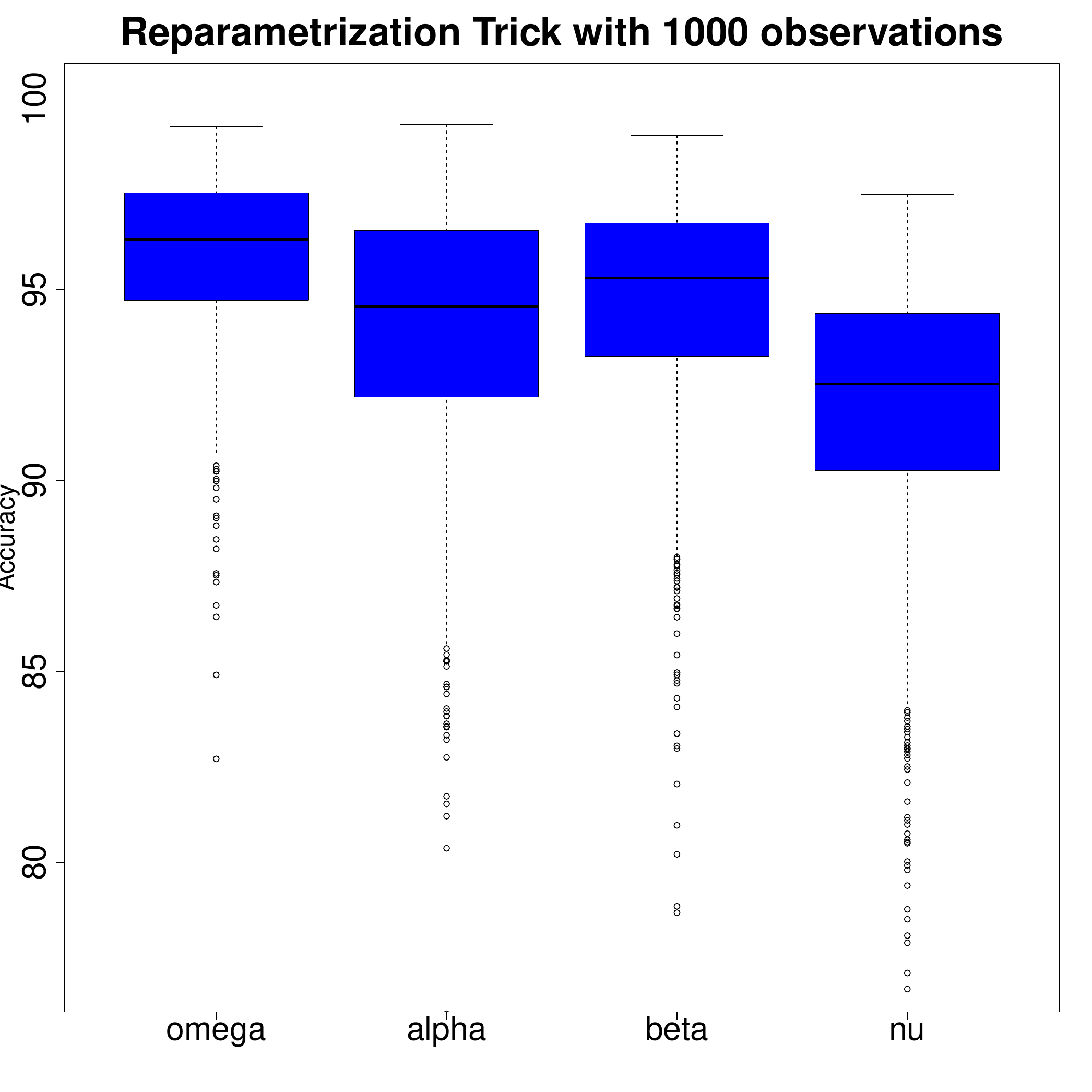}
    \label{fig:plot2}
  \end{subfigure}
  \hfill
  \begin{subfigure}[t]{0.24\textwidth}
    \centering
    \includegraphics[width=\textwidth]{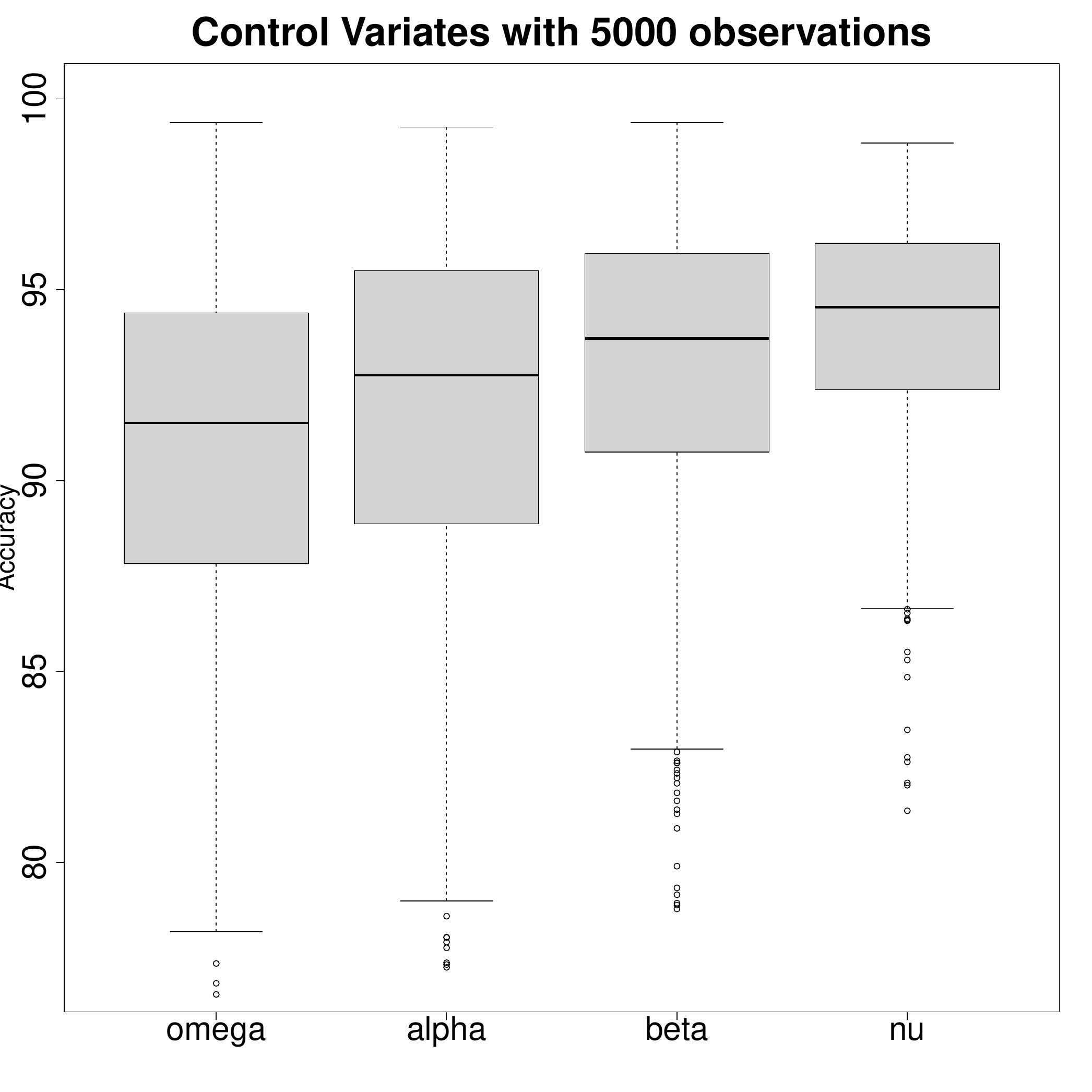}
    \label{fig:plot2}
  \end{subfigure}
  \hfill
  \begin{subfigure}[t]{0.24\textwidth}
    \centering
    \includegraphics[width=\textwidth]{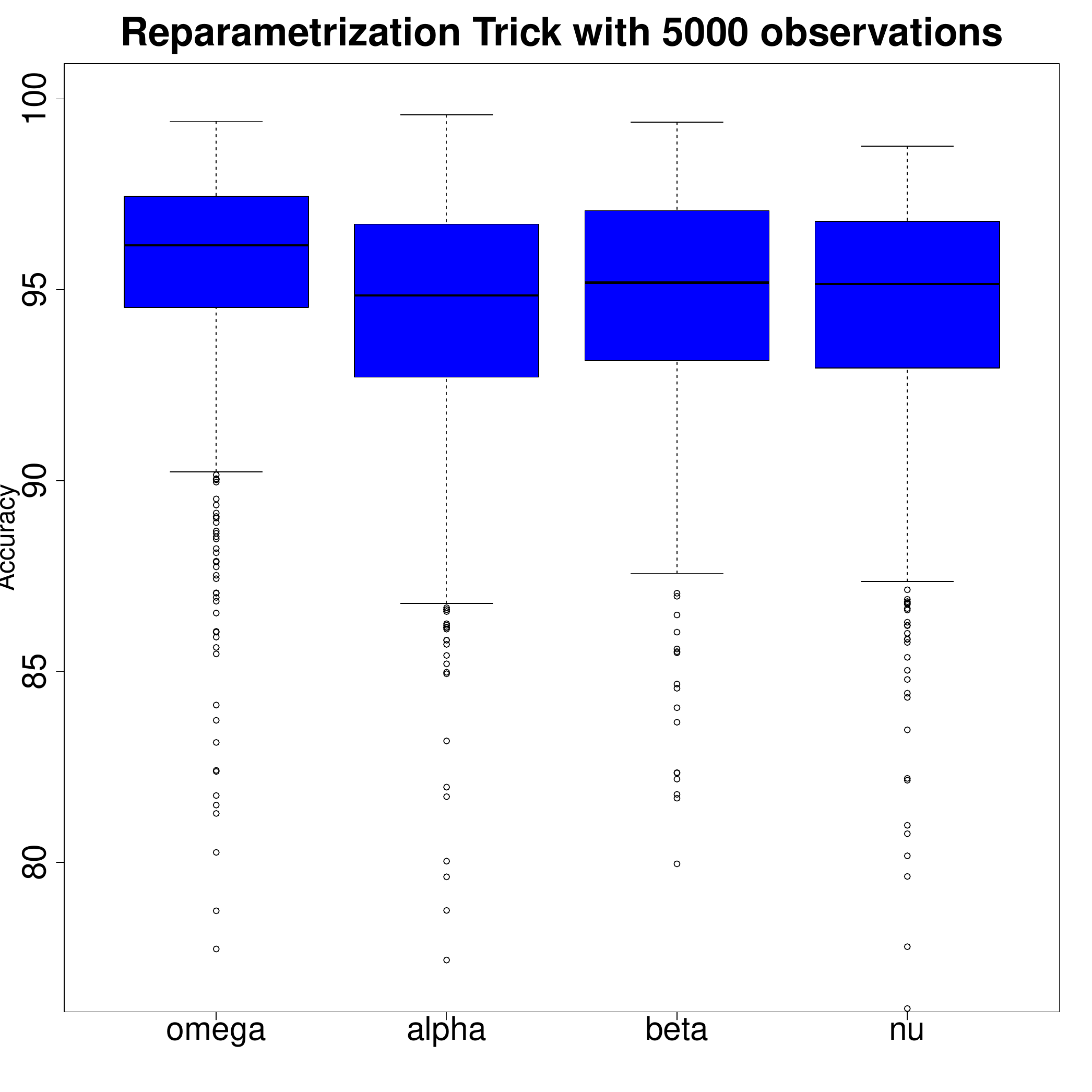}
    \label{fig:plot2}
  \end{subfigure}
  \caption{Boxplots for the accuracy results of the control variates and reparametrization trick methods for the t GARCH model, with 1,000 and 5,000 observations, based on 1,000 replications.}
  \label{fig:Boxplot t GARCH}
\end{figure*}

\begin{figure*}[htbp]
  \centering
  \begin{subfigure}[t]{0.24\textwidth}
    \centering
    \includegraphics[width=\textwidth]{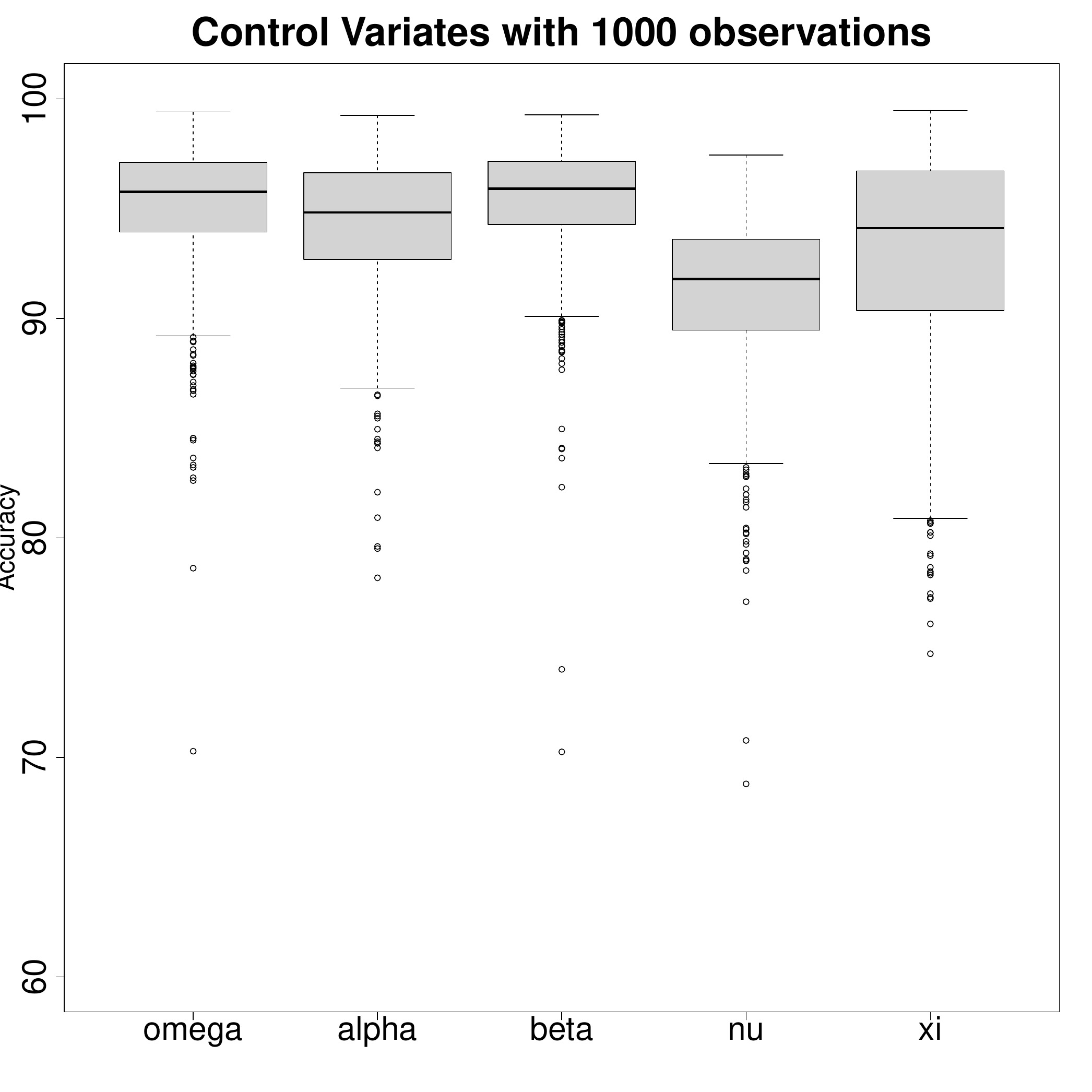}
    \label{fig:plot1}
  \end{subfigure}
  \hfill
  \begin{subfigure}[t]{0.24\textwidth}
    \centering
    \includegraphics[width=\textwidth]{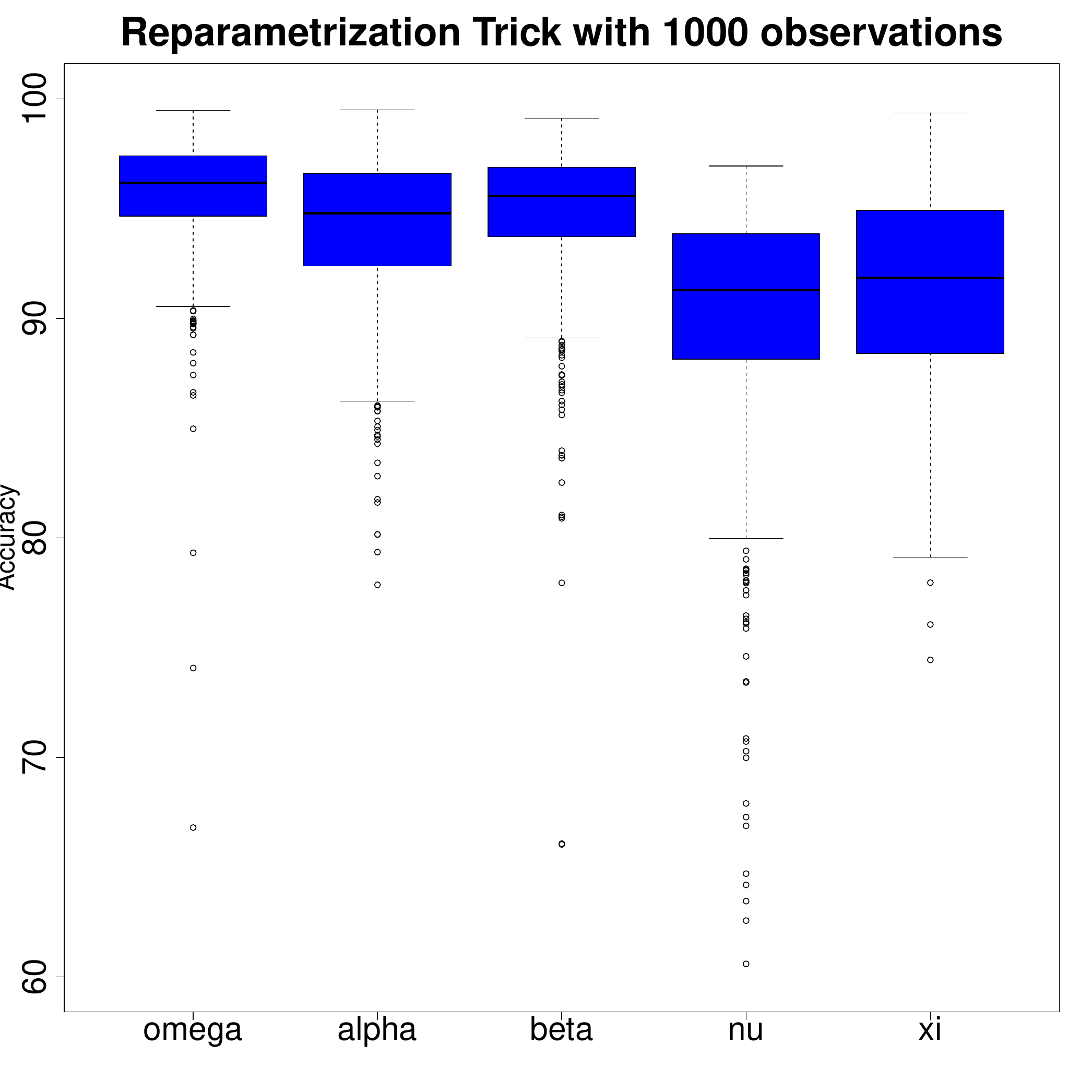}
    \label{fig:plot2}
  \end{subfigure}
  \hfill
  \begin{subfigure}[t]{0.24\textwidth}
    \centering
    \includegraphics[width=\textwidth]{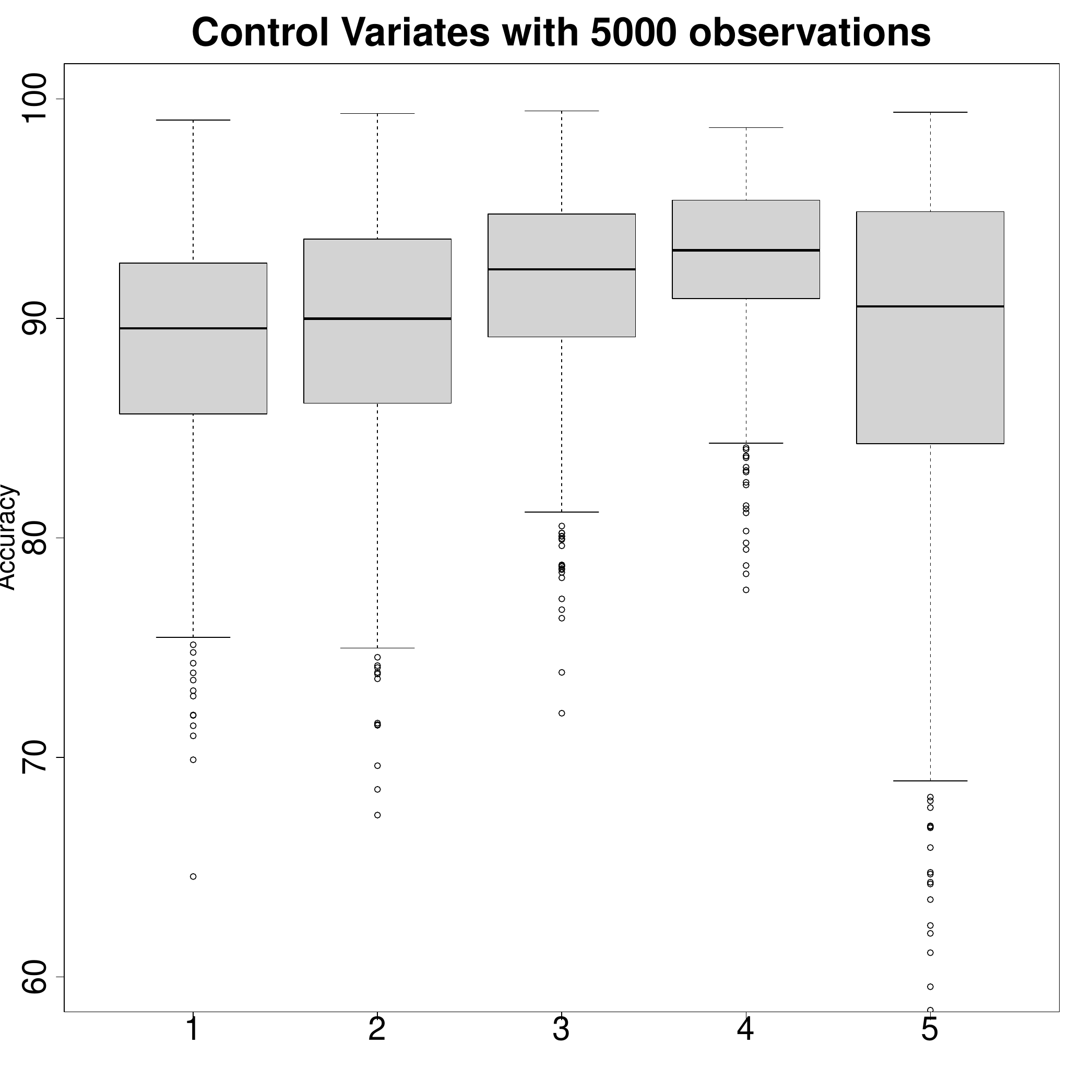}
    \label{fig:plot2}
  \end{subfigure}
  \hfill
  \begin{subfigure}[t]{0.24\textwidth}
    \centering
    \includegraphics[width=\textwidth]{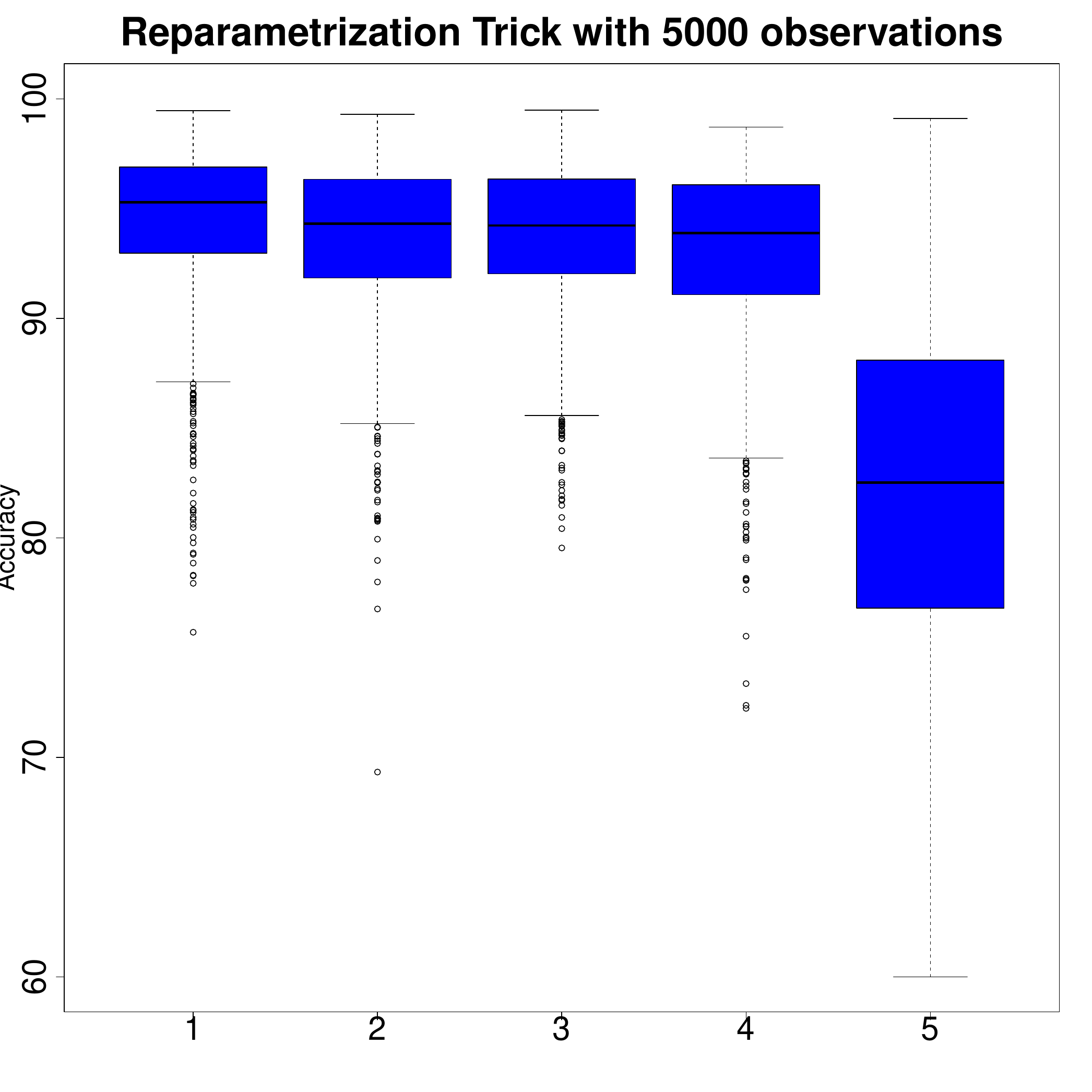}
    \label{fig:plot2}
  \end{subfigure}
  \caption{Boxplots for the accuracy results of the control variates and reparametrization trick methods for the Skewed T GARCH model, with 1,000 and 5,000 observations, based on 1,000 replications.}
  \label{fig:Boxplot Skewed T GARCH}
\end{figure*}

Figure \ref{fig:Boxplot Gaussian GARCH} - \ref{fig:Boxplot Skewed T GARCH} present boxplots comparing the accuracy results of SVB algorithms for a GARCH(1,1) model with different innovations.

\begin{table}[h]
\centering
\begin{tabular}{cccc}
\hline
& $\omega$ & $\alpha$ & $\beta$ \\
\hline
 &\multicolumn{3}{c}{1000 Observations} \\
CV S=5 & 94.67 & 94.82 & 94.92 \\
CV S=10 & 96.62 & 96.35 & 96.52 \\
CV S=50 & 98.09 & 97.53 & 97.57 \\
RT S=5 & 95.94 & 94.75 & 94.99 \\
RT S=10 & 96.43 & 95.31 & 95.63 \\
RT S=50 & 97.15 & 95.98 & 96.24 \\

 &\multicolumn{3}{c}{5000 Observations} \\
CV S=5 & 83.58 & 90.57 & 88.66 \\
CV S=10 & 91.04 & 94.08 & 92.99 \\
CV S=50 & 98.08 & 97.72 & 97.97 \\
RT S=5 & 95.34 & 94.65 & 94.61 \\
RT S=10 & 96.46 & 95.77 & 95.86 \\
RT S=50 & 97.68 & 97.38 & 97.24 \\
\hline
\end{tabular}
\caption{Average accuracy for the parameters of the Gaussian GARCH model. The accuracy results are reported for both the control variates and reparametrization trick approaches with different numbers of Monte Carlo samples and different time lengths.}
\label{tab:accuracy-gaussian}
\end{table}

\begin{table}[!hbt]
\centering
\begin{tabular}{ccccc}
\hline
& $\omega$ & $\alpha$ & $\beta$ & $\nu$ \\
\hline
&\multicolumn{4}{c}{1000 Observations} \\
CV $S=5$ & 93.73 & 93.73 & 94.24 & 91.70 \\
CV $S=10$ & 95.57 & 95.12 & 95.59 & 92.23 \\
CV $S=50$ & 97.35 & 96.21 & 96.78 & 92.40 \\
RT $S=5$ & 95.92 & 93.98 & 94.67 & 91.81 \\
RT $S=10$ & 96.15 & 94.14 & 95.03 & 91.71 \\
RT $S=50$ & 96.50 & 93.81 & 95.54 & 91.32 \\
\hline
&\multicolumn{4}{c}{5000 Observations}\\
CV $S=5$ & 86.32 & 87.21 & 89.92 & 91.73 \\
CV $S=10$ & 90.88 & 91.79 & 93.05 & 94.07 \\
CV $S=50$ & 97.68 & 96.54 & 97.57 & 95.48 \\
RT $S=5$ & 95.64 & 94.36 & 94.81 & 94.42 \\
RT $S=10$ & 96.88 & 94.80 & 96.15 & 94.67 \\
RT $S=50$ & 98.05 & 95.75 & 97.36 & 94.91 \\
\hline
\end{tabular}
\caption{Average accuracy for the parameters of the t GARCH model.}
\label{tab:accuracy-student-t}
\end{table}

\begin{table}[!hbt]
\centering
\begin{tabular}{cccccc}
\hline
& $\omega$ & $\alpha$ & $\beta$ & $\nu$ & $\xi$ \\
\hline
&\multicolumn{5}{c}{1000 Observations} \\
CV $S=5$ & 92.90 & 92.71 & 93.82 & 90.60 & 90.76 \\
CV $S=10$ & 95.20 & 94.42 & 95.40 & 91.22 & 93.04 \\
CV $S=50$ & 97.35 & 95.88 & 96.76 & 91.71 & 96.02 \\
RT $S=5$ & 95.73 & 94.17 & 94.86 & 90.31 & 91.32 \\
RT $S=10$ & 96.19 & 94.19 & 95.52 & 90.07 & 94.12 \\
RT $S=50$ & 96.70 & 93.97 & 96.02 & 90.10 & 96.94 \\
\hline
&\multicolumn{5}{c}{5000 Observations} \\
CV $S=5$ & 84.62 & 85.21 & 88.56 & 90.41 & 84.13 \\
CV $S=10$ & 88.80 & 89.49 & 91.69 & 92.78 & 88.70 \\
CV $S=50$ & 97.52 & 96.28 & 97.38 & 94.91 & 95.13 \\
RT $S=5$ & 94.41 & 93.65 & 93.77 & 93.02 & 82.34 \\
RT $S=10$ & 96.32 & 94.64 & 95.57 & 93.85 & 88.77 \\
RT $S=50$ & 98.03 & 95.76 & 97.57 & 94.29 & 95.76 \\
\hline
\end{tabular}
\caption{Average accuracy for the parameters of the Skewed t GARCH model.}
\label{tab:accuracy-skewed-t}
\end{table}

%%=============================================%%
%% For submissions to Nature Portfolio Journals %%
%% please use the heading ``Extended Data''.   %%
%%=============================================%%

%%=============================================================%%
%% Sample for another appendix section			       %%
%%=============================================================%%

%% \section{Example of another appendix section}\label{secA2}%
%% Appendices may be used for helpful, supporting or essential material that would otherwise 
%% clutter, break up or be distracting to the text. Appendices can consist of sections, figures, 
%% tables and equations etc.

\end{appendices}

%%===========================================================================================%%
%% If you are submitting to one of the Nature Portfolio journals, using the eJP submission   %%
%% system, please include the references within the manuscript file itself. You may do this  %%
%% by copying the reference list from your .bbl file, paste it into the main manuscript .tex %%
%% file, and delete the associated \verb+\bibliography+ commands.                            %%
%%===========================================================================================%%
\bibliography{sn-bibliography}% common bib file
%% if required, the content of .bbl file can be included here once bbl is generated
%%\input sn-article.bbl

%% Default %%
%%\input sn-sample-bib.tex%

\end{document}